\documentclass{naturemodified}

\usepackage[pdftex]{graphicx,hyperref}
\hypersetup{
  colorlinks=true,
  linkcolor=red,
  citecolor=green,
  urlcolor=blue
}
\usepackage{verbatim}
\usepackage{amsbsy}
\usepackage{amssymb}
\usepackage{amsmath}
\usepackage{color}
\usepackage{bm}
\usepackage{marvosym}
\usepackage{amsfonts}
\usepackage{ulem}
\usepackage{anysize}
\usepackage{geometry}
\usepackage[title,titletoc]{appendix}
\usepackage{lipsum}

\renewcommand{\thefigure}{\arabic{figure}}
\renewcommand{\figurename}{\textbf{Figure}}

\newgeometry{left=2.3cm,right=2.3cm,top=2.2cm,bottom=2.2cm}

\usepackage{multibbl}
\newbibliography{refMain}
\newbibliography{refSupp}

\title{Tunable room-temperature ferromagnet using an iron-oxide and graphene oxide nanocomposite}

\author{Aigu L. Lin$^{\dagger 1, 2, 3 \ddagger}$, J. N. B. Rodrigues$^{\dagger 2, 3}$, Chenliang Su$^{2, 4}$, M. Milletari$^{2,3}$, Kian 
Ping Loh$^{2, 4}$, Tom Wu$^{5}$, Wei Chen$^{2, 4}$, A. H. Castro Neto$^{2,3}$, Shaffique Adam$^{2, 3, 6, \star}$ and Andrew T. S. Wee$^{1,2,3}$}

\begin{document}

\maketitle

\begin{affiliations}
\item NUS Graduate School of Integrative Sciences and Engineering,
  National University of Singapore, 28 Medical Drive, Singapore 117456.
\item Graphene Research Centre, Faculty of Science, National
  University of Singapore, 6 Science Drive 2, Singapore 117546.
\item Department of Physics, Faculty of Science, National University
  of Singapore, 2 Science Drive 3, Singapore 117542.
\item Department of Chemistry, Faculty of Science, National University
  of Singapore, 3 Science Drive 3, Singapore 117543.
\item Materials Science and Engineering, King Abdullah University of
  Science and Technology (KAUST), Thuwal, 23955-6900, Saudi Arabia.
\item Yale-NUS College, 6 College Avenue East, 138614, Singapore.
\end{affiliations}
\vspace{-0.39in}

\vspace{0.3cm}

\noindent $^{\dagger}$ These authors contributed equally to this
paper.\\
\noindent $^{\ddagger}$ Present address: Bioengineering Core, Division of Research, Singapore General Hospital, Outram Road, Singapore 169608.\\ 

\vspace{-0.02in}

\vspace{0.3cm}

\begin{abstract}
  Magnetic materials have found wide application ranging from
  electronics and memories to medicine. Essential to these advances
  is the control of the magnetic order. To date, most
  room-temperature applications have a fixed magnetic moment whose
  orientation is manipulated for functionality. Here we demonstrate an
  iron-oxide and graphene oxide nanocomposite based device that acts
  as a tunable ferromagnet at room temperature. Not only can we tune
  its transition temperature in a wide range of temperatures around room
  temperature, but the magnetization can also be tuned from zero to
  0.011 A m$^{2}/$kg through an initialization process with two
  readily accessible knobs (magnetic field and electric current),
  after which the system retains its magnetic properties
  semi-permanently until the next initialization process. We construct
  a theoretical model to illustrate that this tunability originates
  from an indirect exchange interaction mediated by spin-imbalanced
  electrons inside the nanocomposite.
\end{abstract}

Manipulating the properties of a ferromagnet by means other than a magnetic field has had tremendous impact on technology.  
The most prominent example of this is the spin-transfer torque mechanism predicted by Slonczewski\cite{refMain}{Slonczewski_JMMM:1996}
and by Berger\cite{refMain}{Berger_PRB:1996}, 
in which a spin-polarized electrical current transfers angular momentum to the ferromagnet and switches its orientation.  Magnetic memory 
based on this mechanism (ST-MRAM) is already commercially available, is non-volatile, has better energy efficiency, and is more readily 
scalable to smaller devices than most conventional memory\cite{refMain}{Ralph_JMMM:2008}.  Another example is magnetoelectric (or multiferroic) 
materials\cite{refMain}{Fiebig_JPD:2005,Eerenstein_Nature:2006,Vaz_JPCM:2012} where the magnitude of the magnetization can be controlled by an electric 
field.  For room temperature operation, magnetoelectric materials are made by engineering heterostructures combining ferroelectric and 
ferromagnetic materials that are coupled by strain at their interface.  Such materials could also have application in low latency memory. 
Yet another promising mechanism to control ferromagnetic properties include controlling the transition temperature of thin ferromagnetic films using 
an electric field\cite{refMain}{Weisheit_Science:2007,Chiba_NatMaterials:2011,Chiba_JPD:2013}  -- this exploits the sensitivity of magnetic properties 
to the electronic carrier density tuned by the field effect.

In this work, we report on a nanocomposite material that allows for its magnetic properties to be controlled in a new way. The material (discussed 
in detail below) is a nanocomposite of graphene oxide and iron-oxide nanoparticles. We show that using an initialization procedure involving a 
magnetic field and a spin-polarized electric current, we can controllably set the magnetic moment and transition temperature of the ferromagnet that 
then remains stable even after the current and magnetic fields are switched off. Operating at room temperature, this gives an example of a system 
where the magnetism itself can be switched on or off depending on the current and magnetic fields that are applied during the initialization step. 
The mechanism relies on an electron spin-imbalance generated during initialization, that gives rise to an electron mediated ferromagnetic coupling between 
the iron nanoparticles. Starting from a simplified microscopic Hamiltonian, we show theoretically that the coupling is indeed ferromagnetic, 
and provide Monte Carlo simulations for the dependence of the transition temperature on spin-imbalance that is consistent with experimental observations. 
This ability to electrically turn on and off the magnetization might enable applications in nonvolatile memories with novel operation modes and using 
easy processable materials, as well as hybrid devices integrating tunable electric and magnetic components. 

\begin{figure}[htp!]
  \centering
  \includegraphics[width=0.48\columnwidth]{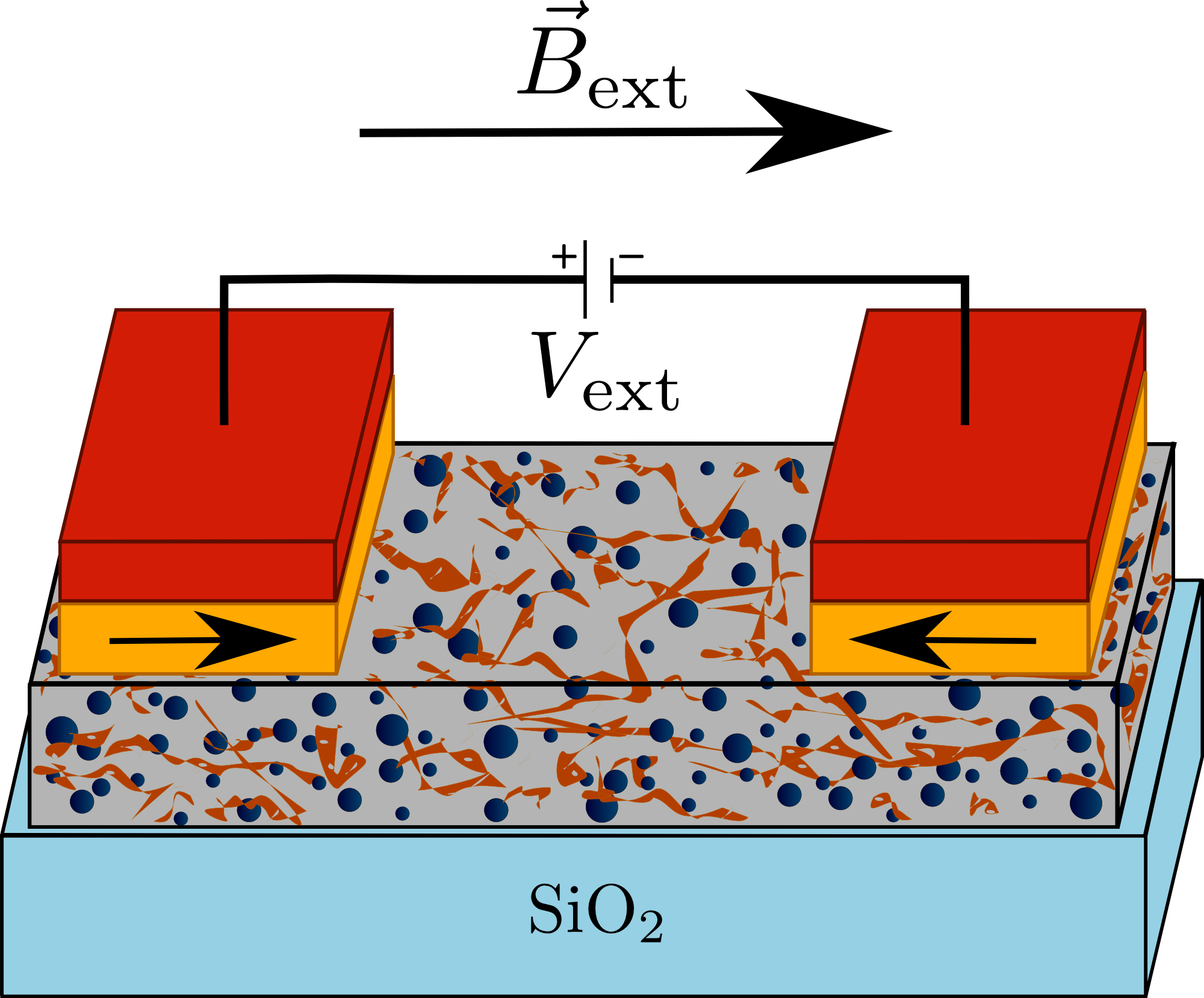}
  \caption{{\bf Schematic of the device geometry and nanocomposite
      composition.} The gray box represents the nanocomposite, with
    the blue spheres representing the iron-oxide nanoparticles and the
    brown strips representing the highly defective graphene oxide
    layers. The nanocomposite's thin film is deposited on top of a
    silicon dioxide substrate (in light blue). Two cobalt
    ferromagnetic electrodes (yellow) are placed on top of the
    nanocomposite. For zero applied magnetic field, these are pinned
    in an anti-parallel configuration by PtMn layers (in red). }
  \label{fig:DeviceScheme}
\end{figure}

The device consists of a nanocomposite of partially reduced (between $18\%$ and $20\%$) and highly defective 
graphene oxide\cite{refMain}{Su_NatComm:2012} mixed up with iron-oxide 
(FeO/Fe$_{3}$O$_{4}$ complex) core shell structure nanoparticles to which one attaches two pinned ferromagnetic cobalt 
electrodes whose configuration is driven by an external magnetic field (see Fig. \ref{fig:DeviceScheme}). The nanoparticles are in a 
canted ferrimagnetic alpha-phase and carry magnetic moments of approximately 3 to 5 $\mu_{B}$ (and 
typical diameter of $6.5$-$9.5$ nm)\cite{refMain}{Aigu_Small:2014}. At room-temperature, 
due to their small dimension, the nanoparticles are in a superparamagnetic state having their magnetic moment thermally flipping between 
their two easy axis directions. The graphene oxide contains a high concentration of nanovoids, vacancies and adatoms which carry magnetic moments 
that are the origin of the paramagnetic response observed in the graphene oxide sheets\cite{refMain}{Su_NatComm:2012} without the iron-oxide nanoparticles. 
The graphene oxide is partially reduced and thus the carbon atoms whose 
p$^{\textrm{z}}$-orbitals are not 
passivated can be regarded as sites where electrons can localize. The hopping electrons moving through the nanocomposite can hop between 
these sites through variable range hopping -- see supplementary information. 

The mixture is strongly disordered: there are nanoparticles of different sizes and thus different magnetic moments, whose position and easy axis 
orientation is random; the partially reduced graphene oxide flakes are also randomly positioned and oriented; thus, from the point of view of a
hopping electron, the sites it can occupy are randomly positioned having random onsite energies. Using the external magnetic field 
to drive the magnetic orientation of the two cobalt electrodes, a spin-imbalance can be generated in the nanocomposite's population of hopping 
electrons whenever an electric current flows across the device at room temperature (throughout the text, we
refer to this as the {\it initialization process}). The source electrode spin-polarizes the current entering the nanocomposite, while the drain 
electrode acts as a filter allowing electrons with one spin orientation to preferentially leak out the nanocomposite. When the 
electrodes are in an anti-parallel (parallel) configuration they generate (destroy) a spin-imbalance in the population of hopping electrons of 
the system. An antiferromagnetic PtMn layer pins the cobalt electrodes magnetic orientation via exchange bias so that their magnetization will 
only be flipped by a sufficiently strong magnetic field.

\begin{figure}[htp!]
  \centering
  \includegraphics[width=0.48\columnwidth]{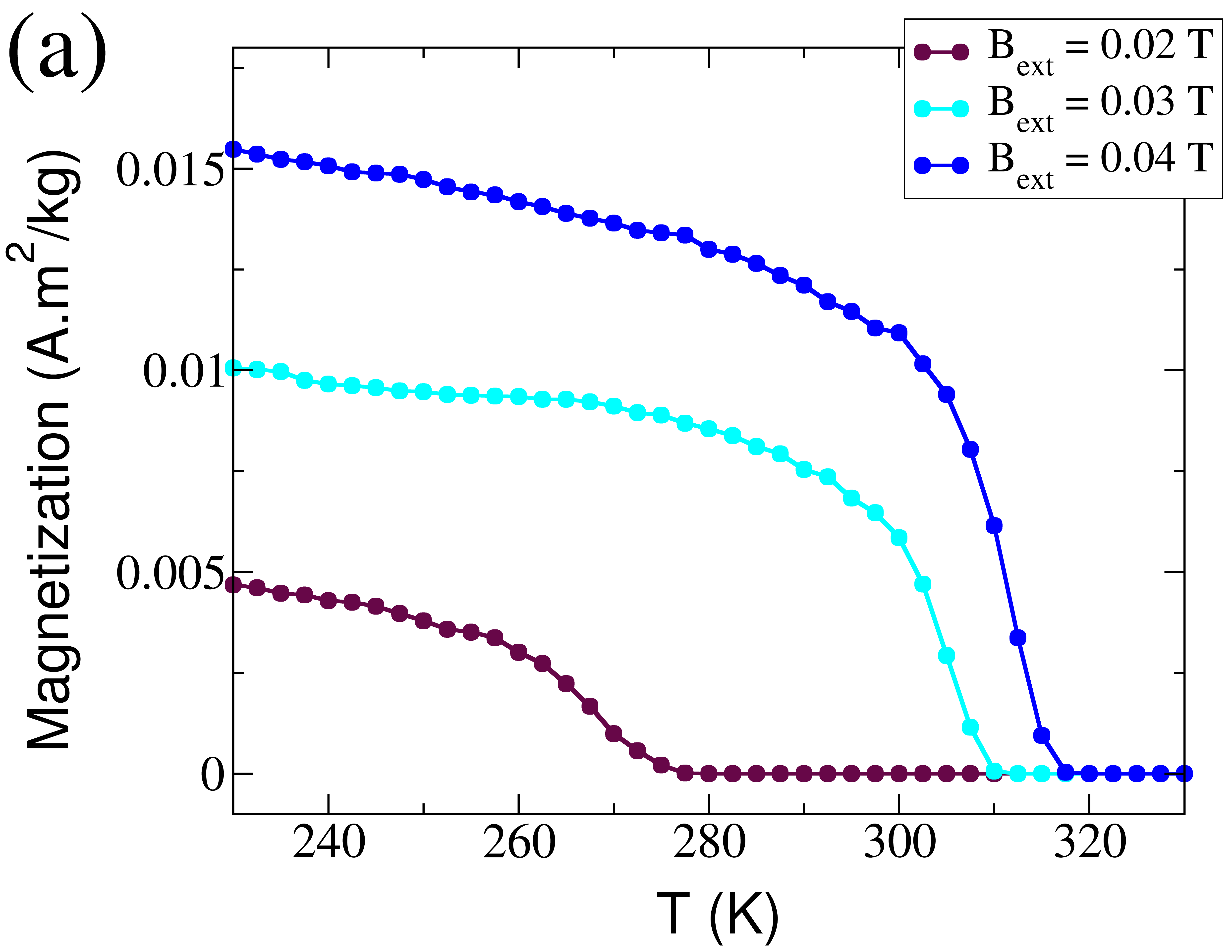}
  \includegraphics[width=0.48\columnwidth]{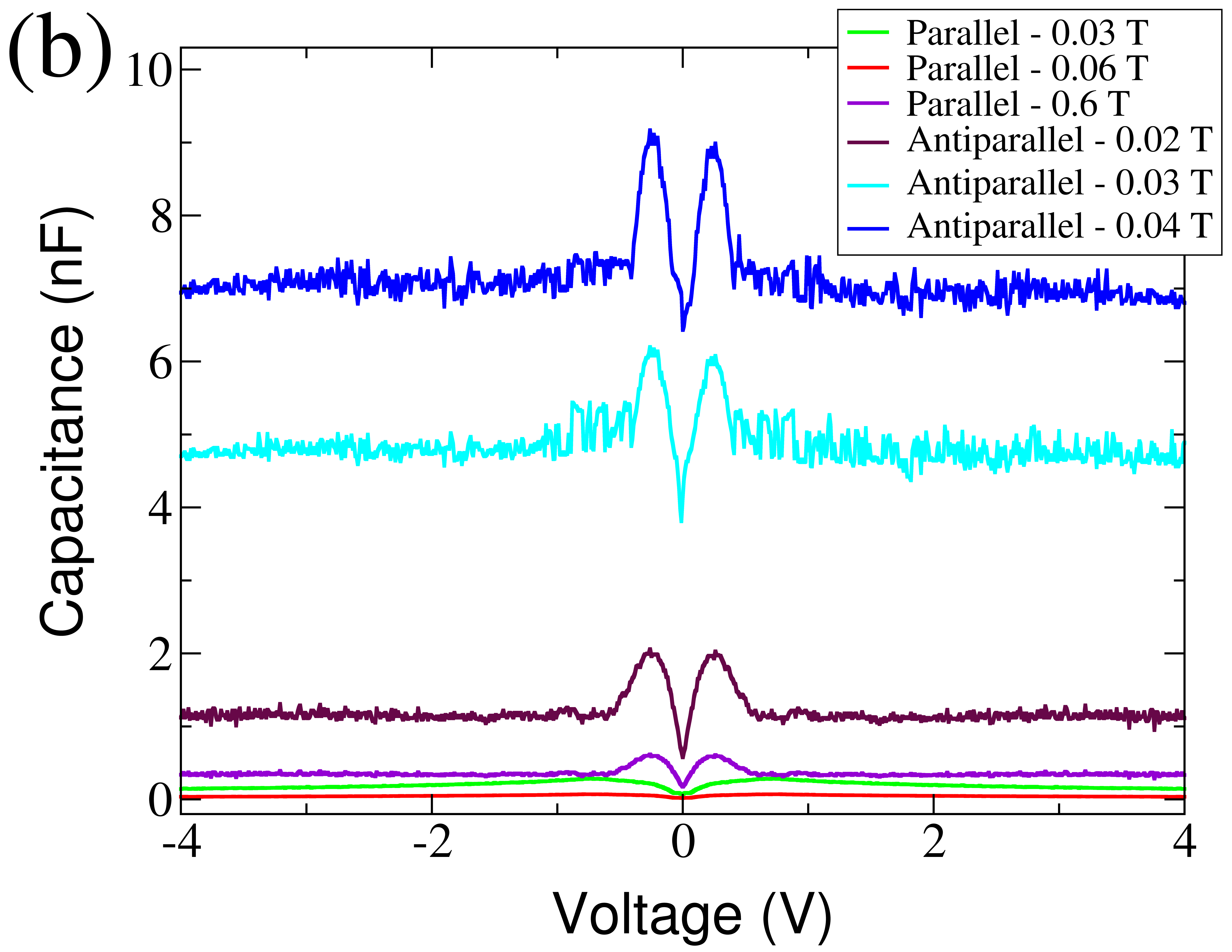}
  \caption{{\bf Magnetization and capacitance for different
      initialization processes.} {\bf (a)} Magnetization as a function
    of temperature for samples {\it initialized} with different
    $B_{\textrm{ext}}$. The transition temperature can be made to vary from
    276 K ($B_{\textrm{ext}} = 0.02$ T) to 317 K ($B_{\textrm{ext}} =
    0.04$ T).  {\bf (b)} Capacitance measurement for several different
    $B_{\textrm{ext}}$ corresponding to distinct electrodes
    configuration (see text and Fig. \ref{fig:DeviceScheme} for
    details).}
  \label{fig:MagTempCapact}
\end{figure}

If no electric current is passed across the device, the nanocomposite is paramagnetic for all tested temperatures. This indicates that the 
nanocomposite's magnetic moments (both from the iron oxide nanoparticles and from the defective graphene oxide) are essentially independent. 
The nanocomposite remains paramagnetic when a spin-unpolarized electric current is passed across it. However, using ferromagnetic electrodes 
to inject a spin-polarized current into the nanocomposite, the system can be made to undergo a ferromagnetic transition depending on the 
particular magnetic configuration of the electrodes. Of practical interest is the fact that this configuration can be controlled by an 
external magnetic field. Has shown in Fig. \ref{fig:MagTempCapact}, the initialization is done with two accessible knobs: a potential bias 
driving an electric current 
that is injected into the nanocomposite through two ferromagnetic electrodes; and an external magnetic field (with a magnitude of the order of 
tens of mT) driving the magnetic configuration of the electrodes. These two knobs determine the device's magnetic properties which 
remain stable for as long as we have measured it (several weeks) after the electric current and magnetic field are turned off.

We argue below that the spin-polarized current injected into the nanocomposite generates a spin-imbalance on the population 
of hopping electrons of the nanocomposite. These spin-polarized hopping electrons effectively couple the magnetic moments of 
the iron-oxide nanoparticles (and of the graphene oxide) through an indirect exchange interaction reminescent of the RKKY 
interaction\cite{refMain}{Ruderman_PR:1954,Kasuya_PTP:1956,Yosida_PR:1957}. The strength of this interaction depends on the degree of 
spin-imbalance in the population of hopping electrons: a greater spin-imbalance gives rise to a stronger interaction. 
The strongly disordered nanocomposite implies that the disorder average of this interaction is exponentially damped and 
effectively ferromagnetic. Thus, it will effectively behave as a disordered array of Heisenberg moments 
constrained to point around their randomly oriented easy axis and give rise to magnetic clusters that, depending on 
the {\it initialization} process, may lead to long range magnetic order. In what follows we will show step-by-step the 
evidence and reasoning leading to this picture.

\begin{figure}[htp!]
  \includegraphics[width=0.48\textwidth]{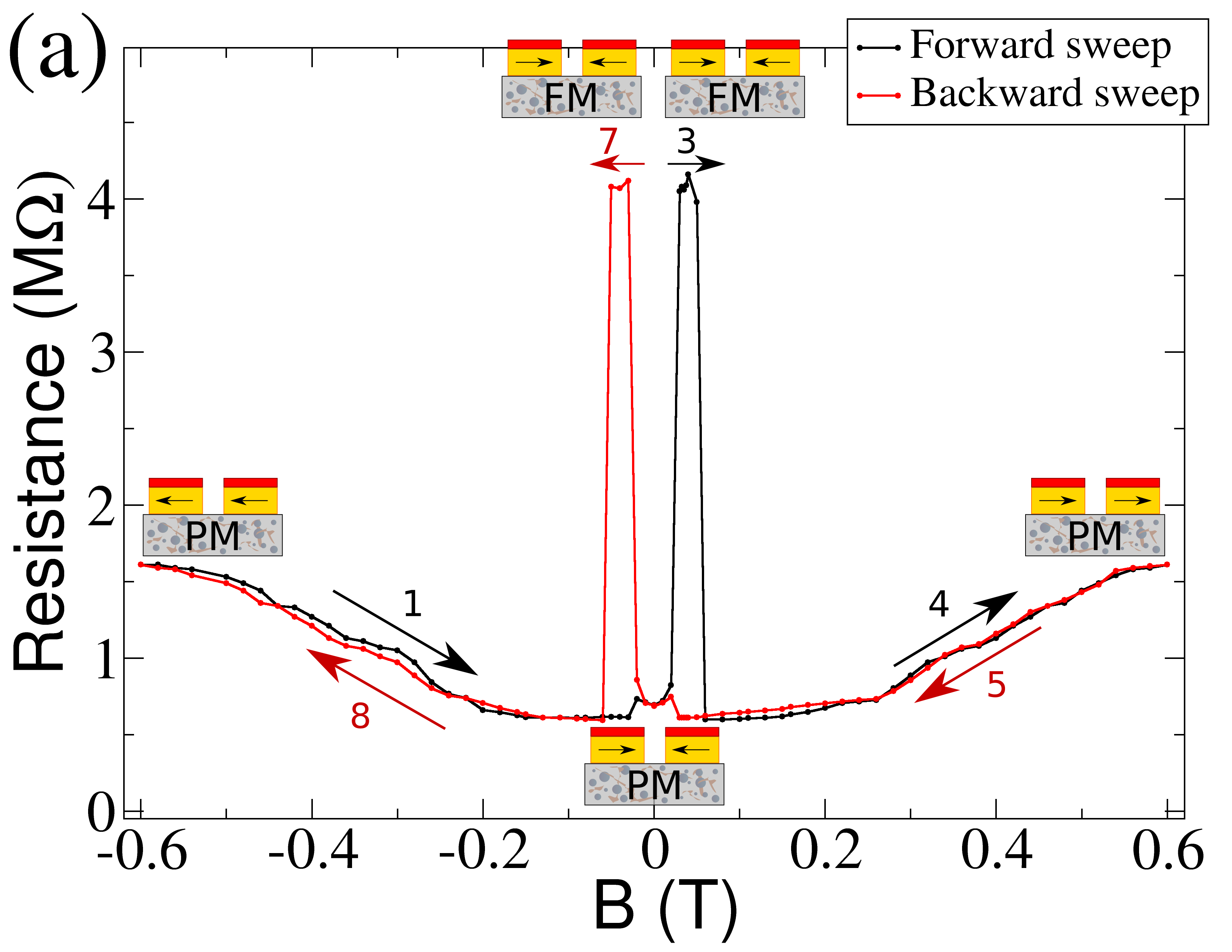}
  \includegraphics[width=0.48\textwidth]{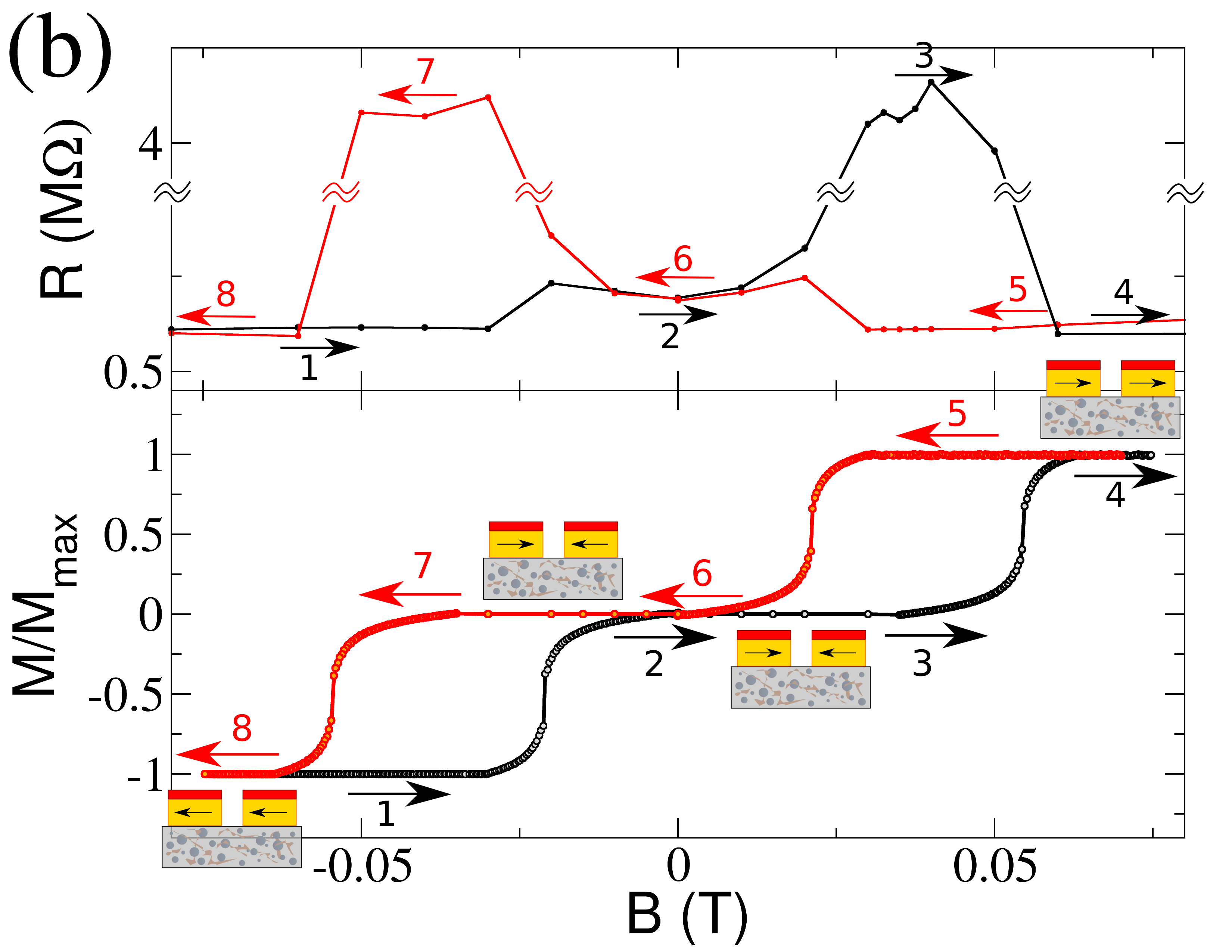}
  \centering
  \caption{{\bf Device's electrical properties.} {\bf (a)} Device's
    electrical resistance in terms of the $B_{\textrm{ext}}$. {\bf
      (b)} Ferromagnetic pinned electrodes response to an external
    magnetic field (bottom) and a blow up of the resistance data in
    the same field range (top). Measurements reveal two distinct jumps
    in resistance, one corresponding to the giant magnetoresistance
    and the other due to a ferromagnetic transition.}
  \label{fig:Res-vs-MagF}
\end{figure}

We first discuss the spin-dependent electronic transport properties of the device. An electrical current was injected on 
the device through the ferromagnetic electrodes while an external magnetic field was applied to the system to drive the configuration 
of the electrodes. We have measured the electrical resistance of the device while gradually varying the strength of the magnetic field. 
Figure \ref{fig:Res-vs-MagF}(a) shows the result of such a measurement.

Starting from the electrodes in a parallel configuration 
($B_{\textrm{ext}} = - 0.6$ T) we first increase the magnetic field (forward sweep, black curve). At $B_{\textrm{ext}} \simeq - 0.02$ T there
is an increase in resistance caused by the switching of one electrode resulting in an antiparallel configuration [see lower panel of Fig. 
\ref{fig:Res-vs-MagF}(b)]. This is the well known giant magnetoresistance effect (GMR)\cite{refMain}{Fert_PRL:1988,Grunberg_PRB:1989} except 
that our high resistance values suggest that we are in the variable range hopping regime (VRH) rather than the metallic 
one\cite{refMain}{Jiang_PRB:1992,Ioffe_JETP:2013}. More interesting is a 
second (and larger) jump in the resistance that occurs when there is no change in the electrode's configuration (between arrow's 2 and 3 in 
Fig. \ref{fig:Res-vs-MagF}). This second jump is related to the ferromagnetic transition that is the main result of this work. Further 
increasing $B_{\textrm{ext}}$, we then observe the expected drop in resistance (between arrow 3 and 4) when the second electrode switches orientation. 
This corresponds to both the usual GMR effect and the loss of the spin-imbalance required for the ferromagnetic state, which brings the
nanocomposite back to a paramagnetic state. The exact same sequence is observed for the backward sweep (red curve, labeled 5-8) where the region 
7 corresponds to the range of $B_{\textrm{ext}}$ for which we find ferromagnetism in the nanocomposite.

In order to investigate the origin of this effect, we have probed the device's magnetic properties after having passed an electric current 
through it at different external magnetic fields to {\it initialize} it. We have checked that whenever the magnetic field is such 
that the sample resistance is 
small [i.e., it is either $B_{\textrm{ext}} < 0.02$ T or $B_{\textrm{ext}} > 0.05$ T -- see Fig. \ref{fig:Res-vs-MagF}(a)], the nanocomposite
is paramagnetic. However, whenever the {\it initialization} process is performed with an external magnetic field in the range of high 
sample resistance (i.e. $B_{\textrm{ext}} \in [0.02, 0.05]$ T), the nanocomposite is found to be in a ferromagnetic state. 
This confirms that the sharp jumps in the electrical resistance of the device are related to the ferromagnetic transition occurring in 
the nanocomposite. Note that the orientation of the electrodes is necessary for this transition, since it only occurs when they are anti-aligned. 
As mentioned before, a strong spin-imbalance in the population of hopping electrons is only generated for the anti-parallel electrodes' configuration.

Capacitance is a direct measure of the spin-imbalance generated in the nanocomposite.
When the device is in the anti-parallel configuration the measured capacitance increases with increasing magnetic field [see Fig. 
\ref{fig:MagTempCapact}(b)]. The peak capacitance increases from 2 nF at $B_{\textrm{ext}} \approx 0.02$ T to 9 nF when the applied field is 
$B_{\textrm{ext}} \approx 0.04$ T. Further increase of $B_{\textrm{ext}}$  did not lead to a noticeable change to the peak capacitance value. 
In contrast, whenever the device is in its parallel configuration, the measured capacitance is invariably one or two orders of magnitude 
smaller than that measured for the anti-parallel electrodes' configuration.

The comparison between the capacitance and the magnetization measurements indicates that the nanocomposite becomes ferromagnetic at room-temperature
whenever the capacitance increases above a critical value of 6 nF.
The sharp decrease of capacitance once the drain electrode is reversed at $B_{\textrm{ext}} \approx 
0.05$ T confirms the intuitive picture that the trapped spin-polarized charges are released when the electrodes become parallel. Moreover,
the fact that the system transitions back to the paramagnetic state, confirms that it is the spin-imbalance that controls the magnetic 
state of the nanocomposite.

Finally, the temperature dependence of magnetization was measured for several samples {\it initialized} under different magnetic fields 
- see Fig. \ref{fig:MagTempCapact}(a). The transition temperature ($T_{b}$) was observed to be strongly affected by the nanocomposite's 
spin-imbalance, as indicated by the sample capacitance: when the capacitance is $9$ nF, $T_{b} \approx 317$ K; $T_{b}$ decreases to 
$309$ K and $276$ K when the capacitance decreases to $6$ nF and $2$ nF respectively; whenever the capacitance is $\lesssim 1.5$ nF, no 
ferromagnetic ordering is observed even when the temperature is decreased to $10$ K. Figure \ref{fig:MagTempCapact} clearly demonstrates that 
the two external knobs present during the {\it initialization} process ($B_{\textrm{ext}}$ and $V_{\textrm{ext}}$) control the spin-imbalance in the 
population of hopping electrons of the nanocomposite and the magnetic properties of the system.

To understand this ferromagnetic transition we first estimate the direct magnetostatic interaction between the iron-oxide nanoparticles.
We find that this is several orders of magnitude smaller than $k_{B} T_{\textrm{room}}$ implying that it can be ruled out as the origin of 
the magnetism in this system. This explains why the system always remains paramagnetic when no current is passed through it. Moreover, 
the localized electron states are necessary to explain the origin of the ferromagnetism, since no ferromagnetism is observed in experiments 
without the partially reduced graphene oxide, e.g. when it is replaced with highly conducting graphene or strongly reduced graphene oxide.

The next logical step is to include a Zeeman-like coupling between the hopping electrons and the iron-oxide nanoparticles. This will
give rise to an effective interaction between the nanoparticles mediated by the sea of spin-polarized hopping electrons [without spin-imbalance,
this is reminescent of the well known Ruderman-Kittel-Kasuya-Yosida (RKKY) interaction\cite{refMain}{Ruderman_PR:1954,Kasuya_PTP:1956,Yosida_PR:1957}].
Basically, an electron in
the vicinity of one nanoparticle will retain information on its orientation that will then be seen by the other nanoparticles. Estimates of
the magnitude of this coupling prove difficult due to uncertainties in several parameters of the system. However, reasonable estimates
for material parameters (see supplementary information) suggest that the energy scale of these mediated interactions can be of the order
of $k_{B} T_{\textrm{room}}$. This is therefore the most plausible explanation for the observed phenomena. In what follows we take $J_{0}$ to be
the scale of the coupling between the hopping electron and the nanoparticle. This will be an input into the theoretical calculations. 

With such a mechanism in mind we can write an effective microscopic Hamiltonian governing a system of hopping electrons with spin and 
localized iron-oxide magnetic moments. We use Ising moments for the model and do not believe that the behavior would be qualitatively 
different for a different choice -- see supplementary information. The Hamiltonian reads
\begin{eqnarray}
H &=& H_{e}^{0} + H_{M}^{0} + H_{e-e} + H_{M-M} + H_{e-M} ,
\end{eqnarray}
where the $H_{e}^{0}$ ($H_{M}^{0}$) stands for the part governing free electrons (Ising moments), $H_{e-e}$ ($H_{M-M}$) stands for the part 
containing electron-electron interactions (dipole-dipole interactions), while $H_{e-M}$ stands for the part containing the Zeeman interaction 
between hopping electrons and Ising moments. In what follows we put aside the terms $H_{M-M}^{0}$ and $H_{M}^{0}$, not relevant to the following 
computations, and, for the sake of simplicity, disregard the term $H_{e-e}$. Following the 
spirit of RKKY\cite{refMain}{Ruderman_PR:1954,Kasuya_PTP:1956,Yosida_PR:1957} interaction, considering terms in $H$ that flip the spin of the 
hopping electrons, and employing several simplifications to the calculation (see supplementary information), 
one finds that integrating out the hopping electrons' degrees of freedom gives rise to the following effective Hamiltonian 
for the Ising moments (see supplementary information)
\begin{eqnarray}
  H^{\textrm{eff}} &=& - K \big( n_{+}-n_{-} \big) \sum_{\alpha} M_{\alpha} - \sum_{\alpha, \beta} J\big(r_{\alpha \beta},n_{+},n_{-}\big) M_{\alpha} M_{\beta} , \label{eq:EffH}
\end{eqnarray}
where $M_{\alpha}$ stands for the magnetic moment indexed by $\alpha$ (expressed in terms of Bohr magnetons, $M_{\alpha} = \mu_{B} m_{\alpha} 
\lambda_{\alpha}$, with $\lambda_{\alpha}=\pm1$), 
$r_{\alpha \beta}$ stands for the distance between the two magnetic moments indexed 
by $\alpha$ and $\beta$, the constant $K$ reads $K \equiv J_{0} \mu_{0} \mu_{B} A$ , while $n_{\sigma}$ stands for the average density 
of hopping electrons with spin $\sigma=+,-$. The first term therefore acts on each 
Ising moment as an effective magnetic field generated by the cloud of spin-imbalanced hopping electrons. The second term is a local indirect 
exchange interaction term between different Ising moments, with the RKKY-like exchange parameter $J\big(r, n_{+}, n_{-}\big)$ given by
\begin{eqnarray}
  J(r,n_{+},n_{-}) &=& C \, \bigg( A^{2} \Big[ 
  \mathbb{J}^{(1)}(r,n_{+}) + \mathbb{J}^{(1)}(r,n_{-}) \Big] + B^{2} \mathbb{G}(r,n_{+},n_{-}) \bigg) \, , \label{eq:JRKKY}
\end{eqnarray}
where $C \equiv J_{0}^{2} \mu_{0}^{2} \mu_{B}^{2} m^{*}/(32 \pi^{3} \hbar^2)$ and $\mathbb{G}(r,n_{+},n_{-})$ is given by
  \begin{eqnarray}
    \mathbb{G}(r,n_{+},n_{-}) &\equiv& \sum_{\lambda=\pm1} \Big(\mathbb{J}^{(2)}_{\lambda}(r,n_{+},n_{-}) 
    + \mathbb{J}^{(3)}_{\lambda}(r,n_{+},n_{-}) \Big) + \mathbb{J}^{(4)}(r,n_{+},n_{-}) .
  \end{eqnarray}
  In the above equation the functions $\mathbb{J}^{(1)}$, $\mathbb{J}^{(2)}_{\lambda}$, $\mathbb{J}^{(3)}_{\lambda}$ and $\mathbb{J}^{(4)}$ read
  \begin{eqnarray}
    \mathbb{J}^{(1)}(r,n_{\sigma}) &\equiv& \frac{\sin\big(2 \sqrt[3]{6 \pi^{2} n_{\sigma}} \, r\big) - 2 \sqrt[3]{6 \pi^{2} n_{\sigma}} \, r \, \cos\big(2 
      \sqrt[3]{6 \pi^{2} n_{\sigma}} \, r\big)}{r^{4}} \, , \\ 
    \mathbb{J}^{(2)}_{\lambda}(r,n_{+},n_{-}) &\equiv& 2 \lambda \, \frac{\sin\big[\vert \Omega_{\lambda} \vert \, r\big] - \vert \Omega_{\lambda} \vert \, r 
      \cos\big[\Omega_{\lambda} \, r \big]}{r^{4}} \, , \\
    \mathbb{J}^{(3)}_{\lambda}(r,n_{+},n_{-}) &\equiv&  \lambda \, \Omega_{-\lambda}^{2} \, \frac{\sin\big[\vert \Omega_{\lambda} \vert \, r \big] + \vert 
      \Omega_{\lambda} \vert \, r \cos\big[ \Omega_{\lambda} \, r\big] }{r^{2}}\, , \\ 
    \mathbb{J}^{(4)}(r,n_{+},n_{-}) &\equiv& \Big( \textrm{sinI}\big[\Omega_{+} \, r\big] + \textrm{sinI}\big[\Omega_{-} \, r\big] \Big) (6 \pi^{2})^{4/3} 
    \Big( {n_{+}}^{2} - {n_{-}}^{2} \Big)^{2} \, ,
  \end{eqnarray}
where we have defined $\Omega_{\lambda} \equiv \sqrt[3]{6 \pi^{2} n_{+}} + \lambda \sqrt[3]{6 \pi^{2} n_{-}}$. In these expressions $\mu_{0}$ ($\mu_{B}$) 
stands for the vacuum permitivity (Bohr magneton), $m^{*}$ for the effective mass of the free hopping electron gas, while $A$ ($B$) stands 
for the amplitude for an electron with spin state $\sigma$ to have its spin unchanged (flipped) when interacting with a nanoparticle. 

\begin{figure}[htp!]
  \includegraphics[width=0.48\textwidth]{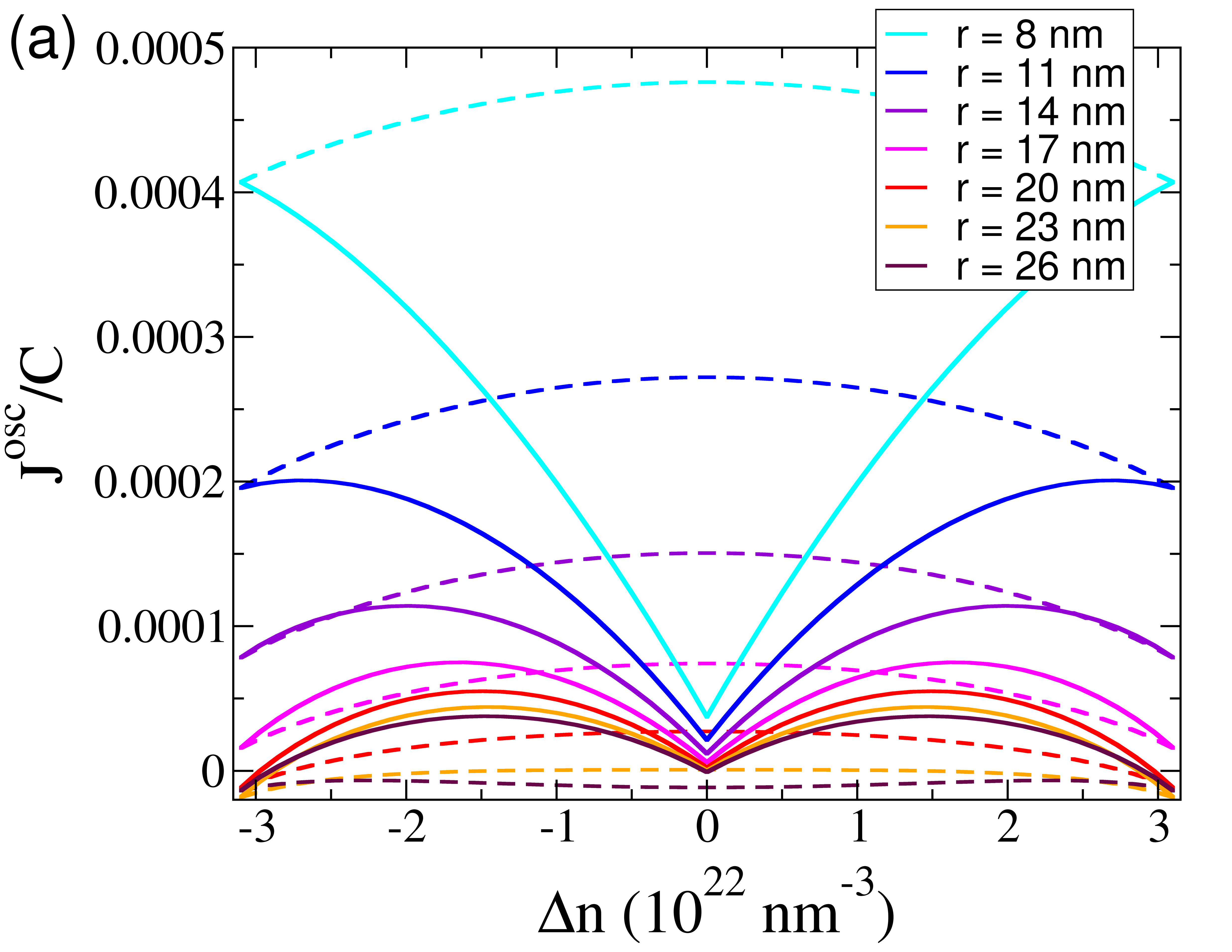}
  \includegraphics[width=0.48\textwidth]{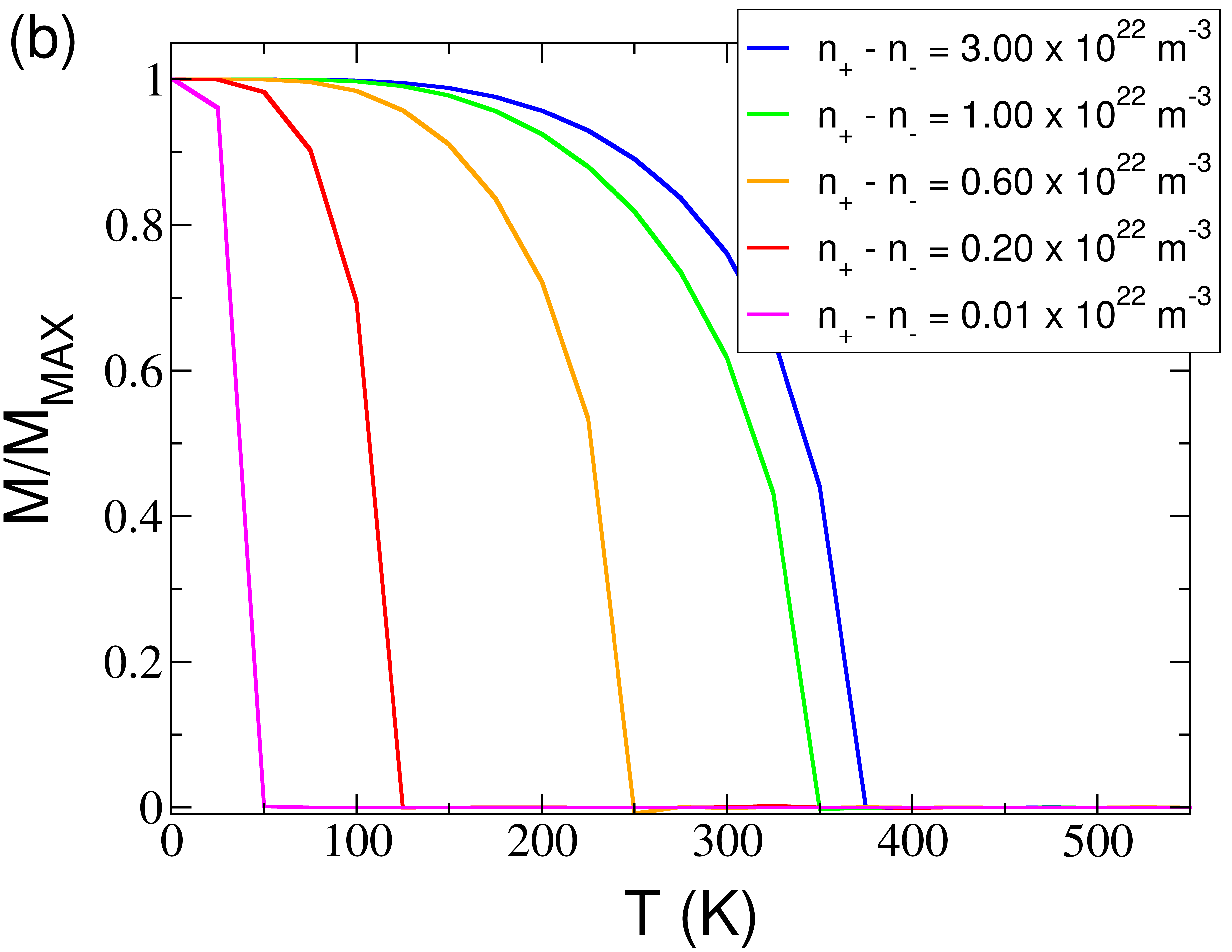}
  \includegraphics[width=0.48\textwidth]{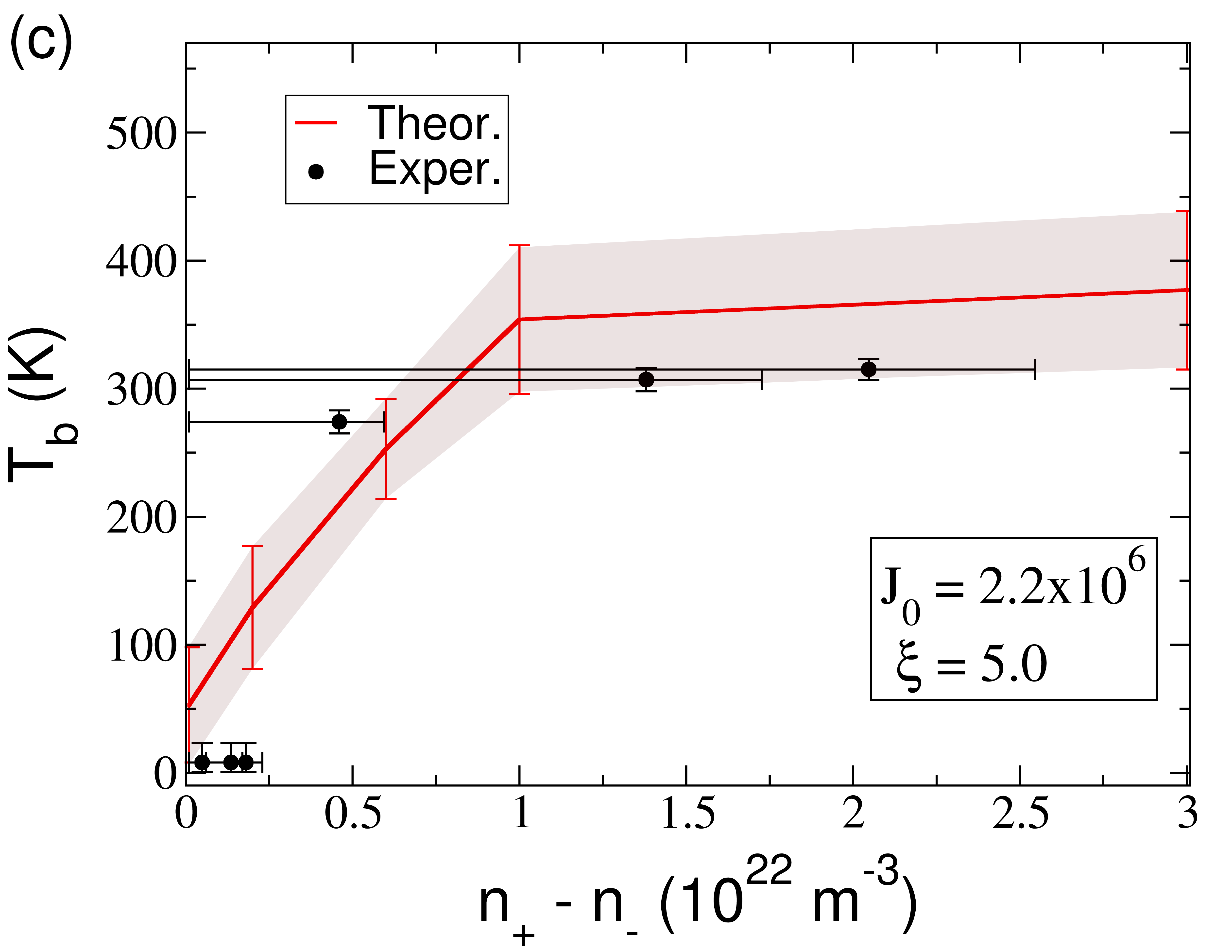}
  \centering
  \caption{{\bf Exchange parameter and Monte Carlo Simulations of a
      disordered 3D Ising model with exponential decaying RKKY
      interactions.} {\bf (a)} Plot of the oscillating part of the
    exchange parameter [see equation (\ref{eq:JRKKY})] for several
    distances between nanoparticles. The constant $C$ dividing
    $J^{\textrm{osc}}$ is the leading factor in equation
    (\ref{eq:JRKKY}). We plot both the case where no electron's
    spin-flips are allowed (dashed curves) and the case where these
    are allowed (full curves). {\bf (b)} Magnetization in terms of the
    temperature for different strengths of the indirect exchange
    coupling. {\bf (c)} Comparison between the experimental and
    theoretical blocking temperature in terms of the strength of the
    spin-imbalance (the theoretical fitting parameters used were
    $J_{0}=2.2\times10^{6}$ and $\xi=5$ nm). The Monte Carlo results
    of panels (b) and (c) were obtained for simulations (with 79507
    Ising moments) starting from a highly ordered state (see
    supplementary information) that use Metropolis algorithm. They
    explore the phase space region in the vicinity of the global
    energy minimum, and indicate that the system can show long range
    order if the clusters are initially aligned by an external
    magnetic field.}
  \label{fig:MonteCarlo}
\end{figure}

From Fig. \ref{fig:MagTempCapact}(b) we estimate the sample's average electronic densities, $n_{\pm}$, finding that they are typically small such 
that first-neighbor interactions are generally ferromagnetic -- see supplementary information. Assuming that $A \gtrsim B$ then 
we conclude that $J(r,n_{+},n_{-})$ is minimal for spin-imbalance zero, growing with increasing spin-imbalance -- see Fig. 
\ref{fig:MonteCarlo}(a). This is in contrast with the typical RKKY result where no spin-flips of the electrons are considered. 
Our analytical result explains how the ferromagnetic coupling increases with spin-imbalance explaining the experimental observation 
that the magnetization vanishes without the spin-imbalance and increases with larger spin-imbalance. 

Strong disorder exponentially suppresses the typical value of the RKKY interaction\cite{refMain}{DeGennes_JPR:1962,Lerner_PRB:1993,Sobota_PRB:2007}
as $J(r,n_{+},n_{-}) \to J(r,n_{+},n_{-}) e^{-r/\xi}$, where in the metallic case $\xi$ is the electron's 
mean free path. Since our system is strongly disordered $\xi$ should be small, and the exponential suppression essentially kills all longer 
ranged interactions, such that the only relevant interactions in our system are those comparable with the first-neighbor ones.
Therefore all the relevant interactions are ferromagnetic. To compare with the experiment we take $\xi$ to be a fitting parameter 
comparable to the spacing between the nanoparticles.

The experimental results strongly suggest that the first order term in equation (\ref{eq:EffH}) is irrelevant when compared with the 
second order one (see supplementary information). This is perfectly compatible with the theoretical model despite the fact that 
the effective Hamiltonian -- see equation (\ref{eq:EffH}) -- arises from a series expansion on the electron-nanoparticle interaction. 
The relative magnitude of the effective Hamiltonian's first and second order terms is determined by the {\it external} parameters 
($n_{+}-n_{-}$, $\xi$, $J_{0}$, $m^{*}$, $A$ and $B$) rather than by the expansion parameter. The parameters used to obtain the 
results of Fig. \ref{fig:MonteCarlo}, yield a second order term at least one order of magnitude greater than the first order one, 
for the range of spin-imbalances estimated from the experimental results. Accordingly, only the second order term was considered 
when performing the Monte Carlo simulations.

For typical values of $\xi$, the exponentially damped coupling gives rise to the ordering of the system in 
magnetic clusters that interact weakly between themselves. Upon decreasing temperature, the magnetic moments inside each cluster start
aligning, with different clusters doing so at slightly distinct temperatures. Moreover, as clusters interact weakly, individual clusters
will generally have different magnetization directions. As a consequence, the system should not in general present long-range order when 
temperature is decreased below the {\it blocking} temperature $T_{b}$ and this is confirmed in our Monte Carlo 
simulations -- see supplementary information. Similarly, 
if we remain at a fixed temperature while turning on the exchange interaction (by generating a spin-imbalance in the system), one should 
not observe long-range order in the system. However, if we start from an ordered state generated, for example, by applying an external 
magnetic field when the spin-imbalance is being generated (as is done in the experiment), then long-range order should be observed since 
the nearly independent clusters were from the beginning aligned by the external magnetic field. This is observed in 
our system: if no magnetic field is applied to the device while the current is flowing across it, no magnetization is observed (see 
supplementary information for a detailed discussion). In Fig. \ref{fig:MonteCarlo} we show 
that, when starting from an ordered state, Metropolis Monte Carlo simulations show a transition between an ordered and a disordered state upon 
variation of temperature. Its blocking temperature depends on the magnitude of the indirect exchange, that we have shown is 
dependent on the spin-imbalance of hopping electrons.

In conclusion, we have demonstrated a tunable magnet where the iron-oxide and graphene oxide nanocomposite undergoes a paramagnetic to ferromagnetic 
transition whenever a critical concentration of spin-polarized electrons are trapped within the nanocomposite such that they can generate 
a sufficiently strong indirect exchange coupling between neighboring iron nanoparticles. This ferromagnetic state is controllable by tuning 
the spin-imbalance of hopping electrons through the external magnetic field and the potential bias that drives the current across the device 
during its {\it initialization} process. Moreover, this state is reversible by elimination of the spin-imbalance, in which case the 
nanocomposite transitions back to a paramagnetic state. Such artificial composite materials with easily processable components and highly 
tunable magnetic/transport properties open doors towards constructing high-performance data storage and spintronic devices operating at room 
temperature.

\begin{methods}

Graphene oxide was synthesized based on the Hummers method. Graphite flakes (3.0 g) were stirred in ice bath. Sodium nitrate 
(3.0 g) and concentrated sulfuric acid (135 ml) were added into the round-bottom flask. Next, potassium permanganate (18 g) was added 
slowly over 2 hours. Once the mixture is homogeneous, the solution was transferred to 35 °C oil bath and stirred for another 1 hour. 
A thick paste was formed and deionized (DI) water (240 ml) was added. The mixture was stirred for 1 hour as the temperature was increased 
to 90 °C. Deionized water (600 ml) was added, followed by slow addition of $30\%$ hydrogen peroxide (18 ml) solution. The color of the 
suspension changed from brown to yellow. The suspension was filtered and washed with $3\%$ HCl solution. It was then repeatedly centrifuged
and decanted until the pH of the supernatant is 7. The as-produced graphene oxide was dispersed in 750 ml DI water at a concentration of 
0.6 mg/mL$^{-1}$. 3 g of NaOH plate was added into graphene oxide solution (0.1 mol/L). The mixture was refluxed in a round bottom flask under constant 
magnetic stirring for 1 hour. Subsequently, the based treated GO were separated by centrifuging at 13000 rpm. It was then repeatedly 
centrifuged and decanted until the pH of the supernatant is 7.The iron oxide nanoparticles were then added to the solution and 
dispersed with the application of ultrasound for 30 seconds before the solution was spin coated on a silicon dioxide substrate at 
a speed of 8000 rpm for 30 seconds. We repeated this spin coating process 3 times before thermally reducing the nanocomposite by 
applying a temperature of 340 K for 15 minutes. Cobalt electrodes 10 nm thick and 200 nm apart were then deposited on the nanocomposite. 
20nm of PtMn was then deposited on one of the cobalt electrodes and was slowly cooled down while an external magnetic field of 0.1 T was 
applied. This is repeated for the other cobalt electrode but with the external magnetic field applied in the opposite direction. I-V 
measurements were done after connecting two probes onto the two electrodes using a KEITHLEY Semiconductor Characterization System with 
voltage varying from 0 to 5V. The magnetic field was generated using a DEXTER Adjustable Pole Electromagnet (Model \# 1607037) and was 
varied from 0T to 0.6 T for each I-V measurement. VRH data obtained from the Quantum Design Physical Property Measurement System (PPMS) 
measurements where the electrical resistance is measured at a fixed magnetic field strength and the temperature is gradually decreased 
at intervals of 5 K from 298 K to 210 K under a constant applied voltage of 0.5 V. The PPMS is also used to measure the change in 
electrical resistance at a fixed temperature but under varying magnetic field strengths from 0 to 0.06 T.  The magnetic characterization 
of the device is done by a superconducting quantum interference device (SQUID) magnetometer. 

In order to investigate the magnetic ordering of the 3D disordered Ising model arising from the integration of the electronic degrees of 
freedom, we have performed Monte Carlo simulations using our own implementation of Metropolis single spin-flip algorithm\cite{refMain}{Metropolis_JChemPHys:1953} 
and Wolf's cluster algorithm\cite{refMain}{Wolff_PRL:1989}.

\end{methods}

\vspace{1.5cm}

\begin{addendum}
\item This work was supported by the Singapore Ministry of Education
  (ARF Grant No. R-398-000-056-112) and by the Singapore National
  Research Foundation through its Fellowship Program (NRF-NRFF2012-01)
  and CRP Program (R-144-000-295-281).  The Graphene Research Centre
  computing facilities and those of Yale-NUS College were used in this
  work. JNBR thanks M. D. Costa, J. E. Santos, J. Viana Lopes,
  M. D. Stiles and J. M. Lopes dos Santos for valuable discussions.
\item[Author Contributions] A.L.L., T.W., W.C. and A.T.S.W. fabricated the device 
 and characterised it electrically and magnetically. C.S. and K.P.L. synthesised and 
 characterised the properties of the GO nanoflakes. J.N.B.R., M.M., A.C.H.N. and S.A. 
 developed the theoretical model for the system.  J.N.B.R., A.L.L., S.A. and A.T.S.W. 
 wrote the manuscript.
\item[Competing Interests] The authors declare that they have no
  competing financial interests.
\item[Correspondence] Correspondence should
  be addressed to S.A.:~ shaffique.adam@yale-nus.edu.sg .
\end{addendum}

\vspace{1.5cm}



\pagebreak


\newgeometry{left=1.7cm,right=1.7cm,top=1.6cm,bottom=1.7cm}

\renewcommand{\thefigure}{S\arabic{figure}}
\renewcommand{\figurename}{\textbf{Supplementary Figure}}

\renewcommand{\theequation}{S\arabic{equation}}

\renewcommand{\thepage}{S\arabic{page}}

\renewcommand\thesection{\Roman{section}}
\renewcommand\thesubsection{\Alph{subsection}}
\renewcommand\thesubsubsection{\arabic{subsubsection}}

\renewcommand{\appendixname}{} 
\renewcommand{\appendixtocname}{Appendices} 
\renewcommand{\appendixpagename}{Appendices} 

\begin{center}
\textbf{\huge Supplementary Information}
\end{center}

\vspace{0.8cm}

\setcounter{equation}{0}
\setcounter{figure}{0}
\setcounter{table}{0}
\setcounter{page}{1}
\setcounter{section}{0}
\setcounter{subsection}{0}
\setcounter{subsubsection}{0}

\Ssection{Additional experimental information}

\label{sec:Experiment}

\noindent\textbf{\underline{The nanocomposite}}

\vspace{0.2cm}

The nanocomposite's graphene oxide was thermally reduced by approximately $18\%$ to $20\%$. Its electrical resistance 
was observed to increase from an initial magnitude of $(1.15 \pm 0.027)$ M$\Omega$ to $(0.94 \pm 0.035)$ M$\Omega$ 
after the reduction process.

The left panel of Supplementary Fig. \ref{fig:TEMmicrograph-Transport} shows a TEM micrograph of the nanocomposite where it is possible 
to see its strongly disordered nature with the nanoparticles randomly positioned and the graphene oxide flakes placed in 
between them. From this figure we can estimate that the typical distance between nanoparticles is roughly $\approx12$ nm.
\begin{figure}[h!]
  \centering
  \includegraphics[height=6.2cm]{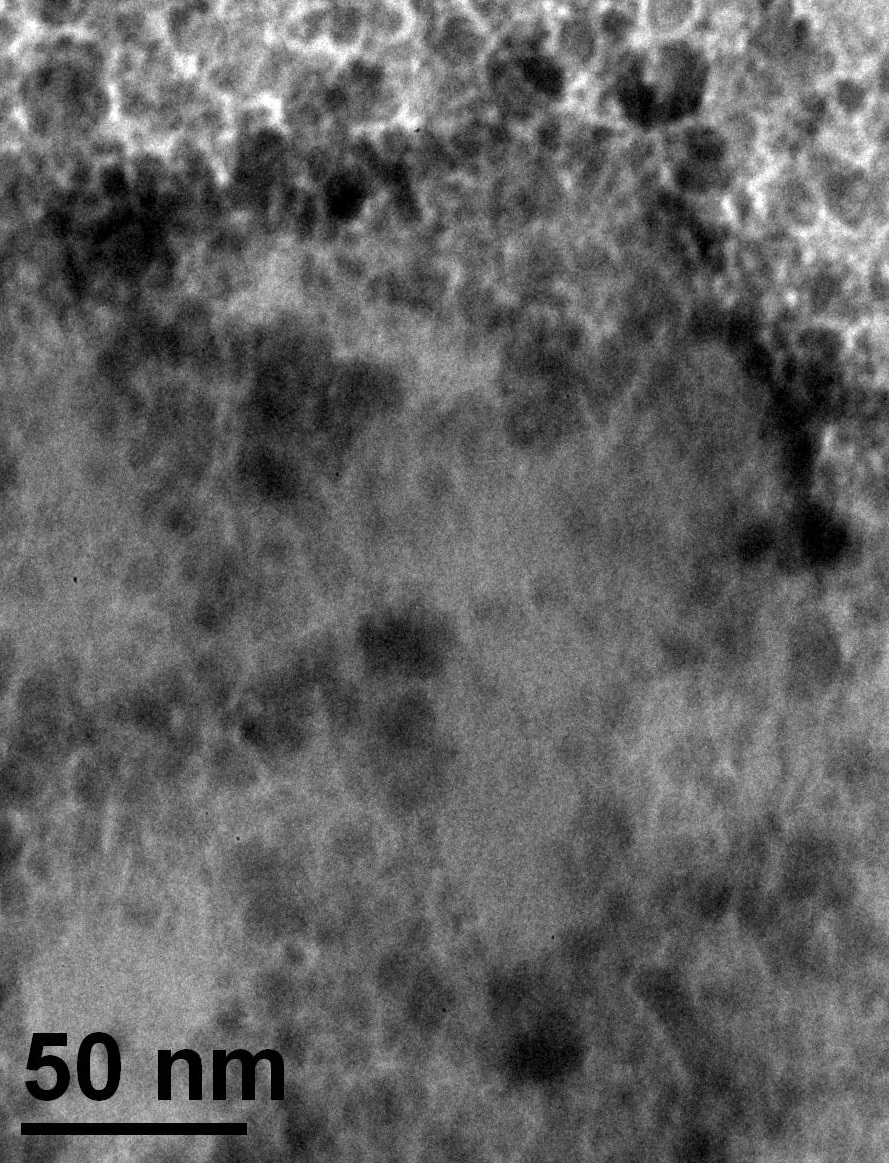}
  \hspace{0.5cm}
  \includegraphics[height=6.2cm]{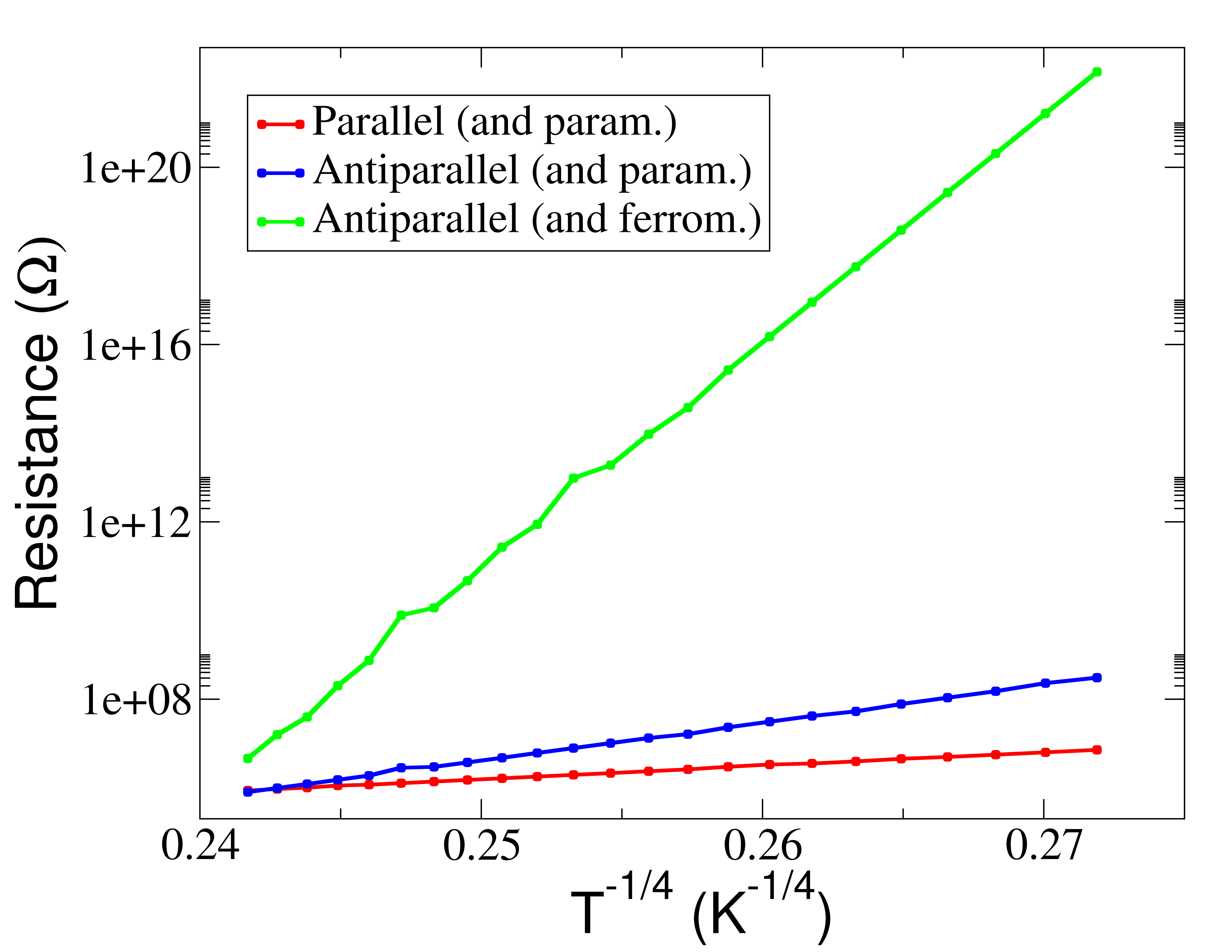}
  \caption{{\bf Left:} TEM micrograph of the nanocomposite, where we can
    see the nanoparticles (spheroidal structures) surrounded by the
    graphene oxide flakes (bright structures). From it we can
    roughly estimate that the typical nanoparticles' radius is
    $\approx5$ nm, while the distance between their centers is roughly
    $\approx12$ nm. {\bf Right:} Plots showing the variable
    range hopping character of the electronic transport in the
    nanocomposite, $R(T) = R_{0} \exp \big[(T_{M}/T)^{1/4}\big]$. The Mott temperature $T_{M}$ in
    each of the cases reads: $T_{M} \approx 2.41\times10^{4}$ K (red);
    $T_{M} \approx 1.48\times10^{9}$ K (blue); $T_{M} \approx
    1.92\times10^{12}$ K (green).}
  \label{fig:TEMmicrograph-Transport}
\end{figure}

\vspace{0.5cm}

\noindent\textbf{\underline{The transport properties of the device}}

\vspace{0.2cm}

The right panel of Supplementary Fig. \ref{fig:TEMmicrograph-Transport} we plot the device's resistance (in logarithmic scale) in terms 
of the fourth root of the temperature to show the variable range hopping nature of the electronic transport in the nanocomposite.

\vspace{0.5cm}

\noindent\textbf{\underline{The magnetization measurements}}

\vspace{0.2cm}

Each one of the curves in main text's Fig. 2(a), showing the temperature dependence of the nanocomposite's magnetization (at zero-field) 
were obtained by {\it initializing} the system (at room temperature) with the indicated $B_{\textrm{ext}}$. After {\it initialization} the 
temperature was decreased to $T \approx 230$ K and from there the system's magnetization was recorded while increasing temperature 
up to $T \approx 330$ K. Whenever the {\it initialization} was done with $0$ T $\lesssim B_{\textrm{ext}} \lesssim 0.015$ T, no magnetization 
was observed upon temperatures of $T \approx 10$ K. Note that the several curves of Fig. 2(a) of the main text were obtained for different 
samples, since after measuring the sample magnetization up to $T \approx 330$ K the graphene oxide layers present in the nanocomposite were 
strongly reduced and thus the nanocomposite properties irreversibly changed (see discussion of Sections \ref{sec:TODO}).

The left panel of Supplementary Fig. \ref{fig:RemMag-vs-Init} shows the ferromagnetic response of the nanocomposite after 
{\it initialization} with 
a magnetic field of $B_{\textrm{ext}} = 0.03$ T (orange curve) and $B_{\textrm{ext}} = 0.04$ T (green curve). Its right panel shows several 
room-temperature magnetization values obtained for an ensemble of successive {\it initialization} processes done (at room temperature) 
with different magnetic fields $B_{\textrm{ext}}$. This set of measurements was done using the same sample.
\begin{figure}
  \centering
  \includegraphics[width=0.4\columnwidth]{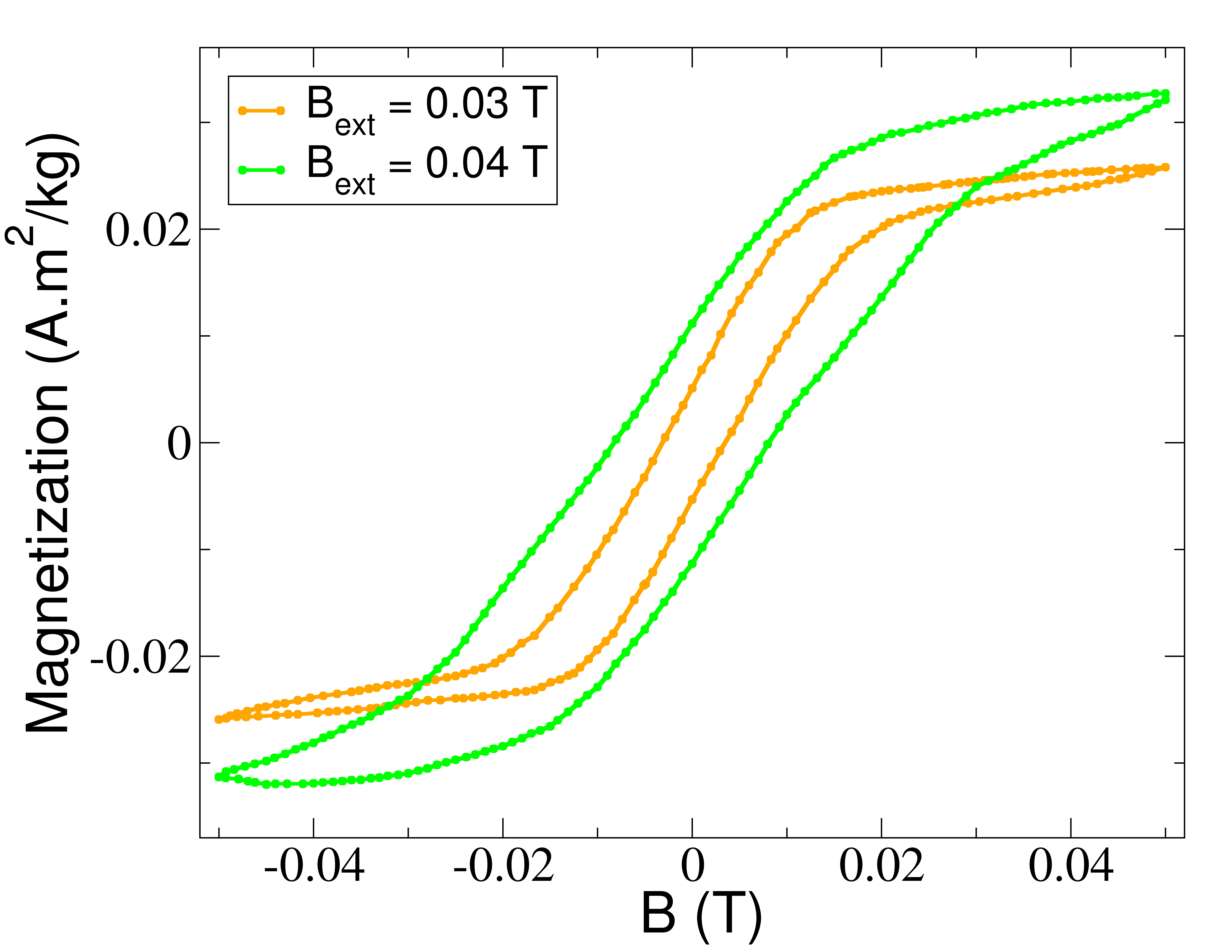}
  \includegraphics[width=0.4\columnwidth]{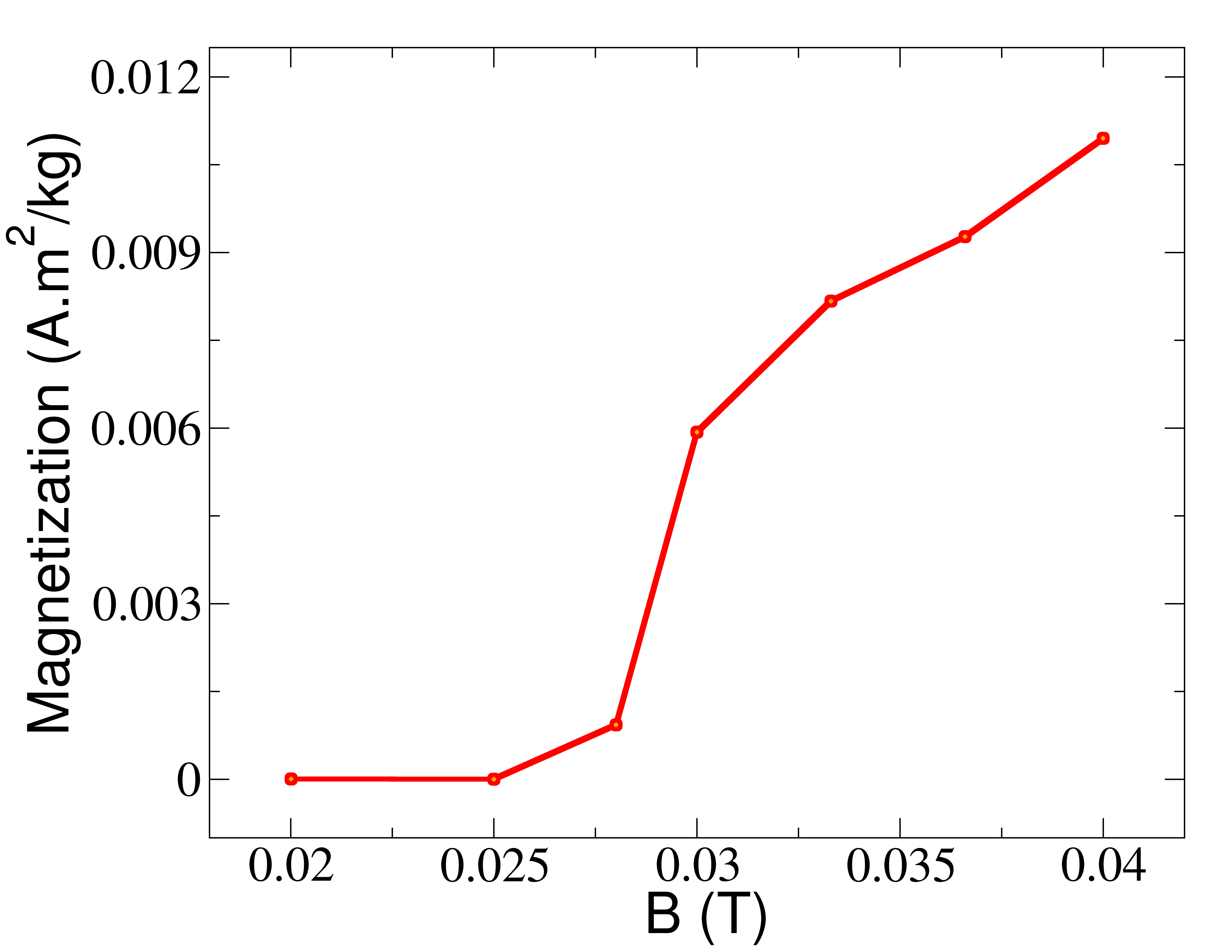}
  \caption{{\bf Left:} Hysteretic response of the nanocomposite after
    {\it initialization} with a spin-polarized current and an external
    magnetic field of $B_{\textrm{ext}} = 0.03$ T (orange curve) and
    of $B_{\textrm{ext}} = 0.04$ T (green curve). {\bf Right:}
    Room-temperature magnetization in terms of the magnetic field
    applied during the {\it initialization} process for a given sample
    that was successively {\it initialized} (at room temperature) with
    different magnetic fields.}
  \label{fig:RemMag-vs-Init}
\end{figure}

\vspace{0.5cm}

\noindent\textbf{\underline{Capacitance measurements}}

\vspace{0.2cm}

The capacitance measurements presented in main text's Fig. 2(b) were performed during the {\it initialization} process, i. e. at the 
same time that the magnetic field was being applied to the device (to drive its electrode's configuration) and the electric current was 
flowing across it. In sub-Section \ref{sec:Density-MMmagnitude-SpinImb} we discuss a simple picture for this capacitance measurement 
that allows us to estimate the spin-imbalance of the hopping electrons' population. 
The precise ingredients involved in the generation and retention of the spin-imbalance are discussed in Section \ref{sec:TODO}.

\vspace{0.5cm}

\noindent\textbf{\underline{Estimates}}

\vspace{0.2cm}

We estimate several parameters and energy scales of the system, such as: the typical density of the nanocomposite, the average magnitude 
of the iron-oxide magnetic moments and the spin-imbalances associated with each measured capacitance (see \ref{sec:Density-MMmagnitude-SpinImb});
the energy scale of the magnetostatic interaction between two iron-oxide nanoparticles (see \ref{sec:Magnetostatic}); the magnitude of the indirect 
exchange interaction mediated by the hopping electrons (see \ref{sec:IndirectExchange}).


\Ssubsection{Nanocomposite's density, magnetic moments' magnitude and spin-imbalance \hfill\space}

\label{sec:Density-MMmagnitude-SpinImb}


To estimate the typical density of the nanocomposite, we note that it is deposited on top of the SiO$_{2}$ 
by a spin-coating process which gives rise to a disc-like structure with typical thickness of $150$ nm (believed to be nearly uniform) 
and some hundreds of micrometers of radius. The typical weight of the samples is of the order of $\mu$g. One of the samples weights 
$M \approx 1.01 \, \mu\textrm{g}$ and has a radius of $\approx 1200 \mu\textrm{m}$, i.e., a volume of $\mathbb{V}\approx6.8 \times 10^{-13}$ 
m$^{3}$. Therefore its density is around $\rho = M/\mathbb{V} \approx 1.5 \times 10^{6}$ g/m$^{3}$. Such an estimate for the nanocomposite's density is 
compatible with an estimate of the density taking into account the density of each of the components, Fe$_{2}$O$_{3}$ nanoparticles, 
graphene oxide, toulene and spacer.

The magnitude of each Fe$_{2}$O$_{3}$ nanoparticle magnetic moment, $m$, depends both on the nanoparticle's size (typical diameter of $6.5$-$9.5$ nm) 
and on its microscopic ordering ($\alpha$-phase nanoparticles with a ferrimagnetic canted ordering). The value usually assumed for their typical 
magnetic moment is in the range $3.2$-$3.6 \mu_{B}$. But we can still try to estimate it 
from the measurements made in the system. From the blue curve of main text's Fig. 2(a) (i.e. {\it initialization} done with $B_{\textrm{ext}} = 0.04$ 
T) we can conservatively estimate the saturation magnetization of the nanocomposite (nanoparticles, graphene oxide, toulene, spacer) to be 
$M_{S} \approx 2.2 \times 10^{-2}$ A.m$^{2}$/kg. Using the estimated value for the nanocomposite mass density, $\rho \approx 1.5 \times 10^{6}$ 
g/m$^{3}$, and the typical volume for a box containing one nanoparticle [from Supplementary Fig. \ref{fig:TEMmicrograph-Transport}(a) 
we estimate the 
typical distance between nanoparticles  to be $\approx12$ nm] $V_{\textrm{box}} \approx 1.7 \times 10^{-24}$ m$^{3}$, we estimate the typical 
magnetic moment associated with that box to be $\bar{M}_{\textrm{box}} \approx 6.1$ $\mu_{B}$. In such a volume box only the nanoparticle and the 
graphene oxide flakes carry magnetic moments. Thus using the mass density and typical volume of the nanoparticles (typical nanoparticle's volume 
of $\approx 2.7 \times 10^{25}$ m$^{3}$ and density $\approx 5.2 \times 10^6$ g/m$^{3}$) we estimate that the typical magnetic moment associated 
with each nanoparticle is $\approx5.1$ $\mu_{B}$, while that of the graphene oxide flakes contained in that volume box is $\approx 1.0$ $\mu_{B}$.

To estimate the spin-imbalance in the hopping electrons' population we adopt the simplistic picture where the device acts as a 
{\it spin capacitor}\cite{refSupp}{Fleetwood_JAP:1993-S,McCarthy_PRL:2003-S,Aigu_APL:2013-S,Zhu_EES:2013-S}: the source electrode spin-polarizes 
the current entering the nanocomposite, while the drain electrode acts as a filter allowing electrons 
with the spin state aligned with its magnetization to leak out the nanocomposite but not those with an opposite one. Anti-parallel (parallel) 
electrodes generate (destroy) a spin-imbalance in the population of hopping electrons of the nanocomposite.
Within such a picture we can straightforwardly estimate the spin-imbalance from the capacitance measurement by $\Delta n \equiv n_{+}-n_{-} = C V 
/ (e \mathbb{V})$, where $C$ stands for the measured capacitance, $V$ stands for the applied bias voltage, $e$ is the electron charge and $\mathbb{V}$
again is the sample's volume. For the three curves in
main text's Fig. 2(b) corresponding to the anti-parallel configuration of the electrodes, $B_{\textrm{ext}}=0.04$ T, $B_{\textrm{ext}}=0.03$ T and 
$B_{\textrm{ext}}=0.02$ T, the pick capacitance of the device at $V\approx0.25$ V is approximately $9$ nF, $6$ nF and $2$ nF respectively. Therefore, the 
corresponding spin-imbalances respectively read $\Delta n\approx2.04\times10^{22}$, $\Delta n\approx1.38\times10^{22}$ and $\Delta n\approx0.46\times10^{22}$ electrons/m$^{3}$. This rules out Pauli paramagnetism as the phenomenon behind the ferromagnetic state since the overall maximal magnetic moment 
contributed by the hopping electrons (when such spin-imbalances are at play) is typically two orders of magnitude smaller than the measured 
nanocomposite's magnetization. Similarly, when the electrodes are parallel aligned [see main text's Fig. 2(b)], the curves obtained 
with $B_{\textrm{ext}}=0.03$ T, $B_{\textrm{ext}}=0.06$ T and $B_{\textrm{ext}}=0.6$ T, show a pick capacitance (at $V\approx0.7$ V, $V\approx0.75$ V and 
$V\approx0.25$ V) of approximately $0.3$ nF, $0.07$ nF and $0.6$ nF respectively. Therefore, the corresponding spin-imbalances read $\Delta n\approx
1.80\times10^{21}$, $\Delta n\approx4.90\times10^{20}$ and $\Delta n\approx1.36\times10^{21}$ electrons/m$^{3}$. 
Note that these estimates for the spin-imbalance are upper bounds, since we expect that the spin-imbalance generated during the {\it initialization} 
process relaxes with time due to thermal activated electron spin-flipping. Whether or not such a relaxation completely eliminates the 
original spin-imbalance will result from the competition between thermal flipping of the hopping electrons, that works to diminish spin-imbalance, and 
the reverse effect originating from the action of the nanocomposite's magnetization -- see discussion of Section \ref{sec:TODO}.


\Ssubsection{The magnetostatic interaction \hfill\space}

\label{sec:Magnetostatic}


In order to estimate the value of the energy associated with the magnetostatic interaction between two iron-oxide nanoparticles, we neglect all 
the shape and finite size effects playing a role in the interaction between two nanoparticles. We consider, in first order approximation, that 
this interaction can be well described by the classical dipole-dipole interaction. The energy of such interaction is given by
\begin{eqnarray}
  U_{\textrm{MS}} &=& - \frac{\mu_{0}}{4 \pi} \mathbf{m}_{1} \cdot \bigg( \frac{3 \mathbf{r} (\mathbf{m}_{2} \cdot \mathbf{r})}{r^{5}} 
  - \frac{\mathbf{m}_{2}}{r^{3}} \bigg) ,
\end{eqnarray}
where $\mathbf{m}_{1}$ and $\mathbf{m}_{2}$ stand for the magnetic moment vectors of the two magnetic moments, while $\mathbf{r}$ 
stands for the distance between the two moments. Note that depending on the relative position and orientation of the two magnetic 
moments, this term may favor ferromagnetism or antiferromagnetism. The interaction energy between two dipoles will at most have a 
magnitude of $\vert U_{\textrm{MS}} \vert \leq \mu_{0} \, m_{1} m_{2} / (2 \pi r^{3})$.
If the two magnetic moments have a magnitude of $m_{1,2} \approx 4.0 \, \mu_{B}$, and are at a distance of $r \approx 12$ nm, 
then the magnetostatic interaction energy reads $\vert U_{\textrm{MS}} \vert \approx 1.6 \times 10^{-28} J$. This energy is several 
orders of magnitude smaller than the room temperature thermal energy $k_{B} T_{\textrm{room}} \approx 4.1 \times 10^{-21}$ J, and
thus this interaction can be discarded as the origin of the room-temperature ferromagnetic state. This fact is in accordance with the paramagnetic 
behavior of the nanocomposite that is observed when no current is passed across the system. As was stated above, the experimental observations 
strongly suggest that it is the spin-imbalance in the hopping electrons' population that is driving the nanocomposite's ferromagnetic transition.


\Ssubsection{The indirect exchange interaction \hfill\space}

\label{sec:IndirectExchange}


To estimate the energy scale of the indirect exchange coupling between the magnetic moments of two iron-oxide nanoparticles, $U_{\textrm{ex}}$, 
is tricky since it is not trivial to picture, in a simple manner, an indirect interaction mediated by the hopping electrons. In addition, there 
are several quantities that are relevant for such an estimate that we can hardly extract from experimental measurements. Nevertheless, using
reasonable system's parameters, we conclude that it is indeed possible that the indirect interaction's 
energy scale is considerably bigger than that of the magnetostatic interaction.

As previously stated, an electron moving through the nanocomposite will first visit a nanoparticle and retain information on its magnetic state,
that will later {\it exhibit} to another nanoparticle that it will subsequently visit. The effective coupling between the two nanoparticles is
going to result from the combined effect of the several hopping electrons scattering off both nanoparticles in a given characteristic 
time scale. Both temperature
and disorder are expected to reduce the indirect coupling since they will likely enhance the decoherence of the hopping electron spin state. 

We will assume that both the nanoparticle's magnetic moments and the hopping electrons' spin are Ising moments. Furthermore, we will consider 
that a hopping electron feels a nanoparticle through a Zeeman-like coupling between the electron's spin and the magnetic field generated by the 
nanoparticle. The energy associated with such an interaction reads
\begin{eqnarray}
  E_{i \alpha} &=& - \boldsymbol{\mu}_{i} \cdot \mathbf{B}_{\alpha} = - \sigma_{i} \mu_{B} B_{\alpha} (\mathbf{r}_{i}) ,
\end{eqnarray}
where $\mathbf{B}_{\alpha}(\mathbf{r}_{i})$ stands for the value, at position $\mathbf{r}_{i}$, of the magnetic field generated by the nanoparticle 
indexed by $\alpha$ (located at $\mathbf{r}_{\alpha}$), while $\sigma_{i}=\pm 1$ stands for the hopping electron spin and $\mu_{B}$ is the Bohr magneton.
For the sake of simplicity we consider that the magnetic field generated by the nanoparticle only has support inside the nanoparticle, and
that it can be approximated by the magnetic field of a uniformly magnetized sphere $B_{\alpha}(r \leq R) = J_{0} \mu_{0} 2 m_{\alpha} 
\mu_{B} /(3 V_{nnp})$, where $\mu_{0}$ stands for the vacuum permitivity, $m_{\alpha}$ stands for the absolute value (in Bohr magnetons) of 
the magnetic moment of the 
nanoparticle positioned at $\mathbf{r}_{\alpha}$, while $V_{nnp}$ is the volume of the nanoparticle. The inclusion of $J_{0}$ accounts for the 
possibility that the scale of the interaction between the electron and the nanoparticle may be stronger than the bare Zeeman coupling.
We can simplistically assume that an electron visiting two nanoparticles with $4 \mu_{B}$ magnetic moments, that are separated by 
$12$ nm, will give rise to an indirect interaction with an average energy scale given by $\langle E \rangle = - J_{0} 4 \mu_{B}^{2} \mu_{0} 
/\big(3 \pi (4 \times 10^{-9})^{3} \big) \approx J_{0} \times 10^{-28}$ J where the last factor of $2/3$ accounts for the average over the 
electron path. Remember that this is the energy of the indirect coupling mediated by a single hopping electron. We need to take all 
the other electrons scattering off the two nanoparticles into account. If we consider that the 
nanoparticles' are thermally flipping with a characteristic time of the order of the nanosecond, and assume that 
the hopping electrons are moving at a typical velocity of $10^{-12}$ nm/s, then, a spin-imbalance of $\Delta n \approx 10^{22}$ 
will correspond to an average of $10^{4}$ electrons scattering off the two nanoparticles during their characteristic flipping time. 
In such a case, the typical energy scale of the indirect exchange interaction would be of the order of $U_{\textrm{ex}} \approx J_{0} \times 
10^{-24}$ J, a value that is considerably bigger than the magnetostatic interaction energy scale. Still, and if $J_{0}$ is of the
order of $10^{3}$, then the $U_{\textrm{ex}}$ becomes comparable to $k_{B} T_{room}$.

\Ssection{Additional theoretical information}

\label{sec:Theory} 

In this section we will show that by considering a model where hopping electrons and nanoparticles interact through a Zeeman-like coupling,
which gives rise to an electron mediated coupling between the nanoparticles magnetic moments, we can qualitatively reproduce the tunability of 
the magnetic properties of the nanocomposite by the manipulation of the knobs that control 
the nanocomposite's hopping electrons spin-imbalance. Starting from a microscopic Hamiltonian and upon integration of the electronic degrees of 
freedom we will end up with an effective Hamiltonian governing the nanoparticles' magnetic moments. This Hamiltonian will contain an exchange 
interaction term that depends on the spin-imbalance of hopping electrons: a stronger spin-imbalance will give rise to a stronger effective coupling 
between the nanoparticles' magnetic moments. Finally, using Monte Carlo simulations, we will show that the variation of the nanocomposite's ordering 
temperature with its spin-imbalance qualitatively mimics the experimental observations of higher nanocomposite's ordering temperature for greater 
spin-imbalance.


\Ssubsection{The model \hfill\space}

\label{sec:TheModel}


Based on the main text's description of the nanocomposite and the device we can idealize the following picture of the system: The 
magnetic nanoparticles can be considered to behave as Heisenberg moments constrained to point around their easy axis. 
The easy axis is randomly oriented and the magnitude of the moments are randomly distributed around an average value of approximately 
$4 \mu_{B}$. The magnetic moments of graphene oxide can also be thought of as Heisenberg moments of smaller magnitude. The hopping 
electrons moving around the nanocomposite through variable range hopping can localize at a variety of sites (both the non-passivated 
$p^{z}$ orbitals of graphene oxide and the iron-oxide nanoparticles) that are randomly distributed in space and energy. Finally, 
we must allow for the possibility that the hopping electrons' population is spin-imbalanced (due to the action of the Cobalt electrodes).


\Ssubsection{Derivation of the effective Hamiltonian \hfill\space}

\label{sec:TheEffectiveHamiltonian}


In this section, starting from a general {\it microscopic} Hamiltonian for the ensemble of hopping electrons and nanoparticles' magnetic moments, 
we will integrate the electronic degrees of freedom so that we end up with an effective Hamiltonian governing the nanoparticles' magnetic moments.
We will see that such integration is going to renormalize both the nanoparticles' kinetic term and the nanoparticle-nanoparticle (magnetostatic)
interaction term. The aim of this calculation is to show that the expression obtained for the indirect exchange interaction is generally 
ferromagnetic and increases with increasing spin-imbalance, which simplistically is tantamount to say that the ordering temperature increases with 
spin-imbalance, qualitatively reproducing the experimental observation. 

We should draw the reader's attention to the fact that the following calculation is going to be done in first approximation by completely 
neglecting all the disorder effects. The disorder effects will be included {\it a posteriori} in a qualitatively manner through an exponential
factor damping the interaction with distance (see sub-Section \ref{sec:DiscussionApprox}). Thus, in what follows we will in fact 
compute the indirect exchange for a system of delocalized electrons in a perfect crystal. However, we must always keep in mind that the problem 
that interests us is that of localized hopping electrons in a strongly disordered system, hence the interpretation of the results from the 
following calculations must be judicious and careful.

We start by writing the partition function for the system composed of hopping electrons and nanoparticles
\begin{eqnarray}
  Z &=& \sum_{\{\alpha\}} \sum_{\{i\}}  e^{-\beta H'[\{i\},\{\alpha\}]} \, , \label{eq:Zinit}
\end{eqnarray} 
where $\{i\}$ ($\{\alpha\}$) indicates a given configuration of the hopping electrons' (nanoparticles') degrees of freedom, while
$\beta = 1 / (k_{B} T)$ and $H'[\{i\},\{\alpha\}] \equiv H'[\{i\},\{\alpha\}]-\sum_{\sigma=\pm1} \mu_{\sigma} N_{\sigma}$ stands for the 
system's Hamiltonian. By integrating the electronic degrees of freedom, $\{i\}$, we will write the system's partition function as
\begin{eqnarray}
  Z &=& \sum_{\{\alpha\}} e^{-\beta H_{\textrm{eff}}[\{\alpha\}]} \, , \label{eq:Zfin}
\end{eqnarray}
where $H_{\textrm{eff}}[\{\alpha\}]$ is the effective Hamiltonian governing the nanoparticles.

As pointed out in the main text, we can write the effective microscopic Hamiltonian as $\hat{H} = \hat{H}_{e}^{0} + \hat{H}_{M}^{0} + \hat{H}_{e-e} 
+ \hat{H}_{M-M} + \hat{H}_{e-M}$, where the $\hat{H}_{e}^{0}$ ($\hat{H}_{M}^{0}$) stands for the part governing free electrons (constrained Heisenberg moments), 
$\hat{H}_{e-e}$ ($\hat{H}_{M-M}$) stands for the part 
containing electron-electron interactions (dipole-dipole interactions), while $\hat{H}_{e-M}$ stands for the part containing the Zeeman interaction 
between hopping electrons and constrained Heisenberg moments. In what follows we put aside 
the terms $\hat{H}_{M-M}^{0}$ and $\hat{H}_{M}^{0}$, not relevant to the following computations, and, for the sake of simplicity disregard the term 
$\hat{H}_{e-e}$. Therefore, from now onward we will thus work with the following Hamiltonian $\hat{H} = \hat{H}_{e}^{0} + \hat{H}_{e-M}$.
The hopping electron's kinetic term can be written as
\begin{eqnarray}
  \hat{H}_{e}^{0} &=& - \sum_{i j} \sum_{\sigma} \big( \gamma_{i j} \hat{c}_{i \sigma}^{\dagger} \hat{c}_{j \sigma} + \textrm{h.c.} \big) \, .
  \label{eq:HamElecKinSimp}
\end{eqnarray}
where the $\hat{c}_{i \sigma}$ stands for the (fermionic) operator annihilating an hopping electron with spin $\sigma$ that is siting at position 
$\mathbf{r}_{i}$. The $\gamma_{i j}$ stands for the hopping amplitude for an electron to hop to the position $\mathbf{r}_{i}$. Note that as a first
approach we do not allow the hopping electrons to flip their spin when hopping between sites, i. e. we impose $\gamma_{i j}^{\sigma' \sigma} \longrightarrow
\gamma_{i j}$. The hopping amplitude is given by the usual phonon assisted variable range hopping law\cite{refSupp}{Mott_PhilM:1969-S} $\gamma_{i j}^{\sigma' \sigma}
= \exp\big[- r_{i j}/\xi - \beta \, \Delta E \big]$, where $\xi$ stands for a characteristic hopping length, while $\Delta E = E_{i}^{\sigma'} - 
E_{j}^{\sigma}$ is the difference between the energy of the two states and $1/\beta = k_{B} T$ stands for the thermal energy. In 3D this gives rise 
to a temperature dependence of resistance with $T^{1/4}$, i. e. $R(T) = R_{0} \exp\big[ (T_{M}/T)^{1/4} \big]$, where $T_{M}$ is usually named as Mott 
temperature -- see Supplementary Fig. \ref{fig:TEMmicrograph-Transport}(b).

While moving around the nanocomposite the hopping electrons feel the local magnetic fields generated by the nanoparticles. For simplicity, 
we model such an effect by a local Zeeman-like interaction as follows
\begin{eqnarray}
  \hat{H}_{e-M} &=& - \sum_{i \alpha} J_{0} \mu_{0} \hat{\mathbf{S}}_{i} \cdot \hat{\mathbf{M}}_{\alpha} \delta(\mathbf{\mathbf{r}_{i} - \mathbf{r}_{\alpha}}) \, , 
\label{eq:HeM}
\end{eqnarray}
where $\mu_{0}$ stands for the vacuum permeability, $\hat{\mathbf{S}}_{i}$ stands for the angular momentum operator of the hopping electron siting 
at position $\mathbf{r}_{i}$, while $\hat{\mathbf{M}}_{\alpha}$ stands for the angular momentum operator of the nanoparticle siting at position 
$\mathbf{r}_{\alpha}$. The delta function enforces the local character of this interaction. The factor $J_{0}$ stands for the scale of the coupling 
between the hopping electron and the nanoparticle.
When writing such an interaction we are assuming the nanoparticle to be a point-like dipole whose magnetic field vanishes everywhere around itself 
except at the exact position where it is sitting. This corresponds to a first order approximation where all the finite size effects are ignored,
which should preserve the general qualitative behavior of the system.

The typical magnitude of the nanoparticles' magnetic moment (between $3$ and $5$ $\mu_{B}$) justifies regarding 
them as classical moments. For simplicity, we consider the nanoparticles' magnetic moments to be Ising (aligned 
along $\mathbf{e}_{z}$). In doing such we believe that, despite the mathematical simplifications, the system's qualitative behavior 
is going to be preserved so that we can still investigate the influence that manipulating the hopping electrons' spin-imbalance has 
on the system's magnetic ordering. We can indeed consider a more realistic model where the nanoparticles' 
moments (that are constrained around their randomly oriented easy axis) are assumed to be Ising moments with a randomly aligned easy 
axis (fluctuations around the nanoparticles' ground states are neglected). In such a case, a generalization of the following 
calculation can still be done (see end of Section \ref{sec:DiscussionApprox} for details) and it is found that the indirect 
exchange coupling depends not only on the distance between nanoparticles and
on the spin-imbalance, but also on the angle of misalignment between the two nanoparticles' moments. Similarly to what happens for
the model considered below (where Ising moments are all aligned along $\mathbf{e}_{z}$), we find that, in average, at typical 
inter-nanoparticle distances (i. e., $r \approx 12$ nm) the indirect exchange coupling is ferromagnetic and increases with 
spin-imbalance. Moreover, it oscillates and decays with inter-nanoparticle distance.

The magnetic moment of the nanoparticle at $\mathbf{r}_{\alpha}$ is going to be identified as $\mathbf{M}_{\alpha} = \mu_{B} m_{\alpha} 
\lambda_{\alpha} \mathbf{e}_{z}$, where $\mu_{B}$ stands for the Bohr magneton, $m_{\alpha}$ gives
the magnetic moment's magnitude in terms of Bohr magnetons, while $\lambda_{\alpha}=\pm1$ defines the orientation of the moment. We can express the 
(Ising) spin operator of a hopping electron sitting at $\mathbf{r}_{i}$, as $\hat{S}^{z}_{i} = \mu_{B} \sum_{\sigma' \sigma} \hat{c}_{i \sigma_{i}}^{\dagger} 
[\sigma^{z}]_{\sigma_{i} \sigma_{i}'} \hat{c}_{i \sigma_{i}'} = \mu_{B} \sigma_{i} \hat{c}_{i \sigma_{i}}^{\dagger} \hat{c}_{i \sigma_{i}}$, where $\sigma^{z}$ stands 
for the corresponding Pauli matrix, while $\sigma_{i}=\pm1$ indicates the spin orientation of the electron sitting at $\mathbf{r}_{i}$. Eq. 
(\ref{eq:HeM}) can thus be written as
\begin{eqnarray}
  \hat{H}_{e-M} &=& - J_{0} \mu_{0} \mu_{B}^{2} \sum_{i \alpha} \sum_{\sigma_{i}} m_{\alpha} \sigma_{i} \lambda_{\alpha} \delta(\mathbf{\mathbf{r}_{i} 
  - \mathbf{r}_{\alpha}}) t_{\eta_{i} \sigma_{i}}^{(\alpha)} \hat{c}_{i \eta_{i}}^{\dagger} \hat{c}_{i \sigma_{i}} \, , 
  \label{eq:HeM-simp-spinflip}
\end{eqnarray}
where by introducing the factor $t_{\eta_{i} \sigma_{i}}^{(\alpha)}$, we allow the hopping electrons to flip their spin when they interact with the 
nanoparticles. This factor stands for the probability amplitude for the electron sitting at the site $\mathbf{r}_{i} = \mathbf{r}_{\alpha}$ to 
have its spin flipped from $\sigma_{i}$ into $\eta_{i}$ when interacting with the nanoparticle at position $\mathbf{r}_{\alpha}$. The introduction of 
an interaction term not conserving the total angular momentum (electron's $+$ nanoparticle's) can be justified with the fact that angular momentum
can be exchanged with the disordered ensemble of components of the nanocomposite surrounding the nanoparticle (graphene oxide, oxide molecules, spacer,
etc.). It is natural to expect that the spin-flip amplitudes will depend on the orientation and magnitude of the magnetic moment of the nanoparticle. 
We thus include a superscript $\alpha$ in $t_{\eta_{i} \sigma_{i}}^{(\alpha)}$.

The full Hamiltonian then reads
\begin{eqnarray}
  \hat{H} &=& - \sum_{i j} \sum_{\sigma_{i}, \sigma_{j}} \big( \gamma_{i j} \delta_{\sigma_{i} \sigma_{j}} \hat{c}_{i \sigma_{i}}^{\dagger} \hat{c}_{j \sigma_{j}} 
  + \textrm{h.c.} \big) 
  - J_{0} \mu_{0} \mu_{B}^{2} \sum_{i \alpha} \sum_{\sigma_{i}} m_{\alpha} \sigma_{i} \lambda_{\alpha} 
  \delta(\mathbf{\mathbf{r}_{i} - \mathbf{r}_{\alpha}}) t_{\eta_{i} \sigma_{i}}^{(\alpha)} \hat{c}_{i \eta_{i}}^{\dagger} \hat{c}_{i \sigma_{i}} \, .
  \label{eq:HeM-full-simp}
\end{eqnarray} 

We now integrate the hopping electrons' degrees of freedom identified by the operators $\hat{c}_{i \sigma_{i}}$, allowing for electronic spin 
imbalances through $\hat{H}' \equiv \hat{H} - \sum_{\sigma} \mu_{\sigma} \hat{N}_{\sigma}$, where $\hat{N}_{\sigma} \equiv \sum_{i} 
\hat{c}_{i \sigma}^{\dagger} \hat{c}_{i \sigma}$. 
From Eqs. (\ref{eq:Zinit}) and (\ref{eq:Zfin}) we can write $Z_{e}^{(\alpha)} = \exp\big[-\beta H_{\textrm{eff}}^{(\alpha)}\big]$, where $Z_{e}^{(\alpha)}$ 
is the partition function for the hopping electrons in a particular (quenched) landscape of the nanoparticles' magnetic 
moments. We can write this quenched partition function in path integral formalism using Grassmann variables $Z_{e}^{(\alpha)} 
= \int \mathcal{D}(\bar{\psi},\psi) \exp\big(-S[\bar{\psi},\psi;\alpha]\big)$
where $S[\bar{\psi},\psi;\alpha]$ stands for the action in a given quenched landscape of nanoparticles' magnetic moments. The $\mathcal{D}(\bar{\psi},
\psi)$ stands for the measure of the path integral, $\mathcal{D}(\bar{\psi},\psi) = \textrm{lim}_{N\to\infty} \prod_{n=1}^{N} d(\bar{\psi}^{n},\psi^{n})$, 
while the Grassmann fields must satisfy the boundary conditions given by $\psi(0) = -\psi(\beta)$ and $\bar{\psi}(0) = -\bar{\psi}(\beta)$. 
By substituting the fields by their time Fourier transform we can express the action in the frequency representation as $S[\bar{\psi},\psi;\alpha] 
=  \bar{\Psi} \mathbb{A} \Psi$, where we have used $\Psi = [\Psi^{(1)}, \Psi^{(2)}, \ldots]^{T}$ and $\Psi^{(n)} = \big[\psi_{1 +}^{(n)}, 
\psi_{1 -}^{(n)}, \psi_{2 +}^{(n)}, \psi_{2 +}^{(n)}, \ldots\big]$, as well as $\mathbb{A} = \textrm{diag}[\mathbb{A}^{{(1)}},\mathbb{A}^{{(2)}},\ldots]$
with
\begin{eqnarray}
  \mathbb{A}_{i \sigma_{i} , j \sigma_{j}}^{{(n)}} &=& \Big[ \delta_{i j} 
  \delta_{\sigma_{i} \sigma_{j}} (-i w_{n} - \mu_{\sigma_{j}}) + \gamma_{i j} \delta_{\sigma_{i} \sigma_{j}} + \delta_{i j} \sum_{\alpha} 
  V_{\alpha} \sigma_{j} \lambda_{\alpha} t_{\sigma_{i} \sigma_{j}}^{(\alpha)} \delta(\mathbf{r}_{i}-\mathbf{r}_{\alpha}) \Big] \, ,
\end{eqnarray}
where $V_{\alpha} = J_{0} \mu_{0} \mu_{B}^{2} m_{\alpha}$ and $\mu_{\sigma}$ stands for the chemical potential associated with the spin state $\sigma=\pm1$,
while the Matsubara frequencies $w_{n} = (2 n + 1) \pi / \beta$ are determined by the boundary conditions for the Grassmann fields,
with $n \in \mathbb{Z}$.

Using the result for multidimensional Grassmann gaussian integrals $\int \mathcal{D}(\bar{\psi},\psi) e^{-\Psi^{T} \mathbb{A} \Psi} = \det{\mathbb{A}}$,
we conclude that the hopping electrons' (quenched) partition function $Z_{e}^{(\alpha)}$ can be restated as
\begin{eqnarray}
  Z_{e}^{(\alpha)} &=& \prod_{w_{n}} \det\big[\mathbb{A}^{(n)}\big] = \exp\bigg[\sum_{w_{n}} \log\Big(\det\big[\mathbb{A}^{(n)}\big]\Big)\bigg] 
= \exp\bigg[\sum_{w_{n}} \textrm{Tr}\Big[\log\big(\mathbb{A}^{(n)}\big)\Big]\bigg] \, . \label{eq:ZeSimp1}
\end{eqnarray}

We can write $\mathbb{A}^{(n)}$ as $\mathbb{A}^{(n)} = - \big[G^{(n)}\big]^{-1} + V$, where $\big[G^{(n)}\big]^{-1}$ stands for the inverse 
propagator of the free hopping electrons' (with Matsubara frequency $w_{n}$) and $V$ for the interaction between them and the nanoparticles. 
Expanding the logarithm in the exponential of Eq. (\ref{eq:ZeSimp1}), $\log\big(\mathbb{A}^{(n)}\big) = \log\big(- \big[G^{(n)}\big]^{-1} \big) 
- \big[ G^{(n)} V + G^{(n)} V G^{(n)} V / 2 + \ldots \big]$ we can rewrite the effective Hamiltonian as
\begin{eqnarray}
  \beta H_{\textrm{eff}}^{(\alpha)} &=& \sum_{w_{n}} \Bigg[ \textrm{Tr}\Big[\log\big(-\big[G^{(n)}\big]^{-1}\big)\Big] - \textrm{Tr}\big[G^{(n)} V\big] - 
  \textrm{Tr}\bigg[ \frac{G^{(n)} V G^{(n)} V}{2} \bigg] + \ldots \Bigg] \, , \label{eq:firstEffH}
\end{eqnarray}
where we should remember that the dependence on the nanoparticles' configuration $\{\alpha\}$ of the hopping electrons' partition function
$Z_{e}^{(\alpha)}$ is contained on the interaction terms between the nanoparticles and the hopping electrons, $V = V(\{\alpha\})$. In this expansion
we are only going to be interested in the first and second order terms, since only these terms renormalize the original Hamiltonian's terms,
$\hat{H}_{M}^{0}$ and $\hat{H}_{M-M}$.

A further simplification of the problem can be done if we both ignore the variable range hopping nature of the electronic dynamics and at the same 
time consider that the electrons move on a perfect 3-dimensional cubic lattice. We can substitute the variable range hopping dynamics by a 
first-neighbor tight-binding one (and thus $\gamma_{i j} = \gamma$ if $i$ and $j$ are first neighbors). Note that by ignoring the variable range hopping 
nature of the electronic dynamics and the strong (position and energy) disorder of the sites at which the electrons can sit, we are fundamentally 
changing the hopping electrons' nature from strongly localized to strongly delocalized. After such a consideration the results of the computations 
below must be carefully interpreted. This simplification is equivalent to consider an average over disorder, where the well defined wave-vectors 
appearing in the following calculations are only meaningful in the scope of the disorder free (i. e. disorder averaged) problem. In order for the
results that follow to be more meaningful physically, in the end we are going to substitute the Fermi wave-vectors, $k_{F}^{\sigma}$, by the average 
value of the density of hopping electrons with spin $\sigma$, i. e. $n_{\sigma}$. Furthermore, and as referred before, we are going to qualitatively 
account for the strong disorder effects {\it a posteriori} by including an exponentially damping 
factor\cite{refSupp}{DeGennes_JPR:1962-S,Lerner_PRB:1993-S,Sobota_PRB:2007-S} (see sub-Section \ref{sec:DiscussionApprox}) in the indirect exchange parameter 
computed for the clean system.

Fourier transforming the fields to momentum space diagonalizes the propagator into
\begin{subequations} \label{eq:PropVk}
\begin{eqnarray}
  G_{\mathbf{k} \sigma , \mathbf{k}' \sigma'}^{(n)} &=& \frac{1}{- (i w_{n} + \mu_{\sigma'}) + E(\mathbf{k}')} \delta_{\sigma \sigma'} \delta(\mathbf{k}-\mathbf{k}') 
  \, . \label{eq:DiagonalProp}
\end{eqnarray}
In this same basis, the interaction term reads
\begin{eqnarray}
  \big[V(\{\alpha\})\big]_{\mathbf{k} \sigma , \mathbf{k}' \sigma'} &=& \sum_{\alpha} V_{\alpha} \sigma' \lambda_{\alpha} 
  t_{\sigma \sigma'}^{(\alpha)} \frac{e^{i (\mathbf{k}'-\mathbf{k}) \cdot \mathbf{r}_{\alpha}}}{\Omega} \, , \label{eq:kVterm}
\end{eqnarray}
\end{subequations}
where $\Omega$ stands for the volume of the crystal (with periodic boundary conditions). Note that the hopping electrons do not conserve 
their spin when interacting with the nanoparticles, as expected after Eq. (\ref{eq:HeM-simp-spinflip}). Their linear momentum $\mathbf{k}$
is also not conserved by these interactions. In the interaction 
term there is no momentum conservation because we are working in a mixed representation, where the hopping electrons' (that move 
around the cubic lattice) second-quantized operators are represented in momentum space, while the nanoparticles ({\it frozen} at fixed 
positions in space) are kept in real space. 

The zeroth order term of the effective Hamiltonian in Eq. (\ref{eq:firstEffH}) is just a constant that does not depend on the nanoparticles' 
configuration. As usually it can be eliminated by a shift of the position of the energy zero. The first order term of the effective 
Hamiltonian [see Eq. (\ref{eq:firstEffH})] is given by $\big[H_{\textrm{eff}}^{(\alpha)}\big]^{(1)} = - \frac{1}{\beta} \sum_{w_{n}} 
\textrm{Tr}\big[G^{(n)} V\big]$, or else
\begin{eqnarray}
  \Big[H_{\textrm{eff}}^{(\alpha)}\Big]^{(1)} &=& - \frac{1}{\beta} \sum_{w_{n}} \sum_{\mathbf{k}} \sum_{\sigma} \big[G^{(n)}\big]_{\mathbf{k} \sigma , 
   \mathbf{k} \sigma} \big[ V(\{\alpha\}) \big]_{\mathbf{k} \sigma , \mathbf{k} \sigma} \, , \label{eq:FirstOrderH}
\end{eqnarray} 
where we have used the fact that the propagator $G^{(n)}$ is diagonal on both the electron's momentum and spin [see Eq. (\ref{eq:DiagonalProp})].
Again using the diagonal character of the propagator we can write the second order term of the effective Hamiltonian 
[see Eq. (\ref{eq:firstEffH})] as $\big[H_{\textrm{eff}}^{(\alpha)}\big]^{(2)} = - \frac{1}{2 \beta} \sum_{w_{n}} \textrm{Tr}\big[ G^{(n)} 
V G^{(n)} V \big]$, or else
\begin{eqnarray}
  \Big[H_{\textrm{eff}}^{(\alpha)}\Big]^{(2)} &=& - \frac{1}{2 \beta} \sum_{w_{n}}  \sum_{\mathbf{k}, \mathbf{p}} \sum_{\sigma , \eta} 
  \big[G^{(n)}\big]_{\mathbf{k} \sigma , \mathbf{k} \sigma} \big[ V(\{\alpha\}) \big]_{\mathbf{k} \sigma , \mathbf{p} \eta} 
  \big[G^{(n)}\big]_{\mathbf{p} \eta , \mathbf{p} \eta} \big[ V(\{\alpha\}) \big]_{\mathbf{p} \eta , \mathbf{k} \sigma} \, . \label{eq:SecondOrderH}
\end{eqnarray}


\Ssubsubsection{First order term of the effective Hamiltonian \hfill\space}

\label{sec:FirstOrderH}


We start by substituting Eqs. (\ref{eq:PropVk}) on the first order term of the effective Hamiltonian in Eq. (\ref{eq:FirstOrderH}),
and then we do the 
sum over the Matsubara frequencies using contour integration in the complex plane, which gives rise to the Fermi-Dirac distribution. 
Doing the integration in $\mathbf{k}$ allows us to write the first order term as
\begin{eqnarray}
  \Big[H_{\textrm{eff}}^{(\alpha)}\Big]^{(1)} &=& - J_{0} \mu_{0} \mu_{B}^{2} \sum_{\alpha} \big( t_{+ +}^{(\alpha)} n_{+} 
  - t_{- -}^{(\alpha)} n_{-} \big) m_{\alpha} \lambda_{\alpha} \, . \label{eq:H1gen}
\end{eqnarray}
where $n_{+}$ and $n_{-}$ stand for the average density of hopping electrons with spin $\sigma=+1$ and $\sigma=-1$. We have also explicitly 
substituted $V_{\alpha}$ by $V_{\alpha} = J_{0} \mu_{0} \mu_{B}^{2} m_{\alpha}$.

This term can be interpreted as the response of the nanoparticles' (Ising) magnetic moment to an effective magnetic field generated by the 
spin-imbalance of the hopping electron gas. Remember that the superscript $\alpha$ in $t_{\sigma \sigma}^{(\alpha)}$, the probability amplitude 
for an hopping electron with spin $\sigma$ to have its spin conserved when interacting with the nanoparticle at position $\mathbf{r}_{\alpha}$, 
indicates that this interaction depends on the state of the nanoparticle. Therefore, and depending on the choice we make for the amplitudes
$t_{\eta \sigma}^{(\alpha)}$, we may have different behaviors arising from Eq. (\ref{eq:H1gen}) -- see Section \ref{sec:DiscussionApprox}.


\Ssubsubsection{Second order term of the effective Hamiltonian \hfill\space}

\label{sec:SecondOrderH}


Again using Eqs. (\ref{eq:PropVk}) we can rewrite the second order term of the effective Hamiltonian [see Eq. (\ref{eq:SecondOrderH})] as
\begin{eqnarray}
  \Big[H_{\textrm{eff}}^{(\alpha)}\Big]^{(2)} = - \frac{1}{2 \beta} \sum_{\alpha, \beta} \Bigg[ \sum_{\sigma , \eta} &\Bigg(& \sum_{\mathbf{k}, \mathbf{p}} \bigg[ 
  \sum_{w_{n}} \frac{1}{- (i w_{n} + \mu_{\sigma}) + E(\mathbf{k})} \frac{1}{- (i w_{n} + \mu_{\eta}) + E(\mathbf{p})} \bigg] . \nonumber \\ 
  && . e^{i (\mathbf{p}-\mathbf{k}) \cdot (\mathbf{r}_{\alpha} - \mathbf{r}_{\beta})} \Bigg) \eta \sigma  
  \frac{t_{\sigma \eta}^{(\alpha)} t_{\eta \sigma}^{(\beta)}}{\Omega^{2}} \Bigg] V_{\alpha} V_{\beta} \lambda_{\alpha} \lambda_{\beta} \, . \label{eq:SecondOrderH2}
\end{eqnarray}
The sum over the Matsubara frequencies can again be computed using contour integration in the complex plane. It results in the following expression
\begin{eqnarray}
  \sum_{w_{n}} \frac{1}{- (i w_{n} + \mu_{\sigma}) + E(\mathbf{k})} \frac{1}{- (i w_{n} + \mu_{\eta}) + E(\mathbf{p})} &=& 
  - \beta \frac{f_{\sigma}(\mathbf{k}) - f_{\eta}(\mathbf{p})}{E(\mathbf{k})-E(\mathbf{p})-\Delta\mu_{\sigma \eta}} \, ,
\end{eqnarray}
where $\Delta \mu_{\sigma \eta} \equiv \mu_{\sigma} - \mu_{\eta}$ and $f_{\sigma}(\mathbf{k})$ stands for the Fermi-Dirac distribution function of hopping 
electrons with spin $\sigma$, i. e. $f_{\sigma}(\mathbf{k}) \equiv 1 / \big[ \exp [\beta (E(\mathbf{k}) - \mu_{\sigma} ) ] + 1 \big]$.
Therefore, the second order term of the effective Hamiltonian in Eq. (\ref{eq:SecondOrderH2}) can now be written as an exchange interaction
between two different nanoparticles' Ising magnetic moments
\begin{eqnarray}
  \Big[H_{\textrm{eff}}^{(\alpha)}\Big]^{(2)} &=& - \sum_{\alpha, \beta} J(r_{\alpha \beta},k_{F}^{+},k_{F}^{-},\lambda_{\alpha},\lambda_{\beta}) 
  M_{\alpha} M_{\beta} \, , \label{eq:H2nonfinala}
\end{eqnarray}
where $M_{\alpha} = \mu_{B} m_{\alpha} \lambda_{\alpha}$ stands for the magnetic moment indexed by $\alpha$ (in Bohr magnetons) aligned along 
$\lambda_{\alpha, \beta} = \pm1$, while the exchange parameter reads 
\begin{eqnarray}
  J(r_{\alpha \beta},k_{F}^{+},k_{F}^{-},\lambda_{\alpha},\lambda_{\beta}) &=& \frac{J_{0}^{2} \mu_{0}^{2} \mu_{B}^{2}}{2 \Omega^{2}} 
  \sum_{\sigma,\eta=\pm1} \Bigg[ \eta \sigma t_{\sigma \eta}^{(\alpha)} t_{\eta \sigma}^{(\beta)} \sum_{\mathbf{k}, \mathbf{p}} \Bigg( - \frac{f_{\sigma}(\mathbf{k}) 
  - f_{\eta}(\mathbf{p})}{E(\mathbf{k})-E(\mathbf{p})-\Delta\mu_{\sigma \eta}} e^{i (\mathbf{p}-\mathbf{k}) \cdot \mathbf{r}_{\alpha \beta}} \Bigg) \Bigg] 
  \, . \nonumber \\ \label{eq:H2nonfinal}
\end{eqnarray}
The above exchange parameter depends on the orientation of the two nanoparticles through the hopping electron flipping amplitudes 
$t_{\sigma \eta}^{(\alpha)}$ and $t_{\eta \sigma}^{(\beta)}$ and thus we can see it as matrix 
exchange parameter $J_{\lambda_{\alpha} \lambda_{\beta}}$. All the difficulty of computing the exchange parameter is contained on the evaluation of the 
sums in $\mathbf{k}$ and in $\mathbf{p}$. The expression for the exchange parameter in Eq. (\ref{eq:H2nonfinal}) is a sum of four terms arising 
from the different combinations of $\sigma=\pm1$ and $\eta=\pm1$. For $\eta=\sigma$, i.e., if there is no spin-flipping of the hopping electrons 
when they interact with the nanoparticles, we end up with the typical RKKY result\cite{refSupp}{Ruderman_PR:1954-S,Kasuya_PTP:1956-S,Yosida_PR:1957-S}. 
However, if $\eta=-\sigma$, then we have new terms coming directly from the spin-flip of the electron when it interacts with the 
nanoparticles\cite{refSupp}{Fullenbaum_PR:1967-S}.

Let us start by computing the sum in $\mathbf{k}$ and $\mathbf{p}$ when $\eta=\sigma$. In such a case, $\Delta\mu_{\sigma \eta} = 0$ and the double 
sum simplifies into
\begin{eqnarray}
  I_{\sigma \sigma}(r_{\alpha \beta}) &\equiv& - \sum_{\mathbf{k}, \mathbf{p}} \frac{f_{\sigma}(\mathbf{k}) - f_{\sigma}(\mathbf{p})}{E(\mathbf{k})-E(\mathbf{p})} 
  e^{i (\mathbf{p}-\mathbf{k}) \cdot \mathbf{r}_{\alpha \beta}} \, , \label{eq:Iss}
\end{eqnarray}
which we show in Appendix \ref{app:ssCase}, is equal to
\begin{eqnarray}
I_{\sigma \sigma}(r) &=& \frac{m^{*} \Omega^{2}}{2 (2\pi)^{3} \hbar^{2}} \,\frac{\sin (2 k_{F}^{\sigma} r) - 2 k_{F}^{\sigma} r \cos(2 k_{F}^{\sigma} r)}{r^{4}} \, ,
\label{eq:Isss2}
\end{eqnarray} 
where we have assumed both that we are at zero temperature and that the hopping electrons behave as a free electron gas with effective mass $m^{*}$
and Fermi wave-vector $k_{F}^{\sigma}$ (see sub-Section \ref{sec:DiscussionApprox} for a discussion of these assumptions).

Similarly, if we compute the sum in $\mathbf{k}$ and $\mathbf{p}$ when $\eta=-\sigma$,
\begin{eqnarray}
  I_{\sigma,-\sigma}(r_{\alpha \beta}) &\equiv& - \sum_{\mathbf{k}, \mathbf{p}} \frac{f_{\sigma}(\mathbf{k}) - f_{-\sigma}(\mathbf{p})}{E(\mathbf{k})-E(\mathbf{p}) - \Delta \mu_{\sigma,-\sigma}} 
  e^{i (\mathbf{p}-\mathbf{k}) \cdot \mathbf{r}_{\alpha \beta}} \, , \label{eq:Isms}
\end{eqnarray}
we conclude that (see Appendix \ref{app:smsCase}) it reads
\begin{eqnarray}
  I_{\sigma,-\sigma}(r) &=& \frac{m^{*} \Omega}{4 (2 \pi)^{3} \hbar^2} \bigg[
    \big( {k_{F}^{\sigma}}^{2} - {k_{F}^{-\sigma}}^{2} \big)^{2} \Big( \textrm{sinI}\big[(k_{F}^{\sigma} + k_{F}^{-\sigma}) r\big] + \textrm{sinI}\big[(k_{F}^{\sigma} 
    - k_{F}^{-\sigma}) r\big] \Big) \nonumber \\
    &+& \sum_{\lambda=\pm1} \lambda \big( k_F^{\sigma} - \lambda k_F^{-\sigma} \big)^{2} r^{2} \frac{\sin\big[\vert k_{F}^{\sigma} + \lambda k_{F}^{-\sigma} \vert r\big] 
      + \vert k_{F}^{\sigma} + \lambda k_{F}^{-\sigma} \vert \, r \cos\big[(k_{F}^{\sigma} + \lambda k_{F}^{-\sigma}) r\big]}{r^{4}} \nonumber \\  
    &+& 2 \sum_{\lambda=\pm1} \lambda \frac{\sin\big[\vert k_{F}^{\sigma} + \lambda k_{F}^{-\sigma} \vert r\big] - \vert k_{F}^{\sigma} + \lambda k_{F}^{-\sigma} \vert \, r 
    \cos\big[(k_{F}^{\sigma} + \lambda k_{F}^{-\sigma}) r\big]}{r^{4}} \bigg] \, , \label{eq:Isms2}
\end{eqnarray}
where $\textrm{sinI}[x]$ stands for the sine integral function of $x$. Again we have assumed to be at zero temperature and the hopping electrons behave 
as a free electron gas with effective mass $m^{*}$ and Fermi wave-vector $k_{F}^{\sigma}$ (see sub-Section \ref{sec:DiscussionApprox}).

For simplicity we consider that the hopping electron's spin-flip amplitudes at the nanoparticle positioned at $\mathbf{r}_{\alpha}$ do not
depend on the orientation of the nanoparticle, i.e. $t_{\sigma \eta}^{(\alpha)} = t_{\sigma \eta} \,\, \forall \alpha$. Moreover, we consider the
reasonable case where the amplitude for an electron to conserve its spin, $t_{+ +} = t_{- -} = A$, and the amplitude for it to have its spin 
flipped, $t_{- +} = t_{+ -} = B$, are similar, with spin-conservation being slightly more probable than spin-flip, $A \gtrsim B$ (see sub-Section 
\ref{sec:DiscussionApprox}). In this case we have $J_{+ +} = J_{+ -} = J_{- +} =J_{- -} \equiv J$. Finally, if we use the free electron gas 
relation $k_{F}^{\sigma} = \sqrt[3]{6 \pi^{2} n_{\sigma}}$ to express the 
exchange parameter in terms of the average density of hopping electrons in the $\sigma$ spin state, $n_{\sigma}$, then we obtain the result 
stated in Eqs. (3)-(8) of the main text.

In main text's Fig. 4(a) we plot both the indirect exchange parameter for the case where hopping electrons cannot have their spin flipped 
[given by main text's Eqs. (3) when $A\neq0$ and $B=0$] and the indirect exchange parameter for the case where they can have their spin 
flipped when interacting with the nanoparticles [given by main text's Eqs. (3) with $A=1$ and $B=0.96$]. In contrast with the case without 
hopping electrons' spin-flips, when we allow the hopping electrons to have their spin flipped, the exchange parameter acquires a local minimum 
at zero spin-imbalance generally having a maximum at (or near) the maximum possible spin-imbalance. This is in accordance with the experimental
observations, where the ordering temperature (and thus the ferromagnetic coupling) increases with spin-imbalance.
\begin{figure}[!htp]
  \centering
  \includegraphics[width=.55\textwidth]{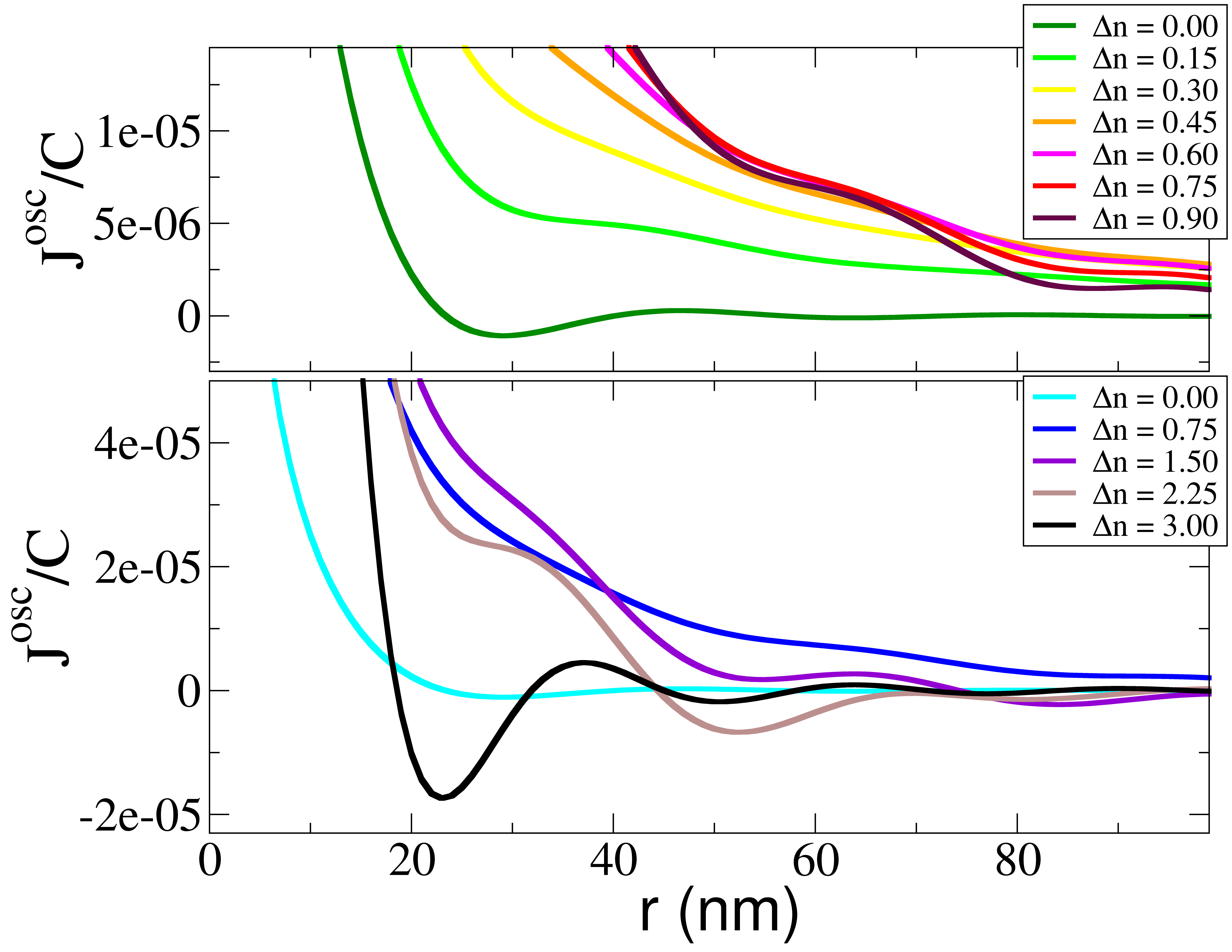}
  \caption{Plot of the $J(r, n_{+},n_{-})$ [given by main text's
    Eq. (3) with $A=1$ and $B=0.96$] in terms of the distance between
    nanoparticles $r$. We plot the $J(r, n_{+},n_{-})$ for several
    spin-imbalances $\Delta n = n_{+}-n_{-}$ (expressed in units of
    $10^{22}$ m$^{-3}$). In the upper panel we plot curves in the
    range $\Delta n \in [0.0,0.9] \times 10^{22}$ m$^{-3}$, while in
    the lower panel we plot curves in the range $\Delta n \in
    [0.0,3.0] \times 10^{22}$ m$^{-3}$. Note that the periodicity of
    the oscillation of $J$ depends on the spin-imbalance. For example,
    if the spin-imbalance has a value such that $\Delta n \in
    [0.15,0.9] \times 10^{22}$ m$^{-3}$, then the indirect exchange
    remains ferromagnetic for at least $r\gtrsim100$ nm. The constant
    $C$ dividing $J^{\textrm{osc}}$ reads $C = J_{0} \mu_{0}^{2}
    \mu_{B}^{2} m^{*} / (4 (2 \pi)^{3} \hbar^2)$.}
  \label{fig:Jrkky}
\end{figure}

In Supplementary Fig. \ref{fig:Jrkky} we can see the dependence of the exchange parameter [given by main text's Eqs. (3)-(8) with $A=1.$ and 
$B=0.96$] in the distance $r$, for several spin-imbalances. From it we conclude that the exchange parameter period of oscillation is also
strongly dependent on the spin-imbalance. Moreover, for some values of the spin imbalance (namely, $\Delta n \in [0.15,0.9] \times 10^{22}$ 
m$^{-3}$) the exchange coupling remains ferromagnetic for distances greater than $100$ nm.     

As previously referred, in the above calculation we have formally considered the hopping electrons to be delocalized electrons instead of
localized, i. e. the system's strong disorder was neglected. Nevertheless, we can qualitatively account for the average disorder effects 
(see sub-Section \ref{sec:DiscussionApprox}) by including an exponential damping and thus modifying the indirect exchange parameter computed 
in main text's Eq. (3) into 
\begin{eqnarray}
  J(r, n_{+},n_{-}) &\longrightarrow& J(r, n_{+},n_{-}) e^{-r/\xi} \, , \label{eq:JrkkyDamped}
\end{eqnarray}
where in the metallic case, $\xi$ is the electron's mean free path. The stronger the disorder, the stronger the damping of $J(r)$. The strong
disorder in our system will render $\xi$ to be small and thus the exponential suppression essentially kills all longer ranged interactions. 
Only those with distances comparable with the first-neighbor separation will be relevant. In the comparison with the 
experiment we take $\xi$ to be a fitting parameter comparable to the spacing between the nanoparticles.


\Ssubsubsection{Discussion of the approximations employed in the above calculations \hfill\space}

\label{sec:DiscussionApprox}


Several approximations were done to arrive at the result for the electron mediated exchange coupling term written 
in main text's Eq. (3): we have assumed the interaction between the hopping electrons and the nanoparticles to be 
local; the variable range hopping character of the hopping electrons was neglected; the effects arising from the 
system's strong disorder were initially ignored and then qualitatively reintroduced later leading to Eq. 
(\ref{eq:JrkkyDamped}); in computing the $\mathbf{k}, \mathbf{q}$-sums in Eq. (\ref{eq:H2nonfinal}) we have both 
considered the hopping electrons spectrum to be that of a free electron gas, and the system to be at zero temperature; 
we have made a particular choice of the electronic spin-flip amplitudes $t_{\sigma \eta}^{(\alpha)}$; and we have considered
the nanoparticles' magnetic moments to be parallel oriented Ising moments. Several of these approximations were already
discussed above, and thus in the following paragraphs we are going to comment on those not yet discussed.

As referred above, to ignore the strong (position and energy) disorder of the sites at which the electrons can sit, amounts to change the 
hopping electrons' nature from strongly localized to strongly delocalized. This simplification is equivalent to consider an average over 
disorder and to assume that the electrons move on a perfect (cubic) lattice. The natural way of describing such a system is in terms of 
well defined wave-vectors. However, these are only meaningful in the scope of a disorder free problem. Since the system we study is strongly 
disordered, we opt by expressing the effective Hamiltonian (computed after the disorder average approximation) in terms of the average density 
of hopping electrons with spin $\sigma=\pm1$, $n_{\sigma}$, than in terms of Fermi wave-vectors, $k_{F}^{\sigma}$.

In addition, Eq. (\ref{eq:JrkkyDamped}) rests upon the works of De Gennes\cite{refSupp}{DeGennes_JPR:1962-S}, Lerner\cite{refSupp}{Lerner_PRB:1993-S} and
Sobota et al.\cite{refSupp}{Sobota_PRB:2007-S}. De Gennes\cite{refSupp}{DeGennes_JPR:1962-S} has shown that, in a weakly disordered metal (i. e., with a random scalar potential)
the average indirect RKKY exchange interaction is exponentially damped at distances greater than the electron mean free path. Using field theoretical 
techniques Lerner\cite{refSupp}{Lerner_PRB:1993-S} has shown that, in the strong disorder limit, despite the fact that the average RKKY interaction is still 
exponentially damped at distances greater than the electron mean free path, the magnitude of the actual interaction is strongly dependent on the disorder 
configuration. In fact it is better characterized by a broad log-normal distribution that indicates that fluctuations that are considerably larger 
than the typical value of the interaction can indeed occur. More recently, in a numerical study, Sobota et al.\cite{refSupp}{Sobota_PRB:2007-S} found out that
strong disorder and localization give rise to a RKKY interaction with a distribution function that develops a strongly non-Gaussian form with long tails. 
They found out that the typical value of the interaction is better characterized by the geometric average of this distribution and that this average
is exponentially suppressed in the presence of strong localization. For the sake of simplicity, we account for the influence of strong disorder in 
our system by including an exponentially damping factor on the indirect exchange parameter computed for the clean system.

When computing the sums over momentum in Eq. (\ref{eq:FirstOrderH}) and Eq. (\ref{eq:SecondOrderH}) we have assumed for simplicity that the
energy dispersion of the hopping electrons moving on the perfect cubic lattice (under a first-neighbor tight-binding model) was that of a free 
electron gas (with spin $\sigma$) with an effective mass $m^{*}$, $E_{\sigma}(\mathbf{k})=\hbar^{2} \big(\mathbf{k}^{\sigma}\big)^{2}/(2 m^{*})$. 
It is acceptable to do such an approximation for electrons on a $3$D lattice if the system's Fermi level is at the {\it bottom} of the 
co-sinusoidal band, i.e. if $k_{F}^{\sigma} a \ll 1$, where $a$ is the cubic 
lattice spacing. Despite the fact that we have no good way of estimating a lattice spacing for the (idealized) cubic lattice, we can consider 
that a value of the order of the nanometer is reasonable. Within the free electron gas approximation and from the estimate of the density of 
hopping electrons with spin $\sigma$ (see sub-Section \ref{sec:Density-MMmagnitude-SpinImb}), we find that $k_{F}^{\sigma} \lesssim 0.12$ nm$^{-1}$,
which indicates that such an approximation is acceptable. We have however no simple way to determine the free electron gas effective mass 
from the experimental data. 

The zero-temperature approximation employed in the computation of the sums in Eq. (\ref{eq:FirstOrderH}) and Eq. (\ref{eq:SecondOrderH}), 
greatly simplifies the sum argument by eliminating the Fermi-Dirac factor while imposing a cut-off on the sum (see 
Appendices \ref{app:ssCase} and \ref{app:smsCase}), but it is only reasonable when $k_{B} T \ll E_{F} = \hbar^{2} k_{F}^{2} / (2 m^{*})$. In our case, 
the condition $k_{B} T \ll E_{F}$ holds if the free electron gas effective mass, $m^{*}$, is at least two orders of magnitude smaller 
than the bare electron mass (since $T = 300$ K and $k_{F} \approx 0.12$ nm$^{-1}$). Kim et al.\cite{refSupp}{Kim_PRB:1995-S,Kim_PRB:1999-S} have computed
the RKKY interaction (without spin-flips) for finite temperature and found out that it decreases both the magnitude of the interaction and 
its period of oscillation between positive and negative values. To extend this computation to the case where electron spin-flips are present
would deserve a publication of its own. However, we do not think that temperature is going to make the $J(r)$ 
not clearly ferromagnetic for first-neighbor distances, since in Supplementary Fig. \ref{fig:Jrkky} we can see that there are big ranges of spin-imbalances 
(for example, $0.15 \times 10^{22} \lesssim n_{+} - n_{-} \lesssim 1.50 \times 10^{22}$ electrons/m$^{3}$) for which the exchange coupling is 
ferromagnetic (i. e. $J(r)>0$) up to distances between nanoparticles of $\gtrsim 70$ nm.

Remember that in order to write main text's Eq. (3), we have both considered that the hopping electrons' spin-flip amplitudes at different
nanoparticles are all equal, and that the amplitude for an electron to conserve its spin, $t_{+ +} = t_{- -} = A$, is slightly bigger than the amplitude 
for it to have its spin flipped, $t_{- +} = t_{+ -} = B$. This simplification is equivalent to consider that the spin-flip amplitudes are decoupled 
from the nanoparticles at which they occur, namely, from their magnetic moment's magnitude and orientation. This can be regarded as a first order 
approximation to the problem that should give the global tendency of the system's magnetic behavior.

Finally, let us briefly comment on the reasonability of considering a model where the nanoparticles' magnetic moments are 
parallel oriented Ising moments. Recall that the nanoparticles' magnetic moments can be viewed as Heisenberg moments
that are constrained to point around their randomly oriented easy axis. We can ignore fluctuations around their ground 
states and thus assume them to be Ising moments aligned (parallel or anti-parallel) with their randomly oriented easy axis.
In such a case, a generalization of the previous calculation implies that we start by modifying the model's microscopic 
Hamiltonian in Eqs. (\ref{eq:HamElecKinSimp})-(\ref{eq:HeM-full-simp}) so that the Ising moments can have their easy 
axis arbitrarily aligned. This amounts to consider all the terms in the dot product $\hat{\mathbf{S}}_{i} \cdot \mathbf{M}_{\alpha} 
= m_{\alpha} \, \big(\hat{S}^{x}_{i} \, \sin(\theta_{\alpha}) \cos(\gamma_{\alpha}) + \hat{S}^{z}_{i} \, \sin(\theta_{\alpha}) 
\sin(\gamma_{\alpha}) + \hat{S}^{z}_{i} \, \cos(\theta_{\alpha}) \big)$ so that the microscopic Hamiltonian in Eq. 
(\ref{eq:HeM-full-simp}) is modified into
\begin{eqnarray}
  \hat{H} &=& - \sum_{i j} \sum_{\sigma_{i}, \sigma_{j}} \big( \gamma_{i j} \delta_{\sigma_{i} \sigma_{j}} \hat{c}_{i \sigma_{i}}^{\dagger} 
  \hat{c}_{j \sigma_{j}} + \textrm{h.c.} \big) - J_{0} \mu_{0} \mu_{B}^{2} \sum_{i \alpha} \sum_{\sigma_{i}} m_{\alpha} 
  \delta(\mathbf{\mathbf{r}_{i} - \mathbf{r}_{\alpha}}) \bigg[ \Big( \sigma_{i} \, \cos(\theta_{\alpha}) \, t_{\sigma_{i} \sigma_{i}}^{(\alpha)} 
  \nonumber \\ && + \sin(\theta_{\alpha}) \, e^{-i \sigma_{i} \gamma_{\alpha}} \, t_{-\sigma_{i} \sigma_{i}}^{(\alpha)} 
  \Big) \, \hat{c}_{i \sigma_{i}}^{\dagger} \hat{c}_{i \sigma_{i}} + \Big( \sin(\theta_{\alpha}) \, e^{i \sigma_{i} \gamma_{\alpha}} \, 
  t_{\sigma_{i} \sigma_{i}}^{(\alpha)} - \sigma_{i} \, \cos(\theta_{\alpha}) \, t_{-\sigma_{i} \sigma_{i}}^{(\alpha)} \Big) \, 
  \hat{c}_{i, -\sigma_{i}}^{\dagger} \hat{c}_{i \sigma_{i}} \bigg] \, , \nonumber \\ \label{eq:HeM-full-simp-mod}
\end{eqnarray} 
where $\mu_{0}$ stands for the vacuum permeability, $\mu_{B}$ stands for the Bohr magneton, $m_{\alpha}$ gives the magnetic 
moment's (sitting at $\mathbf{r}_{\alpha}$) magnitude in terms of Bohr magnetons, while the angles $\theta_{\alpha} \in [0,\pi]$ 
and $\gamma_{\alpha} \in [0, 2 \pi]$ define the orientation of its Ising moment. The spin operator of a hopping electron 
sitting at $\mathbf{r}_{i}$, was written as $\hat{\mathbf{S}}_{i} = (\hat{S}^{x}_{i},\hat{S}^{y}_{i},\hat{S}^{z}_{i}) = 
\mu_{B} \sum_{\sigma' \sigma} \hat{c}_{i \sigma_{i}}^{\dagger} \big( [\sigma^{x}]_{\sigma_{i} \sigma_{i}'}, \, [\sigma^{y}]_{\sigma_{i} \sigma_{i}'}, 
\, [\sigma^{z}]_{\sigma_{i} \sigma_{i}'} \big) \hat{c}_{i \sigma_{i}'} = \mu_{B} \sigma_{i} \hat{c}_{i \sigma_{i}}^{\dagger} \hat{c}_{i \sigma_{i}}$, 
where $\sigma^{x}$, $\sigma^{y}$ and $\sigma^{z}$ stand for the Pauli matrices, while $\sigma_{i}=\pm1$ indicates the electronic spin 
orientation ($\sigma^{z}$ eigenvalue) of the electron sitting at $\mathbf{r}_{i}$. The factor $t_{\eta_{i} \sigma_{i}}^{(\alpha)}$
stands for a probability amplitude for the hopping electron sitting at site $\mathbf{r}_{i}$ ($= \mathbf{r}_{\alpha}$) to have
its spin flipped from $\sigma_{i}$ into $\eta_{i}$ when interacting with the nanoparticle at $\mathbf{r}_{\alpha}$.

Note that now, due to the presence of $\hat{\sigma}_{i}^{x}$ and $\hat{\sigma}_{i}^{y}$, the effective terms conserving and flipping 
the hopping electron's spin (when these interact with the nanoparticles) are different from those obtained in the case where all 
the nanoparticles' easy axis are parallel -- see Eq. (\ref{eq:HeM-full-simp}). The interaction between the hopping electrons and 
the nanoparticles is now going to also depend on the orientation of the nanoparticle's magnetic moment. It is thus natural to expect 
that the polarization of the electron sea surrounding a nanoparticle is going to be determined by its moment orientation. Therefore, the 
indirect exchange coupling between two nanoparticles is necessarily going to be a function of their orientation.

By integrating the hopping electrons degrees of freedom we end up with a slightly different indirect exchange parameter
between two nanoparticles with magnetic moments given by $\mathbf{M}_{\alpha} = m_{\alpha} \, \big( \sin\theta_{\alpha}
\, \cos\gamma_{\alpha}$, $\sin\theta_{\alpha} \, \sin\gamma_{\alpha},\, \cos\theta_{\alpha} \big)$ and $\mathbf{M}_{\beta} = 
m_{\beta} \, \big( \sin\theta_{\beta} \, \cos\gamma_{\beta},\,\sin\theta_{\beta} \, \sin\gamma_{\beta},\, \cos\theta_{\beta} \big)$.
Such an indirect exchange $J(r,n_{+},n_{-};\gamma_{\alpha},\theta_{\alpha},\gamma_{\beta},\theta_{\alpha})$ will read
\begin{eqnarray}
  J(r,n_{+},n_{-};\gamma_{\alpha},\theta_{\alpha},\gamma_{\beta},\theta_{\alpha}) &=& C \, \sum_{\sigma=\pm} \bigg( 
  \mathbb{X}_{\sigma}^{(\alpha \beta)} \, \mathbb{J}^{(1)}(r,n_{\sigma})
  + \mathbb{W}_{\sigma}^{(\alpha \beta)} \, \mathbb{G}(r,n_{+},n_{-}) \bigg) \, , \label{eq:JRKKY-modified}
\end{eqnarray}
where $C \equiv J_{0}^{2} \mu_{0}^{2} \mu_{B}^{2} m^{*}/(32 \pi^{3} \hbar^2)$, $\mathbb{G}(r,n_{+},n_{-})$ is again given 
by the expression written in main text's Eqs. (4)-(8), while the functions $\mathbb{X}_{\sigma}^{(\alpha \beta)}$ and 
$\mathbb{W}_{\sigma}^{(\alpha \beta)}$ read
\begin{subequations}
  \begin{eqnarray}
    \mathbb{X}_{\sigma}^{(\alpha \beta)} \equiv \mathbb{X}_{\sigma}(\gamma_{\alpha},\theta_{\alpha},\gamma_{\beta},\theta_{\alpha}) 
    &=& \Big[ \sigma \cos\theta_{\alpha} t_{\sigma \sigma}^{(\alpha)} + e^{-i \sigma \gamma_{\alpha}} \sin\theta_{\alpha} t_{-\sigma, \sigma}^{(\alpha)} 
    \Big] \Big[ \sigma 
    \cos\theta_{\beta} t_{\sigma \sigma}^{(\beta)} + e^{-i \sigma \gamma_{\beta}} \sin\theta_{\beta} t_{-\sigma, \sigma}^{(\beta)} \Big] \,, \nonumber \\ \\
    \mathbb{W}_{\sigma}^{(\alpha \beta)} \equiv \mathbb{W}_{\sigma}(\gamma_{\alpha},\theta_{\alpha},\gamma_{\beta},\theta_{\alpha}) 
    &=& \Big[ e^{i \sigma \gamma_{\alpha}} \sin\theta_{\alpha} t_{\sigma \sigma}^{(\alpha)} - \sigma \cos\theta_{\alpha} t_{-\sigma, \sigma}^{(\alpha)} 
    \Big] \Big[ e^{- i \sigma \gamma_{\beta}} 
    \sin\theta_{\beta} t_{-\sigma,-\sigma}^{(\beta)} + \sigma \cos\theta_{\beta} t_{\sigma,-\sigma}^{(\beta)} \Big] \,. \nonumber \\
  \end{eqnarray}  
\end{subequations}

In order to gain some insight on this expression we sampled it for two nanoparticles at a distance of $r = 12$ nm, arbitrary 
orientations and several spin-imbalances. The value of $J(r,n_{+},n_{-};\gamma_{\alpha},\theta_{\alpha}, \gamma_{\beta},\theta_{\alpha})$ 
is naturally dependent on the angle between the two magnetic moments, $\varphi_{\alpha \beta}$, taking values in a given interval 
for a particular $\varphi_{\alpha \beta}$ depending on the orientation of the two Ising moments relatively to the hopping electrons 
spin polarization direction. Averaging the sampled values of $J(r,n_{+},n_{-};\gamma_{\alpha},\theta_{\alpha},\gamma_{\beta},
\theta_{\alpha})$ and plotting them in terms of the angle $\varphi_{\alpha \beta}$ (see Supplementary Fig. \ref{fig:Modified_JJ}) we conclude that 
the average value of $J(r,n_{+},n_{-},\varphi_{\alpha \beta})$ is going to be ferromagnetic at typical inter-nanoparticle distances 
(i. e., $r \approx 12$ nm), with its magnitude increasing with increasing spin-imbalance -- see panel (a) of Supplementary Fig. 
\ref{fig:Modified_JJ}. Moreover, it will decay and oscillate with the inter-nanoparticle distance -- see panel (b) of Supplementary Fig. 
\ref{fig:Modified_JJ}.
\begin{figure}[htp!]
  \centering
  \includegraphics[width=0.48\textwidth]{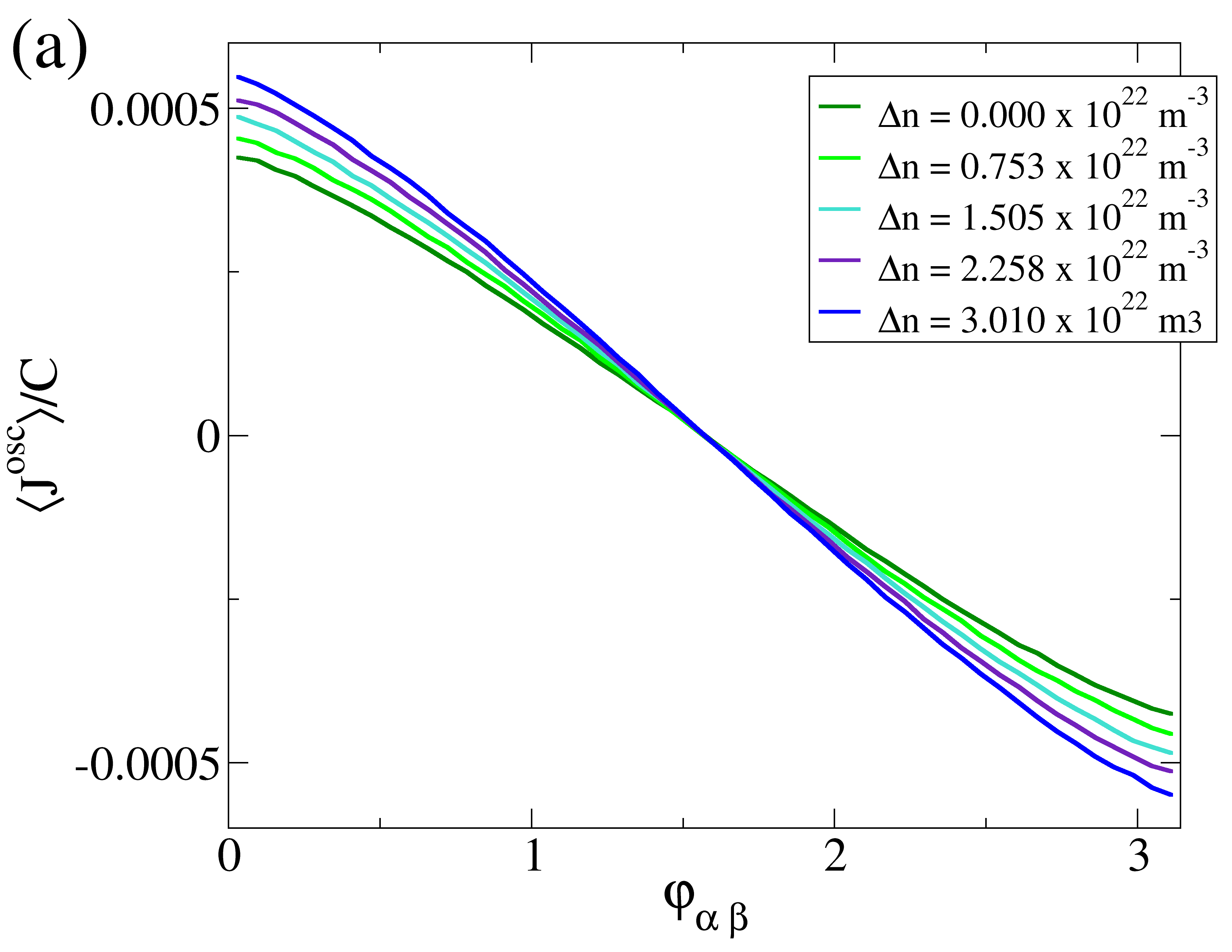}
  \includegraphics[width=0.48\textwidth]{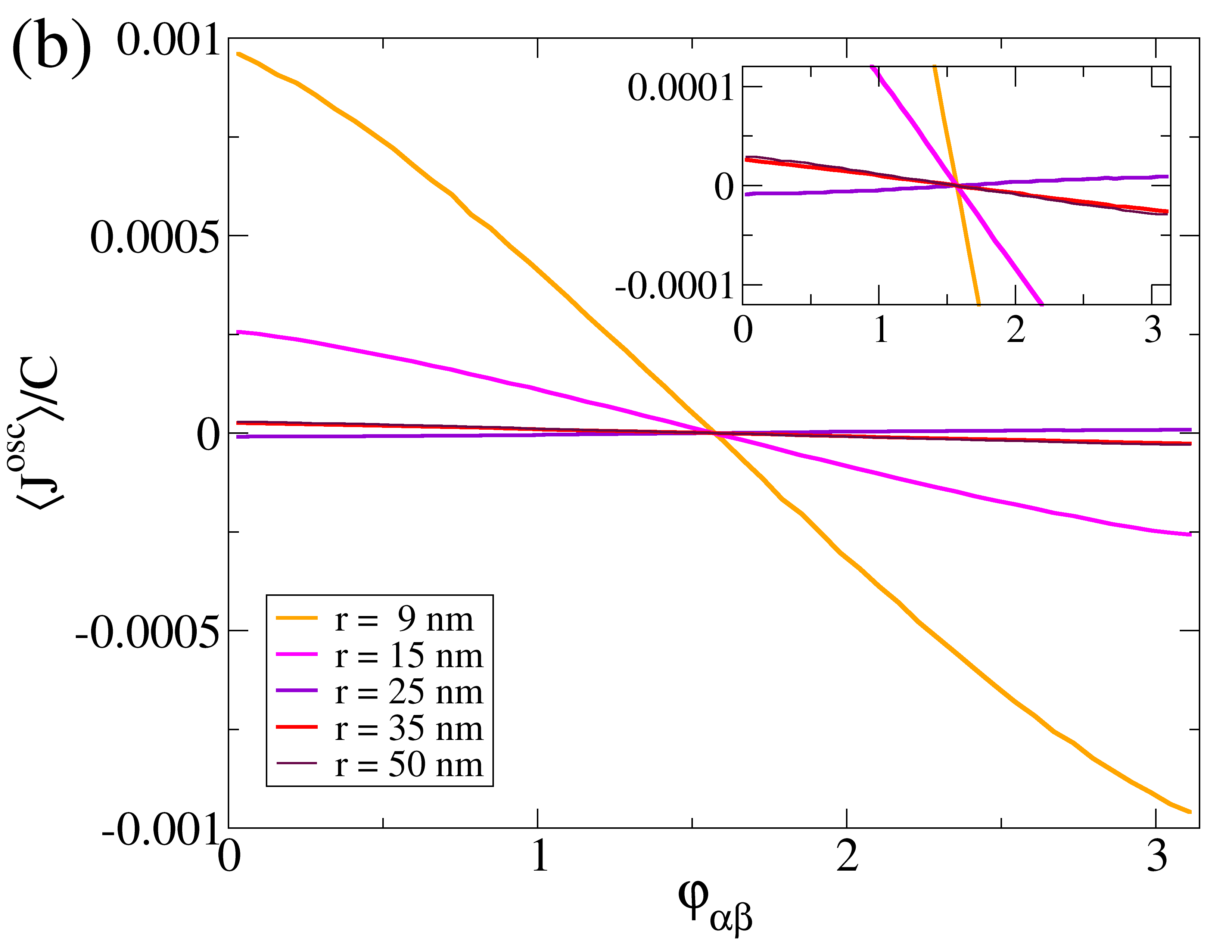}
  \caption{{\bf Average indirect exchange between non-parallel Ising
      moments.} Plot of the oscillating part of the average indirect
    exchange [i. e., the expression in Eq. (\ref{eq:JRKKY-modified})]
    in terms of the angle between the two nanoparticles. {\bf (a)}
    Comparison between the average of the oscillating part of $J$
    between two nanoparticles at $r=12$ nm for different
    spin-imbalances. {\bf (b)} Comparison between the average of the
    oscillating part of $J$ between two nanoparticles (under a
    spin-imbalance of $\Delta n = 2.258 \times 10^{22}$ m$^{-3}$) at
    different distances. The inset is a zoom in around $\langle
    J^{osc} \rangle / C \approx 0$.}
  \label{fig:Modified_JJ}
\end{figure}
This is a strong indication that such a system will qualitatively behave in a very similar manner to that of the model considered
in the main text (parallel Ising moments), thus justifying that simpler approach over this one.


\Ssubsection{Monte Carlo simulations \hfill\space}

\label{sec:MonteCarlo}


Let us start by pointing out that the experimental results strongly suggest that the first order term of main text's
Eq. (2) plays a secondary role on the genesis of the magnetic state of the nanocomposite. The fact that the left panel 
of Supplementary Fig. \ref{fig:RemMag-vs-Init} shows that the nanocomposite's hysteretic response to an external magnetic field 
depends on the initialization field (i. e., on the spin-imbalance), clearly demonstrates that the second order term 
dominates over the first order one. Note that the first order term can be seen as a local effective magnetic field 
proportional to the spin-imbalance of hopping electrons surrounding a nanoparticle. If, fixing the temperature, the 
first order term was the dominant one, then the two hysteretic curves (obtained from systems initialized with different 
magnetic fields, i. e., different spin-imbalances) would only differ on their initial part, collapsing into each other 
thereafter. In a system of independent nanoparticles' magnetic moments, the hopping electrons' spin-imbalance should 
be tightly linked to the system's magnetization (see supplementary information's Section III). Therefore, whenever 
the nanocomposite's magnetization is saturated by a sufficiently strong magnetic field, the spin-imbalance should also 
increase to a maximum (saturated) value. Accordingly, if two such systems (each one of them sustaining distinct 
spin-imbalances at an initial time) were subjected to a strong magnetic field saturating their magnetization, 
then their spin-imbalances would be brought to similar values and from then onward they would present a similar 
hysteresis. This does not happen if the system is controlled by the second order term, since its local spin-imbalance 
will be preserved by the intrinsic magnetization (originating from the coupling between nanoparticles) of the nanocomposite's 
magnetic domains.

To have a first order term that is negligible when compared with the second order one [multiplied by 
the exponential damping factor -- see main text's Eq. (2)] is perfectly 
compatible with the theoretical model. Despite the fact that these two terms are obtained after expanding on the coupling 
constant [see Eq. (\ref{eq:HeM-full-simp})] and integrating the electronic degrees of freedom, their relative magnitude
is determined by the external parameters: spin-imbalance, $n_{+}-n_{-}$; distance between the interacting nanoparticles, $r$; 
magnitude of the nanoparticles' magnetic moments; strength of the exponential suppression of the RKKY interaction due to 
disorder, $\xi$; scale of the hopping electron-nanoparticle interaction, $J_{0}$; effective mass parameter of the model, 
$m^{*}$; electron spin-flip amplitudes $A$ and $B$. Interestingly, whenever the chosen parameters are compatible with the 
logarithm series expansion [see Eq. (\ref{eq:firstEffH})], i. e., when they fulfill the condition $\gamma_{i j} \gg J_{0} 
\mu_{0} \mu_{B}^{2} m_{\alpha} t_{\eta_{i} \sigma_{i}}^{(\alpha)}$, for all $i,j,\alpha,\eta_{i},\sigma_{i}$, there is always a region 
of the external parameter space for which the first order term is irrelevant when compared with the second order one.

Therefore, from here onward we will consider that the indirect exchange term dominates the system's physics. With the intent 
to check if such model holds as a suitable explanation for the tunability of the nanocomposite's ordering temperature upon 
manipulation of its spin-imbalance (see main text's Fig. 2), we have performed Monte Carlo simulations of the following model: 
randomly positioned Ising moments in $3$ dimensions; their magnitudes are random and uniformly distributed around $4$ $\mu_{B}$; 
the moments interact via an exchange coupling as depicted in Eq. (\ref{eq:H2nonfinala}), where the coupling parameter is given 
by main text's Eqs. (3)-(8) with $A=1.00$ and $B=0.96$. The free parameters of the system are: the electron gas effective 
mass, $m^{*}$; the characteristic length controlling the exponential damping of the exchange coupling, $\xi$; and the constant 
$J_{0}$ controlling the scale of the electron-nanoparticle interaction. We keep $m^{*}$ at all times equal to 
$100 \, m_{e}$, where $m_{e}$ stands for the bare electron mass, using $\xi$ and $J_{0}$ as fit parameters that will be fixed 
by the requirement that the dependence of the ordering temperature with spin-imbalance obtained from the Monte Carlo simulations 
reproduces the experimental one -- see main text's Fig. 4(c).

Qualitatively, we expect that for values of $\xi$ smaller than the inter-nanoparticle distance ($r\approx 12$ nm) the exponentially damped 
coupling should give rise to the ordering of the system in magnetic clusters that interact weakly between themselves. Upon decreasing 
temperature, the magnetic moments inside each cluster align, with different clusters doing so at slightly distinct temperatures. 
Moreover, as clusters interact weakly, individual clusters will generally have different magnetization directions. As a consequence, the 
system should not in general present long-range order when temperature is decreased below the {\it blocking} temperature $T_{b}$. We have
confirmed this by Monte Carlo simulations using the Cluster algorithm\cite{refSupp}{Wolff_PRL:1989-S}.

However, if we start from an ordered state 
generated, for example, by applying an external magnetic field when the spin-imbalance is being generated (as is done in the experiment), 
then long-range order should be observed since the nearly independent clusters were from the beginning aligned by the external magnetic 
field. We have used Metropolis Monte Carlo algorithm\cite{refSupp}{Metropolis_JChemPHys:1953-S} to investigate this.

The use of Metropolis Monte Carlo at very low temperatures is highly inefficient in exploring the complete phase space of the system.
At low temperatures the single-flip Monte Carlo dynamics cannot escape from the region of the phase space around a local energy minimum. A random
initial configuration will very fast relax to its nearest local energy minimum and get stuck there {\it ad eternum}. This is an extreme form
of critical slowing down that is characteristic of single spin-flip Metropolis dynamics (but that is avoided by the Cluster algorithm dynamics). 

To use Metropolis single-flip dynamics starting from an initial configuration very close to the global energy minimum will thus imply that
at low temperatures the Monte Carlo sweep will only explore the phase space region around the global minimum. Increasing the temperature will
allow the Monte Carlo sweep to explore increasingly larger regions of the phase space. Therefore, Monte Carlo averages (using this dynamics) 
at low temperatures will not result in correct statistical averages (indicative of the existence of spontaneous phase transitions) due to 
the partial exploration of the phase space. However, such averages will nevertheless give a good indication of the phase space accessible to the
system (or else, of the low-temperature system's magnetic state) when a magnetic field is initially applied to the sample (as is done in the 
experiment) ordering most of its independent clusters in the same direction. 

In main text's Fig. 4(b) we demonstrate exactly this: when starting from an ordered state, single-flip Metropolis Monte Carlo 
simulations suggests that a long range ordered state is possible below a given blocking temperature, $T_{b}$. 
Moreover, this blocking temperature is shown to depend on the magnitude of the indirect exchange, that we know is dependent on the 
spin-imbalance of the hopping electrons. In main text's Fig. 4(c) we compare the dependence of $T_{b}$ with the spin-imbalance
measured in the experiment and the dependence arising from the numerical simulations (with $\xi=5$ nm and 
$J_{0} = 2.2 \times10^{6}$).

Finally, note that in these Monte Carlo studies we have investigated the temperature dependence of the sample magnetization (after 
an initial magnetic field was applied to it). We are at all temperatures using the same temperature-independent ($T=0$) expression 
for the indirect exchange coupling. However, as we have previously referred, the exchange coupling should depend on 
temperature -- see discussion of sub-Section \ref{sec:DiscussionApprox}. In particular, as we know that increasing temperature decreases 
the magnitude of the indirect coupling computed at $T=0$,\cite{refSupp}{Kim_PRB:1995-S,Kim_PRB:1999-S} then, if temperature would be
taken into account in computing the indirect exchange of main text's Eqs. (3)-(8), we expect that the numerical results would 
show a less pronounced slope in main text's Fig. 4(c).

\Ssection{Avenues for future investigations}

\label{sec:TODO}

Although beyond the scope of the current work, here we discuss some aspects of this novel system that deserve future investigation, namely
related to the spin-imbalance generation and the magnetization measurements. 

\vspace{0.5cm}

\noindent\textbf{\underline{The spin-imbalance generation}}

\vspace{0.2cm}

The hopping electrons' spin-imbalance is generated by the {\it initialization} process where we use an external magnetic field to drive 
the magnetic orientation of the two cobalt electrodes that control the flow of the electric current across the device (at room temperature). 
We expect the biggest spin-imbalance to occur for the case where the electrodes are perfectly anti-parallel aligned, i. e. at 
$B_{\textrm{ext}} \approx 0$. However, as 
seen in main text's Fig. 2(b), the highest spin-imbalance occurs for an {\it initialization} process done with $B_{\textrm{ext}} \gtrsim 0.04$ T, 
when the second electrode is already partially reversed [see main text's Fig. 3(b)].

An explanation for this observation invokes the combined action of the magnetic field applied during the {\it initialization} process, 
$B_{\textrm{ext}}$, and the (spin-imbalance dependent) indirect exchange interaction between the nanoparticles. When $B_{\textrm{ext}} = 0$ T, the 
electrodes are anti-parallel aligned and, if no other factor would be playing a role, we would expect that the spin-imbalance generated would be 
maximal. However, we must remember that the hopping electrons' are prone to have their spin flipped both in the VRH events and the interactions 
with the system's nanoparticles, which we expect to be an important factor at room temperature. When no magnetic field is being applied, the 
nanoparticles' clusters are oriented in random directions, no global magnetization exists in the system, and then nothing counteracts the thermal 
activated electron spin-flips. As a consequence, the hopping electron's spin-imbalance will progressively vanish in the electrons' path between 
the source and the drain electrode. This would explain the observation of a low capacitance for anti-parallel electrodes and $B_{\textrm{ext}} \approx 
0$ T.

When we {\it initialize} the system with a non-zero $B_{\textrm{ext}}$, the thermal spin-flips of the hopping electrons will be progressively 
counterbalanced by the action of the nanocomposite's magnetization (arising from the combined action of the indirect exchange coupling, that
will magnetize each cluster of nanoparticles, and the external magnetic field, that will progressively orient the different clusters in a 
given direction), since the hopping electrons will rather align along the direction of magnetization of the nanocomposite. The competition 
between the spin-flips and the effect of the nanocomposite's magnetization will thus determine the measured spin-imbalance.

It is thus reasonable to expect that a stronger magnetic field generates a greater spin-imbalance. Note however that stronger magnetic fields also 
lead to electrodes that are not completely anti-parallel, and thus to a smaller {\it potential} spin-imbalance. This explains why the maximal 
spin-imbalance is not at $B_{\textrm{ext}} \lesssim 0.05$ T (immediately before the electrodes become parallel), but instead somewhere in between this 
value and $B_{\textrm{ext}} = 0$ T. 

Despite the fact that this is not central to the understanding of the ferromagnetic transition observed in the system, it is an issue that deserves
further investigations in order to improve the understanding and control of the system.

\vspace{0.5cm}

\noindent\textbf{\underline{The spin-imbalance after initialization}}

\vspace{0.2cm}

It is likely that the capacitance (measured during the {\it initialization} process) is not going to exactly correspond to the actual capacitance 
of the nanocomposite after the {\it initialization} process is terminated and the system relaxes to equilibrium (through thermal activated 
spin-flips of the hopping electrons). In fact, the measured capacitance is going to fix an upper bound on the spin-imbalance of the nanocomposite. 
Therefore, one other relevant question to ask is how the spin-imbalance relaxes to its equilibrium value after the {\it initialization} process 
is finished, i. e. after the {\it initialization} current and magnetic field are turned off. 

We expect that the spin-imbalance equilibrium value originates from a competition between the thermal activated electron spin-flips and the 
nanocomposite's magnetization after the magnetic field is turned off. If the magnetization is zero, the spin-imbalance should completely 
vanish since nothing counterbalances the action of the thermal activated spin-flips of the hopping electrons. However, if the nanocomposite is 
magnetized, an equilibrium situation should be achieved where the action of the thermal activated spin-flips and the counteraction of the 
nanocomposite's magnetization cause the electron's spin-imbalance to relax to a non-zero value. This is one of the reasons for the big error 
bars in the experimental points of main text's Fig. 4(c).

\vspace{0.5cm}

\noindent\textbf{\underline{The role of the initialization magnetic field}}

\vspace{0.2cm}

As referred in sub-Section \ref{sec:MonteCarlo}, the system's strong disorder gives rise to a short-range exponential damping of the electron 
mediated indirect exchange between nanoparticles. This produces an ordering of the nanoparticles in nearly independent magnetic clusters with 
distinct ordering temperatures (that depend on the distances between them and the strength of the exchange coupling), and thus no spontaneous 
long-range magnetic order is expected upon decreasing temperature. However, when an external magnetic field is applied to the system during the 
{\it initialization} process (as done in the experiment), we expect that in average the nearly independent clusters will end up oriented along 
the same direction. Thus, after the {\it initialization} magnetic field is turned off, the magnetic ordering should be preserved as long as the 
coupling between nanoparticles does not vanish, i. e. as long as the spin-imbalance survives.

In trying to test this hypothesis we have found that no ferromagnetism is observed in the nanocomposite if no magnetic field is applied 
to the system when the spin-polarized current is flowing across the device. However, we must stress that this observation is not a definitive 
proof that the system is arranged in magnetic clusters. It can also be due to the two issues discussed just above: either to the fact that in 
the absence of a magnetic field the electrodes' configuration may not be able to generate a sufficiently big spin-imbalance so that the 
nanoparticles become ferromagnetically coupled; or to the fact that no spin-imbalance can be sustained if the nanocomposite is in a zero 
magnetization state (see above), and thus the indirect coupling between nanoparticles rapidly vanishes after the {\it initialization}.

\vspace{0.5cm}

\noindent\textbf{\underline{The magnetization vs. temperature curves}}

\vspace{0.2cm}

We are now going to further comment on the special case of the maroon magnetization curve in main text's Fig. 2(a) of the main text. After what we have 
just said about the conditions that can generate and preserve a non-zero spin-imbalance in the nanocomposite, the maroon curve in main text's
Fig. 2(a) is puzzling: no magnetization is seen at room temperature, but only for $T < 280$ K.

If immediately after the {\it initialization} process is terminated, the generated spin-imbalance is not sufficiently strong to sustain a
macroscopic magnetization (as experimentally observed), then we expect the spin-imbalance to rapidly and completely vanish. If the spin-imbalance 
is nearly zero, then we expect that no macroscopic magnetization would ever be observed upon decreasing the temperature (provided we do not go 
to sufficiently low temperatures as to make the magnetostatic interaction between nanoparticles relevant).

A possible explanation for such an observation ascribes to the electrodes the
responsibility for this phenomenon. Upon the application of a $B_{\textrm{ext}} = 0.02$ T during the {\it initialization}, the electrodes'
configuration does not give rise to a sufficiently big spin-imbalance so as to generate long range order. Therefore, immediately after the
end of the {\it initialization} process, the spin-imbalance vanishes everywhere except around the electrodes. The local magnetic field
produced by the anti-parallel ferromagnetic electrodes generates a spin-imbalance in their close vicinity. If the electrodes are not
perfectly anti-parallel (since the $B_{\textrm{ext}} = 0.02$ T applied during the {\it initialization} partially reversed one of the electrodes),
then it is possible that upon decreasing temperature this residual spin-imbalance gives rise to a sufficiently strong coupling between the
nanoparticles so that a magnetization is observed.

\vspace{0.5cm}

\noindent\textbf{\underline{High temperatures and the magnetization vs. temperature curves}}

\vspace{0.2cm}

Frequently, after recording the system's magnetization from $T=230$ K up to $T=330$ K (where the nanocomposite's magnetization vanishes), if then the
temperature is decreased while the magnetization is being recorded, we see that the magnetization curve is either shifted to lower temperatures 
or it is completely eliminated. Two phenomena may be contributing to this: the enhanced graphene oxide reduction at higher temperatures; and 
the vanishing of any spin-imbalance as the magnetization becomes zero at $T>T_{b}$.

The temperature enhanced graphene oxide reduction will have as a consequence that the hopping electrons will visit the nanoparticles less
often since they have a lot more sites available at the graphene oxide. This is going to make the indirect exchange between nanoparticles weaker, 
which will then lower the ordering temperature. An indication that this phenomena is occurring is the fact that after the increase in temperature 
the system's conductivity increased several fold, indicating that the graphene oxide reduction is relevant.

Also, and likely more important, when temperature is increased above the blocking temperature so that the nanocomposite's demagnetizes, then 
the spin-imbalance rapidly vanishes since no magnetization exists to counterbalance the thermal activated flipping of the hopping electrons'. 
As a consequence, since there is no spin-imbalance and thus the nanoparticles are independent, then no magnetization will be observed upon
temperature decrease.

\vspace{0.5cm}

\noindent\textbf{\underline{Role of graphene oxide and its degree of oxidation}}

\vspace{0.2cm}

As mentioned in the main text, the graphene oxide used in the nanocomposite is highly defective and partially
reduced (between 18\% and 20\%). Studies have been done where the nanocomposite's graphene oxide was substituted 
by other poorly conducting media such as the non-conducting form of Polyaniline\cite{refSupp}{Aigu_APL:2013-S}. In this
system no ferromagnetism was observed. Nanocomposites with completely oxidized (and highly defective) graphene 
oxide were also studied\cite{refSupp}{Aigu_Small:2014-S} and again no ferromagnetism was observed. We thus may speculate 
that the difference between these results and the ones presented here is linked to the additional 
paramagnetic centers present on the highly defective and partially reduced graphene oxide flakes (which are 
absent or less frequent on the other poorly conducting materials tested so far). Likely the hopping electrons 
will also interact with these additional paramagnetic centers and therefore will effectively couple 
them to the iron oxide nanoparticles' magnetic moments. This may favour the percolation of magnetic 
ordering in between magnetic clusters facilitating long-range magnetic order on the nanocomposite.

Note in addition that the experimental observations are rather sensitive to the degree of reduction of the graphene oxide.
As mentioned above, when the device is subjected to moderately high temperatures the graphene oxide flakes are
usually reduced to such an extent that the ferromagnetic state is irreversibly lost. Despite the obvious complications
that such phenomenon poses, it can also be regarded as potentially interesting: the easy tunability of graphene
oxide's degree of reduction (though unidirectional/irreversible) can be used to manipulate the magnitude and nature of the
coupling between magnetic moments of the nanocomposite.

\begin{subappendices}
  \vspace{1.1cm}
\begin{center}
  \noindent {\Large \bf APPENDICES}
\end{center}
  \addappheadtotoc

\appendix

\renewcommand{\theequation}{S.A\arabic{equation}}


\Ssection{Computation of '$I_{\sigma \sigma}(\mathbf{r}_{\alpha \beta})$'}

\label{app:ssCase}


Let us substitute $\mathbf{q} \equiv \mathbf{k} - \mathbf{p}$ in Eq. (\ref{eq:Iss}) and then write the sum in $\mathbf{k}$ as 
\begin{eqnarray}
  \mathcal{I}_{\sigma}(\mathbf{q}) &\equiv& - \frac{\Omega}{(2\pi)^{3}} \int \textrm{d}^{3}\mathbf{k} \frac{f_{\sigma}(\mathbf{k}) 
  - f_{\sigma}(\mathbf{k}-\mathbf{q})}{E(\mathbf{k}) - E(\mathbf{k}-\mathbf{q})}\, ,
\end{eqnarray}
where we have taken the continuum limit of the sum, and thus $\Omega$ stands for the total volume of the crystal (with periodic boundary
conditions).

To simplify this integral we assume that the energy dispersion of the electrons is that of a free electron gas with an effective mass
$m^{*}$, $E(\mathbf{k})=\hbar^{2} \mathbf{k}^{2}/(2 m^{*})$ (see discussion in sub-Section \ref{sec:DiscussionApprox}). It then reads
\begin{eqnarray}
  \mathcal{I}_{\sigma}(\mathbf{q}) &=& \frac{2 m^{*} \Omega}{(2\pi)^{3} \hbar^{2}} \int \textrm{d}^{3}\mathbf{k} \Bigg( 
  \frac{f_{\sigma}(\mathbf{k})}{(\mathbf{k}-\mathbf{q})^{2}-\mathbf{k}^{2}} 
  + \frac{f_{\sigma}(\mathbf{k})}{(\mathbf{k}+\mathbf{q})^{2}-\mathbf{k}^{2}}\Bigg) 
  \equiv \mathcal{J}_{\sigma}(-\mathbf{q}) + \mathcal{J}_{\sigma}(\mathbf{q}) \, .
\end{eqnarray}
Let us start by focusing on the term $\mathcal{J}_{\sigma}(\mathbf{q})$. 
If we assume to be working at zero temperature (see discussion in sub-Section \ref{sec:DiscussionApprox}) we can simplify it into
\begin{eqnarray}
  \mathcal{J}_{\sigma}(\mathbf{q}) &=& \frac{m^{*} \Omega}{q (2\pi)^{2} \hbar^{2}}  \int_{0}^{k_{F}^{\sigma}} \textrm{d} k \, k \int_{-1}^{1} 
  \textrm{d} u \frac{1}{u + \zeta} \, , \label{eq:intka}
\end{eqnarray}
where we have made the substitution $u=\cos\theta$ and used $\zeta \equiv q/(2 k)$. The Cauchy principal value of the integral in $\textrm{d} u$ reads $\log[\vert (1+\zeta)/(-1+\zeta) \vert]$ and thus $\mathcal{J}_{\sigma}(\mathbf{q})$ can be shown to be equal to
\begin{eqnarray}
 \mathcal{J}_{\sigma}(q) &=& \frac{m^{*} \Omega}{(2\pi)^{2} \hbar^{2}} \frac{k_{F}^{\sigma}}{2} \Bigg( 1 
 + \frac{4 {k_{F}^{\sigma}}^2-q^2}{4 q k_{F}^{\sigma}} \log \bigg[ \bigg\vert \frac{2 k_{F}^{\sigma} + q}{2 k_{F}^{\sigma} - q} \bigg\vert \bigg] \Bigg) \, .
\end{eqnarray}
In a similar manner we can easily verify that $\mathcal{J}_{\sigma}(\mathbf{q}) = \mathcal{J}_{\sigma}(-\mathbf{q})$ and then conclude that 
$\mathcal{I}_{\sigma}(\mathbf{q}) = \mathcal{I}_{\sigma}(q) = 2 \mathcal{J}_{\sigma}(q)$.

Let us now compute the sum $\mathbf{q}$ in Eq. (\ref{eq:Iss}). Taking its continuum limit and doing the variable substitution 
$u=\cos \theta$ we obtain
\begin{eqnarray}
  I_{\sigma\sigma}(r) &=& \frac{\Omega}{(2 \pi)^{2}} \int_{0}^{\infty} \textrm{d} q \, \frac{e^{i q r} - e^{-i q r}}{i r} \, q \, \mathcal{I}_{\sigma}(q) \, .
\end{eqnarray}
As $\mathcal{I}_{\sigma}(-q) = \mathcal{I}_{\sigma}(q)$ we can write $\mathcal{I}_{\sigma}(q)$ as
\begin{eqnarray}
  I_{\sigma\sigma}(r) &=& \frac{m^{*} \Omega^{2} k_{F}^{\sigma}}{i r (2\pi)^{4} \hbar^{2}} \int_{-\infty}^{\infty} \textrm{d} q \, e^{i q r} \, q \, 
  \Bigg( 1 + \frac{4 {k_{F}^{\sigma}}^2-q^2}{4 q k_{F}^{\sigma}} \log \bigg[ \bigg\vert \frac{2 k_{F}^{\sigma} + q}{2 k_{F}^{\sigma} - q} \bigg\vert \bigg] \Bigg) \, .
  \label{eq:intk2}
\end{eqnarray}
The analytic structure of the integrand function in Eq. (\ref{eq:intk2}) is sketched in Supplementary Fig. \ref{fig:IntContour-1}(a): it has two branch points 
on the real axis, $q = \pm a$, connected by a branch cut. We choose a contour as sketched in Supplementary Fig. \ref{fig:IntContour-1}(a) to do the integration.
\begin{figure}[!htp]
  \centering
  \includegraphics[width=.40\textwidth]{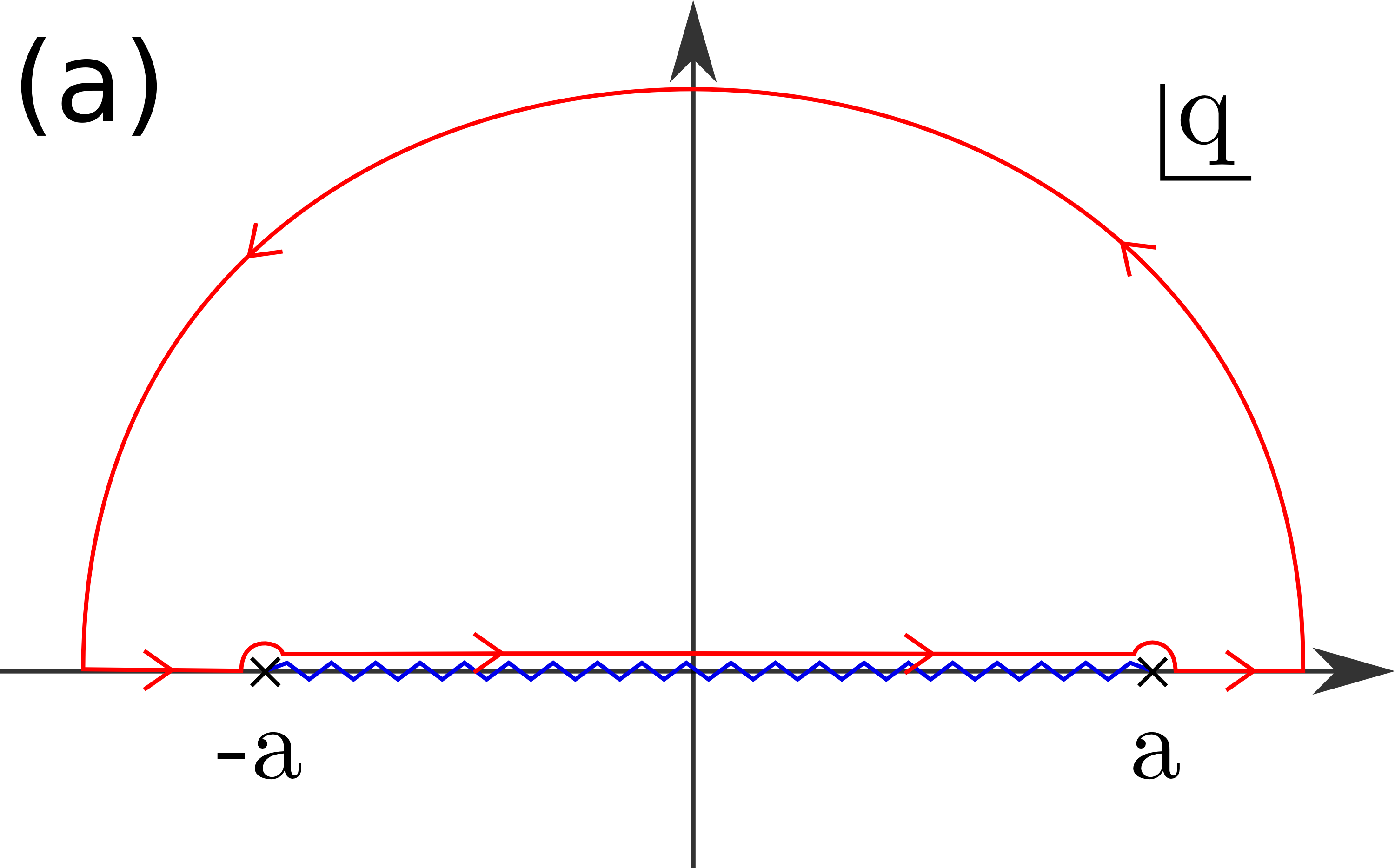}
  \hspace{0.3cm}
  \includegraphics[width=.40\textwidth]{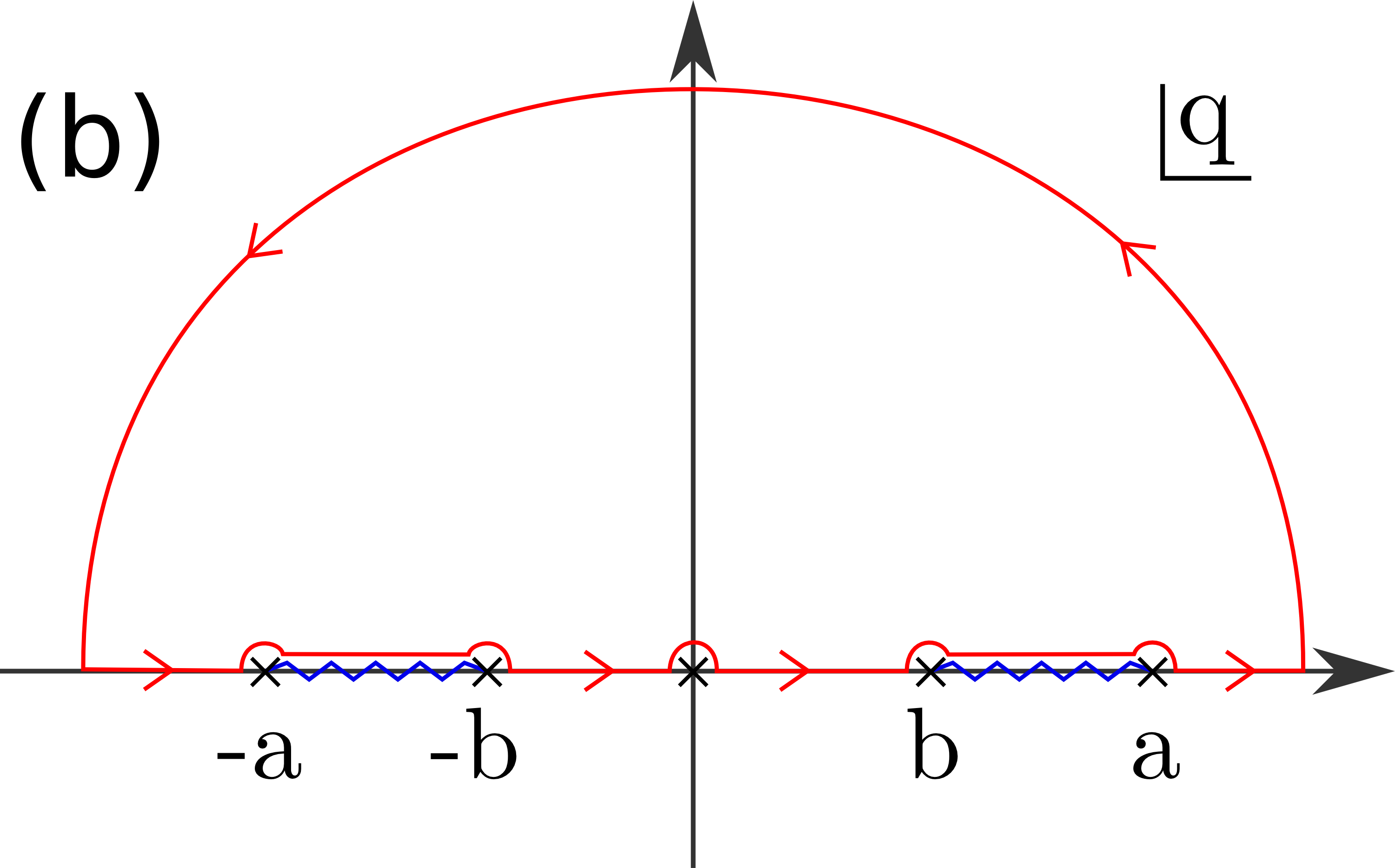}
  \caption{{\bf (a)} Scheme of the analytic structure of the integrand
    function of Eq. (\ref{eq:intk2}), with two branch points $q=\pm a$
    joined by a branch cut. {\bf (b)} Scheme of the analytic structure
    of the integrand function of Eq. (\ref{eq:Bintk2}), with four
    branch points $q=\pm a$ and $q=\pm b$ joined by two branch
    cuts. The integration contour is drawn on red.}
  \label{fig:IntContour-1}
\end{figure}

If we compute a similar integral to that of Eq. (\ref{eq:intk2}), namely an integral where the modulus of the logarithm's argument was dropped,
let us call it $\widetilde{I}_{\sigma\sigma}(r)$, then we can show that it is equal to zero -- the integral over the contour on the upper half 
complex plane vanishes when the contour is taken to infinity; thus the integral over the real axis is equal to zero. As we can write the 
logarithm of a given complex number $z$ as $\log z = log \vert z \vert + i \arg z$, then we can write
\begin{eqnarray}
  \log \bigg[ \frac{2 k_{F}^{\sigma} + q}{2 k_{F}^{\sigma} - q} \bigg] &=& \left \{ \begin{array}{ll} \log \bigg[ \bigg\vert \frac{2 k_{F}^{\sigma} 
  + q}{2 k_{F}^{\sigma} - q} \bigg\vert \bigg] & \textrm{, $\vert q \vert > a$,} \\ \\ \log \bigg[ \bigg\vert \frac{2 k_{F}^{\sigma} + q}{2 k_{F}^{\sigma} 
  - q} \bigg\vert \bigg] + i \pi & \textrm{, $\vert q \vert \leq a$,} \end{array} \right . \, ,
\end{eqnarray}
and thus Eq. (\ref{eq:intk2}) becomes
\begin{eqnarray}
  I_{\sigma\sigma}(r) &=& \frac{m^{*} \Omega^{2}}{8 r (2\pi)^{3} \hbar^{2}} \int_{-a}^{a} \textrm{d} q \, e^{i q r} \, 
  \big( 4 {k_{F}^{\sigma}}^2-q^2 \big) \, . \label{eq:intk3}
\end{eqnarray}
which we can readily compute and obtain the expression in Eq. (\ref{eq:Isss2}).

Note that if we do not allow spin-flips of the hopping electrons, then we will only have the terms with $\zeta=\sigma$ in Eq. 
(\ref{eq:H2nonfinal}). In such a case, the exchange parameter expression will be given by $J(r_{\alpha \beta},k_{F}^{+},k_{F}^{-},
\lambda_{\alpha},\lambda_{\beta}) = J_{0} \mu_{0}^{2} \mu_{B}^{4} / (2 \Omega^{2}) \sum_{\sigma=\pm1}  t_{\sigma \sigma}^{(\alpha)} 
t_{\sigma \sigma}^{(\beta)} I_{\sigma\sigma}(r)$, which is going to result in the usual RKKY 
\cite{refSupp}{Ruderman_PR:1954-S,Kasuya_PTP:1956-S,Yosida_PR:1957-S} expression for the indirect exchange parameter (if the amplitude 
for an hopping electron to have its spin unchanged when interacting with every nanoparticle equal unit, $t_{\sigma \sigma}^{(\alpha)} = 
t_{\sigma \sigma}^{(\beta)}=1$). The two terms for each spin direction arise from the fact that electrons with spin-up and spin-down may have 
different Fermi levels.


\renewcommand{\theequation}{S.B\arabic{equation}}

\Ssection{Computation of '$I_{\sigma, -\sigma}(\mathbf{r}_{\alpha \beta})$'}

\label{app:smsCase}


Upon substituting $\mathbf{q} = \mathbf{k} - \mathbf{p}$ in Eq. (\ref{eq:Isms}), we can write the sum in $\mathbf{k}$ 
(in the continuum limit) as
\begin{eqnarray}
  \widetilde{\mathcal{I}}_{\sigma}(\mathbf{q}) &=& - \frac{\Omega}{(2\pi)^{3}} \int \textrm{d}^{3}\mathbf{k} \frac{f_{\sigma}(\mathbf{k}) 
  - f_{-\sigma}(\mathbf{k}-\mathbf{q})}{E(\mathbf{k}) - E(\mathbf{k}-\mathbf{q}) - \Delta \mu_{\sigma, -\sigma}}\, ,
\end{eqnarray}
where, as before, $\Omega$ stands for the total volume of the crystal (with periodic boundary conditions).

Again we take the free electron gas approximation and define $\Delta_{\sigma} \equiv \Delta \mu_{\sigma,-\sigma} 2 m^{*} / \hbar^{2} 
= \big(k_{F}^{\sigma}\big)^{2}-\big(k_{F}^{-\sigma}\big)^{2}$, such that we can write the integral $\widetilde{\mathcal{I}}_{\sigma}(\mathbf{q})$ as
\begin{eqnarray}
  \widetilde{\mathcal{I}}_{\sigma}(\mathbf{q}) &=& \frac{2 m^{*} \Omega}{(2\pi)^{3} \hbar^{2}} \int \textrm{d}^{3}\mathbf{k} \Bigg( 
  \frac{f_{\sigma}(\mathbf{k})}{(\mathbf{k}-\mathbf{q})^{2}-\mathbf{k}^{2} 
  + \Delta_{\sigma}} + \frac{f_{-\sigma}(\mathbf{k})}{(\mathbf{k}+\mathbf{q})^{2}-\mathbf{k}^{2}+\Delta_{-\sigma}}\Bigg) 
  \equiv \widetilde{\mathcal{J}}_{\sigma}(\mathbf{q}) + \widetilde{\mathcal{G}}_{\sigma}(\mathbf{q}) \, , \nonumber \\
\end{eqnarray}
where $\Delta_{-\sigma} = \big(k_{F}^{-\sigma}\big)^{2} - \big(k_{F}^{\sigma}\big)^{2} = - \Delta_{\sigma}$. And thus one can easily verify that 
$\widetilde{\mathcal{G}}_{\sigma}(\mathbf{q}) = \widetilde{\mathcal{J}}_{-\sigma}(-\mathbf{q})$.

Let us first concentrate on $\widetilde{\mathcal{J}}_{\sigma}(\mathbf{q})$ and again assume that we are at zero temperature (see Appendix 
\ref{app:ssCase}). Making the substitution $u=\cos\theta$ together with the following identification $\xi \equiv (q + \Delta/q)/(2 k)$, 
we find that the Cauchy principal value of the integral in $\textrm{d} u$ reads $\log\big[\big\vert \big(q + \Delta_{\sigma}/q +2 k\big)
\big( q+\Delta_{\sigma}/q -2 k\big) \big\vert \big]$. Computing the remaining integral (in d$k$) we obtain
\begin{eqnarray}
 \widetilde{\mathcal{J}}_{\sigma}(q) &=& \frac{m^{*} \Omega}{(2\pi)^{2} \hbar^{2}} \frac{k_{F}^{\sigma}}{2 q} \bigg( q + \frac{\Delta_{\sigma}}{q} \bigg) \Bigg( 1 
 + \frac{4 {k_{F}^{\sigma}}^2-\big(q + \Delta_{\sigma}/q\big)^2}{4 k_{F}^{\sigma} \big(q + \Delta_{\sigma}/q\big)} \log \bigg[ \bigg\vert \frac{2 k_{F}^{\sigma} 
 + \big(q + \Delta_{\sigma}/q\big)}{2 k_{F}^{\sigma} - \big(q + \Delta_{\sigma}/q\big)} \bigg\vert \bigg] \Bigg) \, . \nonumber \\ \label{eq:BappJ}
\end{eqnarray}
From the above equations it is simple to verify that $\widetilde{\mathcal{J}}_{\sigma}(-\mathbf{q}) = 
\widetilde{\mathcal{J}}_{\sigma}(\mathbf{q}) = \widetilde{\mathcal{J}}_{\sigma}(q)$ and 
$\widetilde{\mathcal{G}}_{\sigma}(\mathbf{q}) = \widetilde{\mathcal{J}}_{-\sigma}(-\mathbf{q})$, and then conclude that the expression 
for $\widetilde{\mathcal{G}}_{\sigma}(\mathbf{q})$ is given by an analogous of Eq. (\ref{eq:BappJ}) where $\sigma \to -\sigma$
everywhere on the LHS.

We can identify $I_{\sigma, -\sigma}(r) = J_{\sigma, -\sigma}(r) + G_{\sigma, -\sigma}(r)$, where $J_{\sigma, -\sigma}(r)$ arises from the Fourier sum in $\mathbf{q}$
of $\widetilde{\mathcal{J}}_{\sigma}(q)$, while $G_{\sigma, -\sigma}(r)$ arises from the Fourier sum in $\mathbf{q}$ of $\widetilde{\mathcal{G}}_{\sigma}(q)$. 
Again, note that $G_{\sigma, -\sigma}(r) = J_{-\sigma, \sigma}(r)$. Taking the continuum limit of the Fourier sum in $\mathbf{q}$ of 
$\widetilde{\mathcal{J}}_{\sigma}(\mathbf{q})$ and doing the variable substitution $u=\cos \theta$ we obtain
\begin{eqnarray}
  J_{\sigma,-\sigma}(r) &=& \frac{m^{*} \Omega^{2} k_{F}^{\sigma}}{2 i r (2\pi)^{4} \hbar^{2}} \int_{-\infty}^{\infty} \textrm{d} q \, e^{i q r} \, \bigg( q 
  + \frac{\Delta_{\sigma}}{q} \bigg) \Bigg( 1 
  + \frac{4 {k_{F}^{\sigma}}^2-\big(q + \Delta_{\sigma}/q\big)^2}{4 k_{F}^{\sigma} \big(q + \Delta_{\sigma}/q\big)} \log \bigg[ \bigg\vert \frac{2 k_{F}^{\sigma} 
  + \big(q + \Delta_{\sigma}/q\big)}{2 k_{F}^{\sigma} - \big(q + \Delta_{\sigma}/q\big)} \bigg\vert \bigg] \Bigg)  \, . \nonumber \\ 
  \label{eq:Bintk2}
\end{eqnarray}
after using $\widetilde{\mathcal{J}}_{\sigma}(-q) = \widetilde{\mathcal{J}}_{\sigma}(q)$ and the expression in Eq. (\ref{eq:BappJ}).

The analytic structure of the integrand function in Eq. (\ref{eq:Bintk2}) is sketched in Supplementary Fig. \ref{fig:IntContour-1}(b): it has four branch points 
$q = \pm a$ and $q = \pm b$ which read $a = k_{F}^{\sigma} + \sqrt{{k_{F}^{\sigma}}^{2}-\Delta_{\sigma}} = k_{F}^{\sigma} + k_{F}^{-\sigma}$ and
$b = k_{F}^{\sigma} - \sqrt{{k_{F}^{\sigma}}^{2}-\Delta_{\sigma}} = k_{F}^{\sigma} - k_{F}^{-\sigma}$.
These four branch points are pairwise connected by a branch cut [see Supplementary Fig. \ref{fig:IntContour-1}(b)]. Additionally, there is a simple pole at $q=0$. 
In order to compute the integral in Eq. (\ref{eq:Bintk2}) we choose a contour as sketched in Supplementary Fig. \ref{fig:IntContour-1}(b) to do the integration.

Again we can make use of the trick used in Appendix \ref{app:ssCase}: instead of computing the integral in Eq. (\ref{eq:Bintk2}), we start by 
computing a similar one, where the modulus of the logarithm's argument was dropped; let us call it $\widetilde{J}_{\sigma,-\sigma}(r)$. Once
more we can show that (the Cauchy principal value of) this integral is equal to zero, since the  integral over the contour on the upper half 
complex plane vanishes when the contour is taken to infinity (as well as the contours around the pole and branch points). From the complex logarithm properties we can write
\begin{eqnarray}
  \log \bigg[ \frac{(q+a) (q+b)}{(q-a) (q-b)} \bigg] &=& \left \{ \begin{array}{ll} \log \bigg[ \bigg\vert \frac{(q+a) (q+b)}{(q-a) (q-b)} 
  \bigg\vert \bigg] & \textrm{, $\vert q \vert > a \land \vert q \vert < b$,} \\ \\ \log \bigg[ \bigg\vert \frac{(q+a) (q+b)}{(q-a) (q-b)} 
  \bigg\vert \bigg] + i \pi & \textrm{, $-a \leq q \leq -b \land b \leq q \leq a$,} \end{array} \right . \, ,
\end{eqnarray}
where we have used the expressions for $a$ and $b$ to rewrite the argument of the logarithm in Eq. (\ref{eq:Bintk2}) in the left hand side of 
the above equation. With this equality we can write $J_{\sigma,-\sigma}(r)$
\begin{eqnarray}
  J_{\sigma,-\sigma}(r) &=& \frac{m^{*} \Omega^{2}}{16 r (2\pi)^{3} \hbar^{2}} \Bigg\{ \int_{-a}^{-b} \textrm{d} q \, e^{i q r} \, 
  \bigg[ 4 \big(k_{F}^{\sigma}\big)^2-\bigg(q + \frac{\Delta_{\sigma}}{q} \bigg)^2 \bigg] + \int_{b}^{a} \textrm{d} q \, e^{i q r} \, 
  \bigg[ 4 \big(k_{F}^{\sigma}\big)^2-\bigg(q + \frac{\Delta_{\sigma}}{q} \bigg)^2 \bigg] \Bigg\} \, , \nonumber \\ \label{eq:Bintk3a} 
\end{eqnarray}
while $G_{\sigma,-\sigma}(r)$ is obtained from substituting $\sigma \to - \sigma$ on the LHS of Eq. (\ref{eq:Bintk3a}).

Finally, we can show that computing the above integrals and summing the $J_{\sigma,-\sigma}(r)$ and the $G_{\sigma,-\sigma}(r)$ [remember that 
$I_{\sigma,-\sigma}(r) = J_{\sigma,-\sigma}(r) + G_{\sigma,-\sigma}(r)$] we obtain the result stated in Eq. (\ref{eq:Isms2}). From it is straightforward to 
verify that $I_{-\sigma,\sigma}(r) = I_{\sigma,-\sigma}(r)$, as it should be since $G_{\sigma,-\sigma}(r)=J_{-\sigma,\sigma}(r)$ and $I_{\sigma,-\sigma}(r) = 
J_{\sigma,-\sigma}(r)+G_{\sigma,-\sigma}(r)=J_{\sigma,-\sigma}(r) + J_{-\sigma,\sigma}(r)$.

\end{subappendices}



\end{document}